\documentclass[11pt,a4paper]{article}
\usepackage[margin=26mm]{geometry}
\usepackage{amsmath,amssymb,amsthm,mathtools,bm,booktabs}
\usepackage{ascmac}     
\usepackage{etoolbox}   
\usepackage{xcolor}
\usepackage[normalem]{ulem}
\usepackage{iftex}
\ifpTeX
  \usepackage[dvipdfmx]{graphicx}
  \usepackage[dvipdfmx,hidelinks]{hyperref}
\else
  \usepackage{graphicx}
  \usepackage[hidelinks]{hyperref}
\fi
\usepackage[numbers,sort&compress]{natbib}
\hypersetup{pdftitle={Multiple Stopping and Exercise Boundaries for American Put Options},
pdfauthor={Katsunori Ano},
pdfkeywords={multiple stopping, American put option, refractory period, random termination, free boundary}}
\usepackage{enumitem}
\usepackage[hypcap=false]{caption}
\allowdisplaybreaks

\newtheorem{theorem}{Theorem}[section]
\newtheorem{proposition}[theorem]{Proposition}
\newtheorem{lemma}[theorem]{Lemma}
\newtheorem{corollary}[theorem]{Corollary}
\theoremstyle{remark}
\newtheorem{remark}[theorem]{Remark}
\newtheorem{assumption}[theorem]{Assumption}

\newcommand{\E}{\mathrm E}
\newcommand{\Pp}{\mathrm P}
\newcommand{\cT}{\mathcal T}
\newcommand{\cL}{\mathcal L}
\newcommand{\cS}{\mathcal S}
\newcommand{\cC}{\mathcal C}
\newcommand{\dd}{\,\mathrm d}
\newcommand{\one}{\mathbf 1}
\newcommand{\R}{\mathbb R}

\title{Multiple Stopping and Exercise Boundaries \\ for American Put Options}
\author{Katsunori Ano}
\date{October 22, 2026}
\begin{document}
\maketitle

\begin{abstract}
This article studies a finite-horizon American put option with up to $m$
	exercise rights under the risk-neutral Black--Scholes model, where consecutive
	exercises must be separated by a prescribed refractory period $\delta>0$
	and unused rights may expire.
For each number of remaining rights, we show that optimal exercise is
	characterized by a single exercise boundary and that these boundaries
	are ordered by the number of remaining rights.
We prove optimality of the first boundary-hitting time and establish local
	Lipschitz continuity of the exercise boundaries within each open deadline strip.
Away from the deadlines, we derive smooth fit, the
	associated quasi-variational inequality and free-boundary formulation,
	an early-exercise-premium representation, recursive Volterra-type boundary
	equations, and a verification theorem.
We also extend the recursive formulation to an independent random termination
	time, thereby allowing for exogenous termination of the option contract.
\end{abstract}

\medskip

\section{Introduction}
The finite-horizon American put is a classical example of an optimal stopping problem; 
	see, for example, \citet{Merton1973,Jacka1991,LambertonLapeyre2008,PeskirShiryaev2006}. 
The purpose of this paper is to develop a corresponding optimal-boundary analysis  
	for the multiple-exercise (multiple-stopping) setting, 
	characterizing the optimal exercise rule when the holder has several exercise rights 
	and successive exercises must be separated by a deterministic refractory period.

The general multiple-stopping framework is well established.
The recursive reduction of a multiple-stopping problem to a sequence of single-stopping problems
	already appears in classical discrete-time theory; see \citet{Haggstrom1967}
	and the historical discussion in \citet{Bender2011}.
A systematic account of discrete-time multiple stopping problems and their applications
	is given in the monograph by \citet{SofronovSzajowski2025}.
For a Japanese-language introduction to optimal stopping problems that also treats multiple stopping,
	see \citet{AnoTiming2000}.
\citet{CarmonaTouzi2008} proved existence 
	and recursive construction of multiple-exercise rules in a broad Snell-envelope setting and, 
	in the Black--Scholes model, gave a constructive perpetual solution together with a numerical 
	approximation procedure for finite maturity. 
Related potential-theoretic methods for linear diffusions were developed by \citet{CarmonaDayanik2008}. 
For finite-horizon problems with a deterministic refractory period, \citet{Bender2011}
	established a continuous-time reduction principle and a Doob--Meyer-type decomposition
	allowing for right-hand discontinuities of the multiple-exercise value process.
For another class of finite-horizon optimal stopping problems arising in mathematical finance,
	formulated for exponential L\'evy processes, see \citet{AnoIvanov2012}.
Our question is more specific: what can be proved analytically about the geometry and regularity of the 
	finite-horizon exercise region for an American put when the exercise rights are optional and 
	successive exercises are separated by a fixed refractory period?

A closely related finite-horizon swing-put problem was analyzed by \citet{DeAngelisKitapbayev2018}, 
	where all rights must be exercised before maturity. 
That obligation can make exercise optimal even when the put payoff is zero and leads, for each right 
	except the last, to a stopping region bounded by two time-dependent curves. 
Here unused rights may expire, so there is no contractual reason to exercise an out-of-the-money right 
	merely to preserve later opportunities. 
For every finite number of rights we prove, under the standing model assumptions,
	that the stopping region is a single lower interval determined by one exercise boundary.

This paper develops a continuous-time counterpart of the discrete-time framework in \citet{AnoGRW2026},
	replacing the geometric random walk by the Black--Scholes diffusion while retaining the multiple-exercise
	and refractory-period structure.
That companion study considers American put, Russian, and floating-strike geometric-average Asian contracts,
	as well as independent random maturity. 
Broader Japanese-language lecture notes covering the surrounding optimal multiple-stopping and
	mathematical finance framework are \citet{AnoJapanese2026}.
The present article develops in English, and in research-article form, the material of Chapters 3--5
	of those notes, reorganized around the one-right American put, the multiple-exercise American put
	with refractory periods, and the extension to random termination, and supplemented by several new
	proofs developed for the present article.
The presentation here is self-contained.

Finite maturity creates another important feature of the multiple-exercise problem.
Because two successive exercises must be separated by the refractory period $\delta$,
	there are deterministic times at which one of the remaining exercise opportunities
	ceases to be usable.  
Define
	\begin{align*}
		d_j:=T-j\delta,\qquad j=0,1,2,\ldots,\qquad d_0=T.
	\end{align*}
At time $d_j$, there is still just enough time to accommodate one additional exercise
	after waiting for the required refractory period, whereas immediately after $d_j$
	there is not.  
We call these times \emph{deterministic feasibility deadlines}.
They divide the time horizon into strips on which only a certain number of exercise rights
	can still be used before maturity, so the original multiple-right problem reduces
	to the corresponding lower-right problem.
When a deadline is crossed from left to right, the effective number of usable rights drops by one.  
As a consequence, the value function may have a downward right-hand
	jump at a deadline, and the optimal exercise boundary may jump as well.
Right-hand discontinuities of the value process at the times $T-j\delta$
	are already treated in the general framework of \citet{Bender2011}.
Here we study their implications for the exercise boundaries and the associated transmission conditions.
Away from the deadlines, the standard quasi-variational inequality (QVI) and smooth-fit relations
	are recovered below from the standing Black--Scholes assumptions.
At a deadline, however, the usual time-continuous matching is replaced by 
	a max-type transmission condition reflecting the loss of one effective exercise right.

The main contribution of this paper is the single-crossing analysis and the resulting one-sided
	exercise boundaries, ordered by the number of remaining rights, for the optional multiple-exercise
	American put, with the feasibility-deadline structure kept explicit.
We first establish the recursive single-stopping representation, stagewise reduction across deadline 
	strips, and basic monotonicity, convexity, Lipschitz, and stripwise time-regularity properties of 
	the value functions. 
We then combine a Bermudan approximation with the \emph{marginal-right values}, 
	which serve as a key comparison device in the present analysis and are also developed 
	in the companion discrete-time study \citet{AnoGRW2026}.
The standard Bermudan approach restricts exercise to a finite time grid and studies convergence
	as the mesh tends to zero; see \citet[Section~2.2]{BouchardWarin2012}
	and \citet[Chapter~11]{Pages2018} for general accounts.
For the present multiple-exercise problem,
	we instead use deadline-aligned grids with
	$h=\delta/N$ that contain maturity and every relevant feasibility deadline,
	so that the refractory period is represented exactly by $N$ steps.
This alignment is essential for the backward-induction argument, because an unaligned recursion
	can cross a time at which the effective number of usable rights changes.

The aligned scheme yields a slope bound for the marginal-right value
	$f_\delta^{[m]}$, defined later for the $m$-right problem, and hence a
	\emph{single-crossing property}: the excess value
	$f_\delta^{[m]}(t,x)-(K-x)^+$ is nondecreasing on $(0,K)$, so its zero set is a lower interval.
This single-crossing property is the key structural step in the boundary analysis.
Together with the strict waiting-loss argument, which excludes optimal exercise at or above the strike,
	it yields a one-sided stopping region for every finite number of rights.
Thus we obtain the threshold characterization
	\begin{align*}
		V^{[m]}(t,x)=g_\delta^{[m]}(t,x)\Longleftrightarrow
		f_\delta^{[m]}(t,x)=(K-x)^+\Longleftrightarrow
			x\le b^{[m]}(t),
	\end{align*}
	where $V^{[m]}$ denotes the value function of the $m$-right problem,
	$g_\delta^{[m]}$ is the stopping reward, and $b^{[m]}$ is the exercise boundary.
The comparison of marginal-right values across exercise levels also yields the ordering of the exercise boundaries by the number
	of remaining rights, and we prove that the first boundary-hitting time is optimal.

The remaining regularity issue is resolved by a localized level-set argument in log-price coordinates,
	adapted from the probabilistic method of \citet{DeAngelisStabile2019} and combined with the weak
	Duhamel representation proved earlier in the paper.  
This yields local Lipschitz continuity of every
	exercise boundary on each open deadline strip and closes the classical delayed-value regularity
	recursively at every finite level.  
Smooth fit then follows away from deterministic deadlines.
The classical QVI relations in the open stopping and continuation regions, together with the associated
	free-boundary formulation, consequently hold under the standing Black--Scholes assumptions.
For the early-exercise-premium representation we use the weak Duhamel formula directly, rather than
	a local-time-space formula; no additional boundary monotonicity or continuation-side boundary
	regularity is required.  
This leads to a recursive Volterra-type integral equation for the exercise boundary in which 
	the deadline jumps enter explicitly, extending the classical one-right integral-equation characterization.

Finally, we extend the recursive architecture to an independent random termination time. 
A general conditional survival function weights future rewards, and under a multiplicative 
	survival condition the survival factors combine consistently along the recursion. 
Exponential termination with intensity $\lambda$ is equivalent, for valuation purposes, 
	to replacing the discount rate $r$ by $r+\lambda$ in the stopping functional, 
	while the risk-neutral drift of the underlying remains $r$; 
	in this case we prove that the recursive and original formulations coincide. 
For a general survival law, a complete proof of this equivalence is not pursued here. 

The remainder of the paper is organized as follows. Section~\ref{sec:model} recalls the Black--Scholes 
	model and the one-right American put. 
Sections~\ref{sec:multiple}--\ref{sec:boundary} develop the recursive multiple-stopping problem, the 
	deadline structure, marginal-right values, the single-crossing property, and the optimal boundary. 
Section~\ref{sec:qvi} gives the QVI and free-boundary formulations, the verification result,
	the early-exercise premium representation, and recursive Volterra equations for the exercise boundary.
Section~\ref{sec:termination} treats random termination, 
Section~\ref{sec:numerics} reports numerical illustrations. 

\section{Model and American put}\label{sec:model}
Work on a filtered probability space carrying a Brownian motion $B$. 
Under the risk-neutral measure, the asset price is
	\begin{align*}
  		\dd X_s=rX_s\dd s+\sigma X_s\dd B_s,\qquad X_t=x>0,
	\end{align*}
	where $r>0$, $\sigma>0$, the strike is $K>0$, and the finite maturity is $T$. 
The put payoff is
	\begin{align*}
		g(x):=(K-x)^+.
	\end{align*}
For $u\ge0$, the multiplicative representation is
	\begin{align*}
 		X_{t+u}^{t,x}=xR_u,\qquad
 		R_u:=\exp\!\left(\left(r-\frac12\sigma^2\right)u+\sigma(B_{t+u}-B_t)\right)>0,
	\end{align*}
	with
	\begin{equation}\label{eq:martfactor}
 		\E[e^{-ru}R_u]=1.
	\end{equation}
The infinitesimal generator is
	\begin{align*}
 		\cL\phi(x)=rx\phi_x(x)+\frac12\sigma^2x^2\phi_{xx}(x).
	\end{align*}

Let $(\mathcal F_s)$ denote the augmented filtration generated by $B$. 
For $t\le T$, $\cT_{[0,T-t]}$ denotes the set of stopping times 
	with respect to $(\mathcal F_{t+u})_{u\ge0}$ 
	taking values in $[0,T-t]$ (relative stopping times), 
	and $\E_{t,x}$ denotes expectation given $X_t=x$.

The one-right American put value is
	\begin{equation}\label{eq:V1}
 		V^{[1]}(t,x)=\sup_{\tau\in\cT_{[0,T-t]}} \E_{t,x}\bigl[e^{-r\tau}g(X_{t+\tau})\bigr].
	\end{equation}
We use the standard finite-horizon exercise-boundary results in the following form.
\begin{theorem}[One-right American put; see \cite{Jacka1991,PeskirShiryaev2006}]\label{th:one-right}
For every $t<T$ there is a boundary $b^{[1]}(t)\in(0,K)$ 
	such that the stopping set $\cS_t^{[1]}:=\{x>0:V^{[1]}(t,x)=g(x)\}$ 
	and the continuation set $\cC_t^{[1]}:=(0,\infty)\setminus\cS_t^{[1]}$ are
	\begin{align*}
 		\cS_t^{[1]}=(0,b^{[1]}(t)],\qquad
 		\cC_t^{[1]}=(b^{[1]}(t),\infty).
	\end{align*}
The boundary is nondecreasing and continuous on $[0,T)$ and satisfies
	$\lim_{t\uparrow T}b^{[1]}(t)=K$. 
The minimal optimal stopping time is
	\begin{align*}
 		\tau_t^{[1]*}:=\inf\{u\in[0,T-t):X_{t+u}\le b^{[1]}(t+u)\}\wedge(T-t).
	\end{align*}
Here $\inf\varnothing=+\infty$, and we set $b^{[1]}(T):=K$ as a terminal convention.
Moreover, value matching and smooth fit hold:
	\begin{align*}
 		V^{[1]}(t,b^{[1]}(t))=K-b^{[1]}(t),\qquad
 		V_x^{[1]}(t,b^{[1]}(t)+)=-1.
	\end{align*}
\end{theorem}
The value admits the familiar early-exercise-premium decomposition
	\begin{align*}
 		V^{[1]}(t,x)=P^E(t,x)+rK\int_t^T e^{-r(u-t)}\Pp_{t,x}\bigl(X_u\le b^{[1]}(u)\bigr)\dd u,
	\end{align*}
	where $P^E$ is the European put price. 
This identity will be used below to quantify the value lost by being unable to exercise during a refractory interval.

\section{Multiple stopping}
\label{sec:multiple}
\subsection{Multiple stopping and recursive formulations}
Let $m\ge1$ be the number of remaining exercise rights. 
An admissible multiple-stopping rule is an ordered vector
	$\boldsymbol\tau^{(m)}=(\tau_1,\ldots,\tau_m)$ of stopping times 
	$\tau_i\in\cT_{[0,T-t]}$ satisfying
	\begin{align*}
 		0\le\tau_1\le\tau_2\le\cdots\le\tau_m\le T-t
	\end{align*}
	and, whenever another exercise remains feasible,
	\begin{align*}
 		\tau_{i+1}-\tau_i\ge\delta\quad\text{on }\{\tau_i+\delta\le T-t\},
 		\qquad i=1,\ldots,m-1.
	\end{align*}
If after some exercise less than $\delta$ time remains, later rights are infeasible and contribute zero. 
We adopt the convenient extension
	\begin{equation}\label{eq:extension}
 		V^{[m]}(s,\cdot)=0\qquad(s>T,\ m=0,1,2,\ldots),
 		\qquad V^{[0]}\equiv0.
	\end{equation}
The direct value is
	\begin{equation}\label{eq:Vm-direct}
 		V^{[m]}_{\mathrm{dir}}(t,x)=\sup_{\boldsymbol\tau^{(m)}}
 		\E_{t,x}\left[\sum_{i=1}^m e^{-r\tau_i}g(X_{t+\tau_i})\one_{A_i}\right],
	\end{equation}
	where
	\begin{align*}
 		A_1:=\Omega,\qquad
	 	A_i:=\bigcap_{k=1}^{i-1}\{\tau_k+\delta\le T-t\},\qquad i=2,\ldots,m,
	\end{align*}
	so that abandoned or infeasible later exercises contribute zero.

Following the classical multiple-stopping reduction of \citet{Haggstrom1967}, 
	and its continuous-time formulation for multiple-exercise options 
	by \citet{CarmonaTouzi2008}, we work with a recursive single-stopping formulation.
Define the \emph{delayed residual-right value}
	\begin{equation}\label{eq:Psi}
		 \Psi^{[m-1]}_\delta(t,x)
			 :=\E_{t,x}\left[e^{-r\delta}V^{[m-1]}(t+\delta,X_{t+\delta})\right]
	\end{equation}
	and the \emph{stopping reward} (the obstacle in the associated QVI)
	\begin{equation}\label{eq:obstacle}
		 g_\delta^{[m]}(t,x):=g(x)+\Psi^{[m-1]}_\delta(t,x).
	\end{equation}
Define the value recursively in $m$ by
	\begin{equation}\label{eq:Vm-rec}
 		V^{[m]}(t,x):=\sup_{\tau\in\cT_{[0,T-t]}}
 		\E_{t,x}\left[e^{-r\tau}g_\delta^{[m]}(t+\tau,X_{t+\tau})\right].
	\end{equation}
For $m=1$, $\Psi^{[0]}_\delta\equiv0$ and \eqref{eq:Vm-rec} reduces to \eqref{eq:V1}.
We prove in Lemma~\ref{lem:direct-rec} that the recursive and direct values coincide, 
	$V^{[m]}=V^{[m]}_{\mathrm{dir}}$. 
Until then, $V^{[m]}$ always denotes the recursive value \eqref{eq:Vm-rec}.
We write
	\begin{align*}
 		\cS^{[m]}:=\{(t,x)\in[0,T)\times(0,\infty):V^{[m]}(t,x)=g_\delta^{[m]}(t,x)\}
	\end{align*}
	for the stopping (contact) set, $\cC^{[m]}:=([0,T)\times(0,\infty))\setminus\cS^{[m]}$ 
	for the continuation set, and $\cS_t^{[m]}$, $\cC_t^{[m]}$ for their $t$-sections.
	
\subsection{Exercise deadlines and stagewise reduction}
Set
	\begin{align*}
 		d_j:=T-j\delta,\qquad j=0,1,2,\ldots,
	\end{align*}
	and let
	\begin{align*}
 		q(t):=1+\left\lfloor\frac{T-t}{\delta}\right\rfloor,
 		\qquad q_m(t):=\min\{m,q(t)\}.
	\end{align*}
Thus $q_m(t)$ is the maximum number of rights that can still be used at time $t$.

\medskip
\noindent\emph{Standing convention.}
Throughout we assume $0<\delta\le T$, so that at least the last deadline $d_1=T-\delta$ 
	lies in $[0,T]$. 
Moreover, whenever level $m$ is considered we assume
	\begin{equation}\label{eq:m-feasible}
		(m-1)\delta\le T,\qquad\text{equivalently}\qquad d_{m-1}\ge0,
	\end{equation}
	so that all deadlines $d_1,\ldots,d_{m-1}$ belong to $[0,T]$ and the first strip 
	$[0,d_{m-1}]$ is nonempty. 
This entails no loss of generality: if $(m-1)\delta>T$, then 
	$m_0:=q(0)=1+\lfloor T/\delta\rfloor\le m-1$ and $q_m(t)=q(t)=q_{m_0}(t)$ for all $t\in[0,T]$, 
	so that, by Proposition~\ref{prop:stage} below, $V^{[m]}=V^{[m_0]}$ on $[0,T]\times(0,\infty)$, 
	and $m_0$ satisfies \eqref{eq:m-feasible}. 
In particular, in Lemma~\ref{lem:deadline} and Corollary~\ref{cor:jump-positive} 
	all deadlines involved lie in $[0,T]$.
\begin{proposition}[Stagewise degeneration]
\label{prop:stage}
For every $m\ge1$,
	\begin{equation}\label{eq:stage}
 		V^{[m]}(t,x)=V^{[q_m(t)]}(t,x).
	\end{equation}
In particular, for $j=1,\ldots,m-1$,
	\begin{align}
 		V^{[m]}(t,x)&=V^{[j]}(t,x), &&t\in(d_j,d_{j-1}],\quad m\ge j,\label{eq:strip}\\
 		V^{[m]}(d_j,x)&=V^{[j+1]}(d_j,x), &&m\ge j+1,\label{eq:dvalue}\\
 		V^{[m]}(d_j+,x)&=V^{[j]}(d_j,x), &&m\ge j+1.\label{eq:dright}
	\end{align}
The deadline jump
	\begin{align*}
 		\Delta^{[j+1]}(x):=V^{[j+1]}(d_j,x)-V^{[j]}(d_j,x)\ge0
	\end{align*}
	is nonnegative by Theorem~\ref{th:Vprops}(viii) below, and it is strictly positive on a nonempty interval at the last feasibility deadline $d_1$ 
	(the chronologically last one); see Corollary~\ref{cor:jump-positive}.
\end{proposition}
\begin{proof}
We prove \eqref{eq:strip} and \eqref{eq:dvalue} simultaneously by induction on $m$. 
The case $m=1$ is immediate. 
Assume the statements hold up to level $m-1$.

Fix $u\in(d_i,d_{i-1}]$, $1\le i\le m-1$. 
Then $u+\delta\in(d_{i-1},d_{i-2}]$, with the convention that for $i=1$ we have $u+\delta>T$. 
By the induction hypothesis and the extension \eqref{eq:extension},
	\begin{align*}
 		V^{[m-1]}(u+\delta,\cdot)=V^{[i-1]}(u+\delta,\cdot).
	\end{align*}
Consequently,
	\begin{equation}\label{eq:obstacle-strip}
 		g^{[m]}_\delta(u,\cdot)=g^{[i]}_\delta(u,\cdot),
 		\qquad u\in(d_i,d_{i-1}].
	\end{equation}
The same identity holds with $m$ replaced by any $j\in\{i,\ldots,m-1\}$. 
If $t\in(d_j,d_{j-1}]$, every $u\in[t,T]$ belongs to a strip with index $i\le j$, 
	hence $g^{[m]}_\delta(u,\cdot)=g^{[j]}_\delta(u,\cdot)$ 
	throughout the entire remaining horizon. 
Taking the supremum in \eqref{eq:Vm-rec} gives \eqref{eq:strip}.

Now set $t=d_j$. Since $t+\delta=d_{j-1}$, the induction hypothesis gives
	$V^{[m-1]}(d_{j-1},\cdot)=V^{[j]}(d_{j-1},\cdot)$; 
	for $j=1$ this is simply the terminal identity $V^{[m-1]}(T,\cdot)=V^{[1]}(T,\cdot)=g$. 
Therefore $g^{[m]}_\delta(d_j,\cdot)=g^{[j+1]}_\delta(d_j,\cdot)$. 
For $u>d_j$, \eqref{eq:obstacle-strip} applied to levels $m$ and $j+1$ gives the same identity. 
The stopping rewards therefore agree on $[d_j,T]$, proving \eqref{eq:dvalue}.

Equation \eqref{eq:stage} is the equivalent compact form of \eqref{eq:strip} and \eqref{eq:dvalue}. 
The proof of the right-limit identity \eqref{eq:dright} is deferred until after 
	Theorem~\ref{th:Vprops}; that theorem uses only \eqref{eq:strip}, 
	so no circularity is involved.
\end{proof}

\subsection{Properties of the value function}
\begin{theorem}[Properties of $V^{[m]}$]
\label{th:Vprops}
For every $m\ge1$ and $(t,x)\in[0,T]\times(0,\infty)$:
	\begin{enumerate}[label={\rm(\roman*)}]
		\item $0\le V^{[m]}(t,x)\le mK$;
		\item $t\mapsto V^{[m]}(t,x)$ is nonincreasing;
		\item $x\mapsto V^{[m]}(t,x)$ is nonincreasing;
		\item $x\mapsto V^{[m]}(t,x)$ is convex;
		\item $V^{[m]}$ is $m$-Lipschitz in $x$:
			\begin{align*}
 				|V^{[m]}(t,x_1)-V^{[m]}(t,x_2)|\le m|x_1-x_2|;
			\end{align*}
		\item $V^{[m]}$ is $1/2$-H\"older continuous in time on every deadline strip, 
			namely on $[0,d_{m-1}]$ and on each $(d_j,d_{j-1}]$, $j=1,\ldots,m-1$: 
			there is a constant $C_m<\infty$ such that
			\begin{align*}
				 |V^{[m]}(s,x)-V^{[m]}(s',x)|\le C_m(1+x)|s-s'|^{1/2}
			\end{align*}
			whenever $s,s'$ lie in the same strip; global H\"older continuity generally fails 
			because of the deadline jumps;
				\item $V^{[m]}$ is jointly continuous on $[0,d_{m-1}]\times(0,\infty)$ 
			and on each strip $(d_j,d_{j-1}]\times(0,\infty)$;
		\item $V^{[m]}(t,x)\ge V^{[m-1]}(t,x)$.
	\end{enumerate}
\end{theorem}
\begin{proof}
We argue by induction on $m$, using $X_{t+u}^{t,x}=xR_u$ and \eqref{eq:martfactor}. 
For $m=1$, assertion (viii) reads $V^{[1]}\ge V^{[0]}=0$. 
Assume the assertions hold at level $m-1$. 
By (viii) at level $m-1$, $\Psi_\delta^{[m-1]}\ge\Psi_\delta^{[m-2]}$, hence 
	$g_\delta^{[m]}\ge g_\delta^{[m-1]}$, and \eqref{eq:Vm-rec} gives (viii) at level $m$.
From \eqref{eq:obstacle},
	\begin{align*}
		 g_\delta^{[m]}(s,y)=(K-y)^++\E\!\left[e^{-r\delta}V^{[m-1]}(s+\delta,yR_\delta)\right].
	\end{align*}
The induction hypothesis implies that this stopping reward is nonnegative, nonincreasing and convex in $y$. 
Moreover,
	\begin{align}
		 |g_\delta^{[m]}(s,y_1)-g_\delta^{[m]}(s,y_2)|
 			&\le |y_1-y_2|+e^{-r\delta}(m-1)\E[|y_1-y_2|R_\delta]\nonumber\\
			&=m|y_1-y_2|.\label{eq:gmLip}
	\end{align}

Using \eqref{eq:Vm-rec}, nonnegativity of the stopping reward gives $V^{[m]}\ge0$, 
	while $g_\delta^{[m]}\le K+e^{-r\delta}(m-1)K\le mK$ gives $V^{[m]}\le mK$. 
If $t_1<t_2$, then the admissible stopping set at $t_2$ is contained in that at $t_1$, 
	and the induction hypothesis gives
	$g_\delta^{[m]}(t_1+u,\cdot)\ge g_\delta^{[m]}(t_2+u,\cdot)$; 
	hence time monotonicity follows.

If $x_1<x_2$, then $x_1R_\tau<x_2R_\tau$ pathwise,
	 and the spatial monotonicity of the stopping reward transfers to $V^{[m]}$. 
For convexity, let $x=\alpha x_1+(1-\alpha)x_2$. 
Since $R_\tau$ does not depend on the initial state,
	\begin{align*}
 		xR_\tau=\alpha x_1R_\tau+(1-\alpha)x_2R_\tau.
	\end{align*}
Convexity of the stopping reward and subadditivity of the supremum yield
	\begin{align*}
 		V^{[m]}(t,x)\le \alpha V^{[m]}(t,x_1)+(1-\alpha)V^{[m]}(t,x_2).
	\end{align*}
Likewise, \eqref{eq:gmLip} gives for every stopping time $\tau$
	\begin{align*}
 		|g_\delta^{[m]}(t+\tau,x_1R_\tau)-g_\delta^{[m]}(t+\tau,x_2R_\tau)|
 		\le m|x_1-x_2|R_\tau.
	\end{align*}
After discounting, taking expectations and using \eqref{eq:martfactor}, 
	we obtain the $m$-Lipschitz estimate.

For stripwise time regularity, let $0\le s<s'\le d_{m-1}$ and put $h:=s'-s$. 
Time monotonicity gives $V^{[m]}(s,x)-V^{[m]}(s',x)\ge0$. 
By the dynamic-programming principle on $[s,s']$,
	\begin{align*}
 		V^{[m]}(s,x)\le\E_{s,x}\Bigl[\max\Bigl\{\sup_{0\le u\le h}
			e^{-ru}g_\delta^{[m]}(s+u,X_{s+u}),\ e^{-rh}V^{[m]}(s',X_{s'})\Bigr\}\Bigr].
	\end{align*}
Since $V^{[m]}(s',x)\ge g_\delta^{[m]}(s',x)$ and all functions involved are 
	$m$-Lipschitz in space, each branch minus $V^{[m]}(s',x)$ is bounded by
	\begin{align*}
		 m\sup_{0\le u\le h}|X_{s+u}-x|
 		+\sup_{0\le u\le h}\bigl|\Psi_\delta^{[m-1]}(s+u,x)-\Psi_\delta^{[m-1]}(s',x)\bigr|.
	\end{align*}
For $u\in[0,h]$ the times $s+u+\delta$ lie in $[\delta,d_{m-2}]$, 
	the first strip of level $m-1$ (for $m=1$ the last term vanishes), so the induction hypothesis gives
	\begin{align*}
 		\bigl|\Psi_\delta^{[m-1]}(s+u,x)-\Psi_\delta^{[m-1]}(s',x)\bigr|
 		\le e^{-r\delta}C_{m-1}\E[1+xR_\delta]\sqrt h\le C_{m-1}(1+x)\sqrt h.
	\end{align*}
Together with $\E_{s,x}[\sup_{0\le u\le h}|X_{s+u}-x|]\le c\,x\sqrt h$ for $h\le T$, this yields
	\begin{align*}
 		0\le V^{[m]}(s,x)-V^{[m]}(s',x)\le C_m(1+x)\sqrt h.
	\end{align*}
The comparison is carried out entirely inside the first strip; 
	the estimate is stripwise because crossing a deterministic deadline 
	changes the effective level of the continuation term. 
On a later strip $(d_j,d_{j-1}]$, Proposition~\ref{prop:stage} (only \eqref{eq:strip} is used) 
	identifies $V^{[m]}$ with $V^{[j]}$, and $(d_j,d_{j-1}]$ is contained 
	in the first strip $[0,d_{j-1}]$ of level $j$.

Finally, joint continuity on each strip follows from (v) and (vi), since
	\begin{align*}
		 |V^{[m]}(s,y)-V^{[m]}(s',x)|\le m|y-x|+C_m(1+x)|s-s'|^{1/2}
	\end{align*}
	whenever $s,s'$ lie in the same strip. 
This completes the induction.
\end{proof}
\medskip
\noindent\emph{Completion of the proof of Proposition~\ref{prop:stage}.}
For $s\in(d_j,d_{j-1}]$, \eqref{eq:strip} gives
	$V^{[m]}(s,x)=V^{[j]}(s,x)$. Level $j$ has no deadline at $d_j$, and by
	Theorem~\ref{th:Vprops}(vii), $V^{[j]}$ is jointly continuous on
	$[0,d_{j-1}]\times(0,\infty)$. 
Hence
	\begin{align*}
		V^{[m]}(d_j+,x)=\lim_{s\downarrow d_j}V^{[j]}(s,x)=V^{[j]}(d_j,x),
	\end{align*}
	which proves \eqref{eq:dright} and completes the proof of Proposition~\ref{prop:stage}.

\subsection{Supermartingale bounds and deadline transmission}
\begin{lemma}[Supermartingale bounds]
\label{lem:supermart}
Let $k\ge1$ and fix $(t,x)$. 
Extend $V^{[k]}(s,\cdot)$ by zero for $s>T$ as in \eqref{eq:extension}.
	\begin{enumerate}[label={\rm(\roman*)}]
		\item The two processes
			\begin{align*}
				&\bigl(e^{-ru}V^{[k]}(t+u,X_{t+u})\bigr)_{u\ge0},\qquad
				\bigl(e^{-ru}\Psi_\delta^{[k]}(t+u,X_{t+u})\bigr)_{u\ge0}
			\end{align*}
			satisfy the optional-sampling inequalities for all bounded stopping times and hence are strong 
			supermartingales with respect to $(\mathcal F_{t+u})_{u\ge0}$. 
			In particular,
			\begin{align*}
				\E_{t,x}\bigl[e^{-r\tau}\Psi_\delta^{[k]}(t+\tau,X_{t+\tau})\bigr]\le\Psi_\delta^{[k]}(t,x)
			\end{align*}
			for every bounded stopping time $\tau$, and $V^{[k]}\ge\Psi_\delta^{[k]}$.
		\item For $m\ge2$,
			\begin{align*}
				V^{[m]}\le V^{[1]}+\Psi_\delta^{[m-1]}\le V^{[1]}+V^{[m-1]};
			\end{align*}
			consequently $V^{[m]}\le mV^{[1]}$.
	\end{enumerate}
\end{lemma}
\begin{proof}
(i) Let
	\begin{align*}
		Z_u:=e^{-ru}V^{[k]}(t+u,X_{t+u}),\qquad u\ge0,
	\end{align*}
	with the zero extension beyond maturity. 
The dynamic-programming principle gives directly, 
	for any bounded stopping times $0\le\rho\le\tau$,
	\begin{align}\label{eq:strong-supermart-V}
		\E_{t,x}[Z_\tau\mid\mathcal F_{t+\rho}]\le Z_\rho.
	\end{align}
Indeed, conditional on $\mathcal F_{t+\rho}$, waiting until $\tau$ and then continuing optimally 
	from $(t+\tau,X_{t+\tau})$ is an admissible rule from $(t+\rho,X_{t+\rho})$; 
	the zero extension handles paths for which the shifted time exceeds $T$. 
This is the direction of the dynamic-programming principle that requires 
	$\varepsilon$-optimal continuation rules selected measurably in the state; 
	it holds because on each deadline strip the stopping reward and $V^{[k]}$ are continuous 
	(Theorem~\ref{th:Vprops}(vii) and Remark~\ref{rem:obstacle-usc}), 
	so that the standard Markovian Snell-envelope theory applies strip by strip 
	\citep[Chapter~I]{PeskirShiryaev2006}, while at each deterministic deadline 
	the strips are joined by the dynamic-programming principle at that fixed time. 
Thus $Z$ is a strong supermartingale. 
This formulation is convenient here because deterministic feasibility deadlines may 
	produce downward right-hand jumps, so right continuity of the value process is not assumed.

For its discounted process, set
	\begin{align*}
		Y_u:=e^{-ru}\Psi_\delta^{[k]}(t+u,X_{t+u}).
	\end{align*}
By the strong Markov property and the definition of $\Psi_\delta^{[k]}$, the identity
	$	Y_\tau=\E[Z_{\tau+\delta}\mid\mathcal F_{t+\tau}]
	$ 
	holds for every bounded stopping time $\tau$. Hence, for bounded $\rho\le\tau$, the tower property 
	and \eqref{eq:strong-supermart-V} applied to $\rho+\delta\le\tau+\delta$ give
	\begin{align*}
		\E[Y_\tau\mid\mathcal F_{t+\rho}]
		&=\E[Z_{\tau+\delta}\mid\mathcal F_{t+\rho}]\le
		\E[Z_{\rho+\delta}\mid\mathcal F_{t+\rho}]=Y_\rho.
	\end{align*}
Thus $Y$ is also a strong supermartingale, and the displayed inequality in the statement follows by 
	taking $\rho=0$. Finally, \eqref{eq:strong-supermart-V} with $\rho=0$ and $\tau=\delta$ gives
	$V^{[k]}(t,x)\ge\Psi_\delta^{[k]}(t,x)$.

(ii) For every $\tau\in\cT_{[0,T-t]}$, part (i) gives
	\begin{align*}
				\E_{t,x}\bigl[e^{-r\tau}g_\delta^{[m]}(t+\tau,X_{t+\tau})\bigr]
		&\le\E_{t,x}\bigl[e^{-r\tau}(K-X_{t+\tau})^+\bigr]+\Psi_\delta^{[m-1]}(t,x)\\
		&\le V^{[1]}(t,x)+\Psi_\delta^{[m-1]}(t,x).
	\end{align*}
Taking the supremum in \eqref{eq:Vm-rec} gives the first inequality, 
	and part (i) at level $m-1$ gives the second. 
Induction on $m$ yields $V^{[m]}\le mV^{[1]}$.
\end{proof}

At a feasibility deadline the value is determined by a max-type transmission condition.
\begin{lemma}[Deadline transmission]
\label{lem:deadline}
Let $m\ge2$ and $1\le j\le m-1$. Then, for every $x>0$:
	\begin{enumerate}[label={\rm(\roman*)}]
		\item $V^{[m]}(d_j,x)=\max\bigl\{g_\delta^{[m]}(d_j,x),\,V^{[j]}(d_j,x)\bigr\}$;
		\item $\Delta^{[j+1]}(x)=\bigl(g_\delta^{[j+1]}(d_j,x)-V^{[j]}(d_j,x)\bigr)^+$, and 
			$\Delta^{[j+1]}(x)>0$ implies $x\in\cS^{[m]}_{d_j}$;
		\item $\cS^{[j]}_{d_j}\subseteq\cS^{[m]}_{d_j}$.
	\end{enumerate}
\end{lemma}
\begin{proof}
The first point is to reduce the statement at the general deadline $d_j$ 
	to the first deadline of a lower-level problem.
By \eqref{eq:dvalue} and its proof,
	\begin{align*}
    		V^{[m]}(d_j,\cdot)=V^{[j+1]}(d_j,\cdot),\qquad
    		g_\delta^{[m]}(d_j,\cdot)=g_\delta^{[j+1]}(d_j,\cdot).
	\end{align*}
Hence
	\begin{align*}
   		\cS_{d_j}^{[m]}=\cS_{d_j}^{[j+1]}.
	\end{align*}
It is therefore enough to prove the assertions when $m=j+1$.  
In that case
	\begin{align*}
    		d:=d_{m-1}=T-(m-1)\delta
	\end{align*}
	is the first feasibility deadline for the $m$-right problem.

\emph{Proof of (i).}
At time $d$ the holder has two qualitatively different choices: exercise immediately, 
	or wait for a strictly positive amount of time.  
Immediate exercise gives
    	$V^{[m]}(d,x)\ge g_\delta^{[m]}(d,x).$ 
Also, by Theorem~\ref{th:Vprops}(viii), having one additional right cannot decrease the value, so
    	$V^{[m]}(d,x)\ge V^{[m-1]}(d,x).$ 
Thus
	\begin{align*}
   		 V^{[m]}(d,x)\ge\max\{g_\delta^{[m]}(d,x),V^{[m-1]}(d,x)\}.
	\end{align*}
We now prove the reverse inequality.
Starting from the deterministic state $(d,x)$, 
	the value is unchanged if we work with the filtration generated 
	by the future Brownian increments
   		$(B_{d+u}-B_d)_{u\ge0},$ 
	because these increments are independent of $\mathcal F_d$ and determine the future asset path.  
Let $\tau$ be a stopping time for this shifted Brownian filtration.  
The event $\{\tau=0\}$ belongs to the germ sigma-field at time zero.  
By Blumenthal's zero--one law this sigma-field is trivial, and therefore
	\begin{align*}
   		 \Pp(\tau=0)\in\{0,1\}.
	\end{align*}
Thus either $\tau=0$ almost surely or $\tau>0$ almost surely.
If $\tau=0$ almost surely, the expected reward is simply
   	$ g_\delta^{[m]}(d,x). $ 
If $\tau>0$ almost surely, then $d+\tau\in(d,T]$ almost surely.  
Immediately to the right of the deadline only $m-1$ rights remain effective, 
	and \eqref{eq:obstacle-strip} gives
	\begin{align*}
   		 g_\delta^{[m]}(s,\cdot)=g_\delta^{[m-1]}(s,\cdot),\qquad s\in(d,T].
	\end{align*}
Hence
	\begin{align*}
    		\E_{d,x}\!\left[e^{-r\tau} g_\delta^{[m]}(d+\tau,X_{d+\tau})\right]
    			=\E_{d,x}\!\left[e^{-r\tau}g_\delta^{[m-1]}(d+\tau,X_{d+\tau})\right]
    			\le V^{[m-1]}(d,x).
	\end{align*}
Every admissible first stopping time therefore produces a value not exceeding
	\begin{align*}
    		\max\{g_\delta^{[m]}(d,x),V^{[m-1]}(d,x)\}.
	\end{align*}
Taking the supremum proves
	\begin{align*}
   		 V^{[m]}(d,x)=\max\{g_\delta^{[m]}(d,x),V^{[m-1]}(d,x)\}.
	\end{align*}
Returning to the general deadline $d_j$ gives (i).

\emph{Proof of (ii).}
Apply (i) at level $j+1$.  
From
    	$\Delta^{[j+1]}(x)=V^{[j+1]}(d_j,x)-V^{[j]}(d_j,x),$ 
	it follows that 
	\begin{align*}
    		\Delta^{[j+1]}(x)&=\max\{g_\delta^{[j+1]}(d_j,x),V^{[j]}(d_j,x)\}-V^{[j]}(d_j,x)\\
    				&=\bigl(g_\delta^{[j+1]}(d_j,x)-V^{[j]}(d_j,x)\bigr)^+.
	\end{align*}
If this quantity is strictly positive, the maximum is attained by the stopping-reward branch.  
Therefore
    	$V^{[j+1]}(d_j,x)=g_\delta^{[j+1]}(d_j,x),$ 
	so $x\in\cS_{d_j}^{[j+1]}=\cS_{d_j}^{[m]}$.

\emph{Proof of (iii).}
Take $x\in\cS_{d_j}^{[j]}$.  
Then
    	$V^{[j]}(d_j,x)=g_\delta^{[j]}(d_j,x).$ 
The hierarchy of lower-level values in Theorem~\ref{th:Vprops}(viii) implies
	\begin{align*}
    		\Psi_\delta^{[j-1]}(d_j,x)\le \Psi_\delta^{[j]}(d_j,x),
	\end{align*}
	and hence
	\begin{align*}
    		g_\delta^{[j]}(d_j,x)\le g_\delta^{[j+1]}(d_j,x).
	\end{align*}
Using (i) at level $j+1$ therefore gives
	\begin{align*}
    		V^{[j+1]}(d_j,x)=\max\{g_\delta^{[j+1]}(d_j,x),V^{[j]}(d_j,x)\}
    					=g_\delta^{[j+1]}(d_j,x).
	\end{align*}
Thus $x\in\cS_{d_j}^{[j+1]}=\cS_{d_j}^{[m]}$, which proves the inclusion.
\end{proof}
\begin{remark}[Time regularity of the stopping reward]
\label{rem:obstacle-usc}
Since $V^{[m-1]}$ is continuous on the right-closed strips of Theorem~\ref{th:Vprops}(vii) and has 
	nonpositive right jumps at its deadlines and at $T$, the map $t\mapsto g_\delta^{[m]}(t,x)$ is 
	left-continuous, with nonpositive right jumps at $d_1,\ldots,d_{m-1}$. 
Hence the stopping-reward process $u\mapsto e^{-ru}g_\delta^{[m]}(t+u,X_{t+u})$ is pathwise upper 
	semicontinuous from both sides, which is the regularity under which the Snell-envelope arguments 
	below (flat compensator before contact) are applied on each strip.
\end{remark}

\section{\texorpdfstring{Marginal-right values and the single-crossing property}{Marginal-right values and the single-crossing property}}\label{sec:singlecross}
Define the marginal-right value
	\begin{align*}
		f_\delta^{[m]}(t,x):=V^{[m]}(t,x)-\Psi_\delta^{[m-1]}(t,x),
			\qquad m\ge1.
	\end{align*}
In particular, $f_\delta^{[1]}=V^{[1]}$ because $\Psi_\delta^{[0]}=0$.
Then the stopping decision is determined by comparing
	$f_\delta^{[m]}$ with the one-right payoff $(K-x)^+$.
Indeed, stopping occurs exactly when
	$f_\delta^{[m]}(t,x)=(K-x)^+,$
	while continuation corresponds to
	$f_\delta^{[m]}(t,x)>(K-x)^+.$

Throughout the paper, the term \emph{marginal-right value} is reserved for
$f_\delta^{[m]}$. It should not be confused with the \emph{incremental value of the
$m$th right}, introduced formally in Subsection~\ref{subsec:incremental-values} as
$D^{[m]}:=V^{[m]}-V^{[m-1]}$. The former removes the delayed residual-right value and
therefore converts the stopping condition into a comparison with the one-right payoff,
whereas the latter measures the additional value obtained by adding the $m$th exercise right.

\begin{lemma}[Basic properties of the marginal-right value]
\label{lem:fprops}
For $m\ge2$:
	\begin{enumerate}[label={\rm(\roman*)}]
	\item
		$(K-x)^+\le f_\delta^{[m]}(t,x)\le V^{[m]}(t,x)\le mK,\quad		
			 V^{[m]}(t,x)\ge V^{[m-1]}(t,x)\ge \Psi_\delta^{[m-1]}(t,x);$
	\item $f_\delta^{[m]}(t,\cdot)$ is $(2m-1)$-Lipschitz;
	\item $f_\delta^{[m]}$ is jointly continuous on $[0,d_{m-1}]\times(0,\infty)$;
	\item
		$(t,x)\in\cS^{[m]}\iff f_\delta^{[m]}(t,x)=(K-x)^+,\quad
			 (t,x)\in\cC^{[m]}\iff f_\delta^{[m]}(t,x)>(K-x)^+.$
	\end{enumerate}
\end{lemma}
\begin{proof}
Immediate exercise in \eqref{eq:Vm-rec} gives
	\begin{align*}
 		V^{[m]}(t,x)\ge (K-x)^++\Psi_\delta^{[m-1]}(t,x),
	\end{align*}
	which proves the first lower bound. 
Since $\Psi_\delta^{[m-1]}\ge0$, we also have $f_\delta^{[m]}\le V^{[m]}$. 
The hierarchy $V^{[m]}\ge V^{[m-1]}$ is Theorem~\ref{th:Vprops}(viii), and 
	$V^{[m-1]}\ge\Psi_\delta^{[m-1]}$ is Lemma~\ref{lem:supermart}(i).
The Lipschitz bound follows by subtracting an $(m-1)$-Lipschitz function from an $m$-Lipschitz one. 
Joint continuity follows from Theorem~\ref{th:Vprops} and bounded convergence in \eqref{eq:Psi}. 
Finally,
	\begin{align*}
 		V^{[m]}-g_\delta^{[m]}=f_\delta^{[m]}-(K-x)^+,
	\end{align*}
	which proves the contact characterization.
\end{proof}

\subsection{Residual premium from the remaining exercise right}
The following strict positivity result will be used to show that stopping is never optimal
	when the asset price is above the strike.
\begin{lemma}[Strict positivity of the residual premium]
\label{lem:residual}
For $t\le T-\delta$ define
	\begin{align*}
		 \Theta_\delta(t,x):=rK\E\!\left[\int_0^\delta e^{-rw}
		 \one_{\{xR_w\le b^{[1]}(t+w)\}}\dd w\right].
	\end{align*}
Then
	\begin{equation}\label{eq:residid}
 		V^{[1]}(t,x)-\Psi_\delta^{[1]}(t,x)=\Theta_\delta(t,x)>0.
	\end{equation}
Consequently,
	\begin{align*}
 		f_\delta^{[2]}(t,x)\ge \Theta_\delta(t,x)>0,
	\end{align*}
		and therefore $\cS_t^{[2]}\cap[K,\infty)=\varnothing$ for $t\le T-\delta$.
\end{lemma}
\begin{proof}
From the one-right early-exercise-premium martingale decomposition,
	\begin{align*}
 		S_u=e^{-ru}V^{[1]}(t+u,xR_u)+rK\int_0^u e^{-rw}\one_{\{xR_w\le b^{[1]}(t+w)\}}\dd w
	\end{align*}
	is a square-integrable martingale. 
Since $S$ is a square-integrable martingale and $\delta$ is deterministic,
	the martingale property gives
	\begin{align*}
		V^{[1]}(t,x)=\E[S_\delta]&=\E_{t,x}\!\left[
			e^{-r\delta}V^{[1]}(t+\delta,X_{t+\delta})\right]
				+rK\E\!\left[\int_0^\delta e^{-rw}\one_{\{xR_w\le b^{[1]}(t+w)\}}\dd w\right].
	\end{align*}
By the definitions of $\Psi_\delta^{[1]}$ and $\Theta_\delta$, this becomes
	\begin{align*}
		V^{[1]}(t,x)=\Psi_\delta^{[1]}(t,x)+\Theta_\delta(t,x),
	\end{align*}
	which is exactly \eqref{eq:residid}.
For every $w\in(0,\delta]$, $b^{[1]}(t+w)>0$ and $xR_w$ has a lognormal law 
	with support $(0,\infty)$, so
	\begin{align*}
		\Pp(xR_w\le b^{[1]}(t+w))>0.
	\end{align*}
Since the integrand in the definition of $\Theta_\delta$ is nonnegative,
	Tonelli's theorem allows us to interchange expectation and time integration
	without first verifying integrability. 
Hence
	\begin{align*}
		\Theta_\delta(t,x)&=rK\int_0^\delta e^{-rw}
			\Pp\bigl(xR_w\le b^{[1]}(t+w)\bigr)\,\dd w>0.
	\end{align*}
Since $V^{[2]}\ge V^{[1]}$,
	\begin{align*}
 		f_\delta^{[2]}=V^{[2]}-\Psi_\delta^{[1]}\ge V^{[1]}-\Psi_\delta^{[1]}
			=\Theta_\delta>0.
	\end{align*}
When $x\ge K$, the immediate put payoff is zero, hence contact is impossible.
\end{proof}
\begin{corollary}[Strict positivity of the last deadline jump]
\label{cor:jump-positive}
For every $m\ge2$ and $0<x<Ke^{-r\delta}$,
	\begin{equation}\label{eq:strict-last-jump}
		V^{[m]}(d_1,x)-V^{[m]}(d_1+,x)>0.
	\end{equation}
More precisely,
	\begin{equation}\label{eq:last-jump-lower-bound}
		V^{[m]}(d_1,x)-V^{[m]}(d_1+,x)=\Delta^{[2]}(x)\ge \bigl(Ke^{-r\delta}-x\bigr)^+.
	\end{equation}
\end{corollary}
\begin{proof}
Since $V^{[2]}\ge g_\delta^{[2]}$, Lemma~\ref{lem:residual} gives
	\begin{align*}
		\Delta^{[2]}(x)=V^{[2]}(d_1,x)-V^{[1]}(d_1,x)
			\ge (K-x)^+-\Theta_\delta(d_1,x).
	\end{align*}
Moreover,
	\begin{align*}
		\Theta_\delta(d_1,x)\le rK\int_0^\delta e^{-rw}\dd w=K(1-e^{-r\delta}),
	\end{align*}
	so $\Delta^{[2]}(x)\ge(K-x)^+-K(1-e^{-r\delta})$. Since also $\Delta^{[2]}\ge0$
	by Proposition~\ref{prop:stage}, and $(K-x)^+-K(1-e^{-r\delta})=Ke^{-r\delta}-x$ for $x<K$,
	\begin{align*}
		\Delta^{[2]}(x)\ge \bigl(Ke^{-r\delta}-x\bigr)^+.
	\end{align*}
Hence $\Delta^{[2]}(x)>0$ whenever $0<x<Ke^{-r\delta}$. 
Finally, 
\eqref{eq:dvalue}--\eqref{eq:dright} imply
	\begin{align*}
		V^{[m]}(d_1,x)-V^{[m]}(d_1+,x)=\Delta^{[2]}(x), \qquad m\ge2,
	\end{align*}
	which proves \eqref{eq:strict-last-jump} and \eqref{eq:last-jump-lower-bound}.
\end{proof}

\subsection{Bermudan approximation}
The proof of the single-crossing property uses a Bermudan approximation,
	that is, an approximation in which exercise is restricted to discrete grid times,
	with every deterministic deadline included in the grid.
Only deadline-aligned Bermudan convergence is established in this paper:
	the time step is $h=\delta/N$, the maturity and every relevant deadline
	belong to the exercise grid, and the one-step transition is the exact
	Black--Scholes transition. 
This discretizes exercise opportunities without replacing the diffusion by a binomial process. 
Convergence for general CRR grids with waiting steps $\lceil\delta/\Delta t\rceil$ 
	is reserved for a companion paper. 
In particular, the present argument makes no general CRR
	convergence claim at the deadlines.

Fix $t$ such that $(T-t)/\delta\in\mathbb Q$. 
Choose integers $N\ge2$ and $M$ with $M\delta=N(T-t)$, and set
	\begin{align*}
		 h:=\frac{\delta}{N},\qquad T-t=Mh,
	\end{align*}
	so $t_n=t+nh$ contains every $d_j$. Let $R_1^{(h)},R_2^{(h)},\ldots$ be i.i.d. multiplicative 
	one-step factors with $R_i^{(h)}\stackrel d=R_h$ and $\E[e^{-rh}R_i^{(h)}]=1$.

Set $V_n^{[0],h}\equiv0$ and $V_n^{[k],h}=0$ for $n>M$. For $k=1,\ldots,m$ define
	\begin{align}
 		\Psi_n^{[k-1],h}(x)
		 &:=e^{-rNh}\E\bigl[V_{n+N}^{[k-1],h}(xR_1^{(h)}\ldots R_N^{(h)})\bigr],\label{eq:Psi-h}\\
 		g_n^{[k],h}(x)&:=(K-x)^++\Psi_n^{[k-1],h}(x),\label{eq:g-h}\\
 		V_M^{[k],h}(x)&:=(K-x)^+,\nonumber\\
 		V_n^{[k],h}(x)&:=\max\left\{g_n^{[k],h}(x),
 			e^{-rh}\E[V_{n+1}^{[k],h}(xR_1^{(h)})]\right\},\quad n=M-1,\ldots,0.\label{eq:bermrec}
	\end{align}
Define
	\begin{align*}
 		f_n^{[k],h}(x):=V_n^{[k],h}(x)-\Psi_n^{[k-1],h}(x)
	\end{align*}
	and the one-step exercise premium
	\begin{equation}\label{eq:e-h}
 		e_n^{[k]}(x):=V_n^{[k],h}(x)-e^{-rh}\E[V_{n+1}^{[k],h}(xR_1^{(h)})]
 		=\left(g_n^{[k],h}(x)-e^{-rh}\E[V_{n+1}^{[k],h}(xR_1^{(h)})]\right)^+.
	\end{equation}
At $n=M$, set $e_M^{[k]}=(K-x)^+$, and for $n>M$ set $e_n^{[k]}=0$. 
We also set $e_n^{[0]}\equiv0$, consistently with $V^{[0],h}\equiv0$.
\begin{lemma}[Bermudan identities]
\label{lem:berm-basic}
For every $k=1,\ldots,m$ and every grid index $n$:
	\begin{enumerate}[label={\rm(\roman*)}]
		\item $V_n^{[k],h}$ is nonincreasing and $k$-Lipschitz, and
			$V_n^{[k],h}\ge V_n^{[k-1],h}$;
		\item the correction term
			\begin{equation}\label{eq:correction}
				 c_n^{[k-1]}(x):=\Psi_n^{[k-1],h}(x)-e^{-rh}\E[\Psi_{n+1}^{[k-1],h}(xR_1^{(h)})]
 				=e^{-rNh}\E[e_{n+N}^{[k-1]}(xR_1^{(h)}\ldots R_N^{(h)})]\ge0;
			\end{equation}
		\item the marginal-right value obeys the corrected recursion
			\begin{equation}\label{eq:corrected-rec}
 				f_n^{[k],h}(x)=\max\left\{(K-x)^+,
 					e^{-rh}\E[f_{n+1}^{[k],h}(xR_1^{(h)})]-c_n^{[k-1]}(x)\right\},
			\end{equation}
			and $0\le f_n^{[k],h}(x)\le K$;
		\item for $n\le M$,
			\begin{equation}\label{eq:e-lower}
 				e_n^{[k]}(y)\ge\bigl(K(1-e^{-rh})-y\bigr)^+,
			\end{equation}
			and, for every $n$,
			\begin{equation}\label{eq:e-zero}
 				e_n^{[k]}(x)=0\qquad(x\ge K).
			\end{equation}
	\end{enumerate}
\end{lemma}
\begin{proof}
For (i), use backward induction in $n$ and induction in $k$. 
The terminal payoff is nonincreasing and $1$-Lipschitz. 
If the properties hold one step later and at the lower exercise level, 
	then \eqref{eq:g-h} and $\E[e^{-rh}R_1^{(h)}]=1$ show that both the immediate-exercise branch 
	and the continuation branch in \eqref{eq:bermrec} are nonincreasing and $k$-Lipschitz. 
Their maximum preserves both properties. 
The hierarchy follows because $\Psi^{[k-1],h}\ge\Psi^{[k-2],h}$ implies 
	$g^{[k],h}\ge g^{[k-1],h}$, and backward induction in \eqref{eq:bermrec} 
	then yields $V^{[k],h}\ge V^{[k-1],h}$.

For (ii), the tower property gives
	\begin{align*}
 		e^{-rh}\E[\Psi_{n+1}^{[k-1],h}(xR_1^{(h)})]
 			=e^{-r(N+1)h}\E[V_{n+N+1}^{[k-1],h}(xR_1^{(h)}\ldots R_{N+1}^{(h)})].
	\end{align*}
Subtracting this from \eqref{eq:Psi-h} and conditioning on the first $N$ factors gives 
	exactly \eqref{eq:correction}; nonnegativity follows from \eqref{eq:e-h}.

For (iii), subtract $\Psi_n^{[k-1],h}$ from both branches of \eqref{eq:bermrec}, 
	use $V_{n+1}^{[k],h}=f_{n+1}^{[k],h}+\Psi_{n+1}^{[k-1],h}$, 
	and then \eqref{eq:correction}. 
The upper bound follows by backward induction: 
	at maturity $f_M^{[k],h}=(K-x)^+\le K$, and if $f_{n+1}^{[k],h}\le K$, 
	then \eqref{eq:corrected-rec} and $c_n^{[k-1]}\ge0$ 
	imply $f_n^{[k],h}\le\max\{K,e^{-rh}K\}=K$.

For (iv), rewrite the bracket in \eqref{eq:e-h} as
	\begin{equation}\label{eq:ebracket}
		 (K-y)^+-e^{-rh}\E[f_{n+1}^{[k],h}(yR_1^{(h)})]+c_n^{[k-1]}(y).
	\end{equation}
For $n<M$, using $f\le K$ and $c\ge0$ gives \eqref{eq:e-lower}; 
	for $n=M$ it holds because $e_M^{[k]}=(K-y)^+$. 
(For $n>M$, $e_n^{[k]}=0$, so \eqref{eq:e-lower} is not claimed there.) 
To prove \eqref{eq:e-zero}, define
	\begin{align*}
		W_n^{[k],h}(x):=V_n^{[k],h}(x)-\Psi_n^{[k],h}(x).
	\end{align*}
By the definition of $\Psi_n^{[k],h}$,
	\begin{align*}
		W_n^{[k],h}(x)&=V_n^{[k],h}(x)-e^{-rNh}
		\E\!\left[V_{n+N}^{[k],h}(xR_1^{(h)}\cdots R_N^{(h)})\right].
	\end{align*}
Adding and subtracting the intermediate discounted values gives the telescoping expansion
	\begin{align*}
		W_n^{[k],h}(x)
			&=e_n^{[k]}(x)+e^{-rh}\E\!\left[e_{n+1}^{[k]}(xR_1^{(h)})\right]
			+e^{-2rh}\E\!\left[e_{n+2}^{[k]}(xR_1^{(h)}R_2^{(h)})\right]+\cdots\\
			&\qquad\qquad+e^{-r(N-1)h}\E\!\left[e_{n+N-1}^{[k]}(xR_1^{(h)}\cdots R_{N-1}^{(h)})\right].
	\end{align*}
Equivalently,
	\begin{align*}
		W_n^{[k],h}(x)=\sum_{n'=n}^{n+N-1}e^{-r(n'-n)h}\E\!\left[e_{n'}^{[k]}
			(xR_1^{(h)}\cdots R_{n'-n}^{(h)})\right],
	\end{align*}
	where the product is interpreted as $1$ when $n'=n$.
When $N\ge2$, \eqref{eq:e-lower} and the full support of the lognormal increment 
	imply $W_n^{[k],h}(x)>0$ for $n<M$. If $x\ge K$ and $k\ge2$, 
	the hierarchy from (i) yields
	\begin{align*}
 		V_n^{[k],h}(x)-g_n^{[k],h}(x)
 			\ge V_n^{[k-1],h}(x)-\Psi_n^{[k-1],h}(x)=W_n^{[k-1],h}(x)>0.
	\end{align*}
For $k=1$, the positive probability of ending in the money before maturity gives 
	$V_n^{[1],h}(x)>0=g_n^{[1],h}(x)$. 
Thus immediate exercise is never optimal for $x\ge K$, and \eqref{eq:e-h} gives $e_n^{[k]}(x)=0$.
\end{proof}
\begin{lemma}[Convergence of the Bermudan approximation]
\label{lem:berm-conv}
Let $h=\delta/N\downarrow0$ along grids satisfying $T-t=Mh$. 
Then for each $k\le m$ and $x>0$,
	\begin{align*}
		 V_0^{[k],h}(x)\to V^{[k]}(t,x),\qquad
 		\Psi_0^{[k-1],h}(x)\to\Psi_\delta^{[k-1]}(t,x),\qquad
		f_0^{[k],h}(x)\to f_\delta^{[k]}(t,x).
	\end{align*}
\end{lemma}
\begin{proof}
We prove the convergence by induction on the number of rights $k$.  
To keep the main idea visible, it is useful to separate the proof into three parts:
	\begin{enumerate}[label={\rm(\arabic*)}]
   		 \item compare the discrete and continuous stopping rewards;
   		 \item prove convergence on bounded sets of initial states;
   		 \item control large initial states uniformly.
	\end{enumerate}
Set
	\begin{align*}
    		J_k(h):=\sup_{0\le n\le M}\sup_{y>0}\bigl|V_n^{[k],h}(y)-V^{[k]}(t_n,y)\bigr|.
	\end{align*}
We show $J_k(h)\to0$ as $h\downarrow0$.  
The case $k=0$ is immediate because both values are identically zero, so $J_0\equiv0$.  
Assume that $J_{k-1}(h)\to0$.

\emph{Step 1: the aligned grid gives the same feasibility ranges.}
Because
	\begin{align*}
   		 h=\frac{\delta}{N},\qquad T-t=Mh,
	\end{align*}
	every feasibility deadline is a grid point:
	\begin{align*}
    		d_j=T-j\delta=t+(M-jN)h=t_{M-jN}.
	\end{align*}
Equivalently, for every grid index $n$,
	\begin{align*}
    		t_n+j\delta\le T\quad\Longleftrightarrow\quad n+jN\le M
   			 \quad\Longleftrightarrow\quad   t_n\le d_j.
	\end{align*}
Thus the continuous-time condition ``there is enough time for $j$ further refractory periods'' 
	and its discrete-time counterpart change at exactly the same grid indices.
In particular,
	\begin{align*}
    		t_n+\delta>T\quad\Longleftrightarrow\quad  n+N>M.
	\end{align*}
Hence the continuous delayed residual-right value $\Psi_\delta^{[k-1]}(t_n,\cdot)$ 
	and the discrete delayed residual-right value $\Psi_n^{[k-1],h}$ vanish on exactly the same time range.
When they are nonzero, their definitions differ only through the lower-level value function.  
Therefore
	\begin{align*}
    		\bigl|\Psi_n^{[k-1],h}(y) -\Psi_\delta^{[k-1]}(t_n,y)\bigr|
    			&\le e^{-r\delta}J_{k-1}(h),
	\end{align*}
	and consequently
	\begin{equation}\label{eq:obs-error}
   		 \sup_{n,y}\bigl|g_n^{[k],h}(y)-g_\delta^{[k]}(t_n,y)\bigr|\le e^{-r\delta}J_{k-1}(h).
	\end{equation}
This is the basic error inherited from level $k-1$.

\emph{Step 2: comparison on bounded state sets.}
We first obtain an upper bound for the Bermudan value.  
A grid-valued stopping time is also an admissible stopping time in continuous time.  
If the same grid stopping time is used in the two problems, 
	the stopping rewards differ by at most the quantity in \eqref{eq:obs-error}.  
Taking suprema gives
	\begin{align*}
   		 V_n^{[k],h}(y)\le V^{[k]}(t_n,y)+e^{-r\delta}J_{k-1}(h).
	\end{align*}

For the reverse inequality, fix $Y>0$, $y\le Y$, and $\eta>0$.  
Choose an $\eta$-optimal continuous-time stopping time $\tau$ for $V^{[k]}(t_n,y)$, so that
	\begin{align*}
    		V^{[k]}(t_n,y)\le\E_{t_n,y}\!\left[e^{-r\tau}g_\delta^{[k]}(t_n+\tau,X_{t_n+\tau})\right]+\eta.
	\end{align*}
Round $\tau$ upward to the next grid point:
	\begin{align*}
    		\tau_h:=h\left\lceil\frac{\tau}{h}\right\rceil.
	\end{align*}
Then
	\begin{align*}
   		 \tau\le\tau_h\le\tau+h.
	\end{align*}
Moreover, $\tau_h$ is a stopping time for the sampled filtration because
	\begin{align*}
  		  \{\tau_h\le ih\}=\{\tau\le ih\}.
	\end{align*}
Conditional on the state at a grid time, the future Brownian increments over the grid intervals 
	are independent and identically distributed.
Hence the dynamic-programming recursion \eqref{eq:bermrec} gives the same Bermudan value 
	when the sampled Brownian filtration is used, and $\tau_h$ is admissible for $V_n^{[k],h}(y)$.

The alignment of the deadlines is important at this point.  
Since every $d_j$ is itself a grid point, rounding upward cannot jump across a deadline.  
If $t_n+\tau$ lies below a deadline, the next grid point is still below or at that deadline; 
	if it is exactly at a deadline, it is already a grid point and is unchanged by the rounding.  
Thus $t_n+\tau$ and $t_n+\tau_h$ lie in the same deadline strip.

On a fixed strip, the stopping reward is $k$-Lipschitz in the space variable and is $1/2$-H\"older in time.  
More precisely, Theorem~\ref{th:Vprops}(vi) at level $k-1$, inserted into the definition of $\Psi$, gives
	\begin{align*}
    		\bigl|g_\delta^{[k]}(s,z)-g_\delta^{[k]}(s',z)\bigr| \le C_{k-1}(1+z)|s-s'|^{1/2}
	\end{align*}
	whenever $s$ and $s'$ belong to the same strip.
Put
	\begin{align*}
  		 \Delta_h:=\tau_h-\tau, \qquad 0\le\Delta_h\le h.
	\end{align*}
To compare the rewards at $\tau$ and $\tau_h$, separate the error into three parts:
	the change in the discount factor, the change in time, and the change in the asset state.  
The triangle inequality gives
	\begin{align*}
		&\Bigl|\E\bigl[e^{-r\tau}g_\delta^{[k]}(t_n+\tau,X_{t_n+\tau})\bigr]
			-\E\bigl[e^{-r\tau_h}g_\delta^{[k]}(t_n+\tau_h,X_{t_n+\tau_h})\bigr]\Bigr|\\
		&\quad\le\E\!\left[ |e^{-r\tau}-e^{-r\tau_h}| g_\delta^{[k]}(t_n+\tau,X_{t_n+\tau})\right]\\
		&\qquad+\E\!\left[e^{-r\tau_h} \bigl|g_\delta^{[k]}(t_n+\tau,X_{t_n+\tau})
          		-g_\delta^{[k]}(t_n+\tau_h,X_{t_n+\tau})\bigr|\right]\\
		&\qquad+\E\!\left[ e^{-r\tau_h} \bigl|g_\delta^{[k]}(t_n+\tau_h,X_{t_n+\tau})
         		 -g_\delta^{[k]}(t_n+\tau_h,X_{t_n+\tau_h})\bigr|\right].
\end{align*}
The first term is $O(h)$ because $0\le g_\delta^{[k]}\le kK$ and
    	$	1-e^{-rh}\le rh.$ 
The second term is $O(\sqrt h)$ by the stripwise $1/2$-H\"older estimate.
For the third term, the $k$-Lipschitz property and the strong Markov property at $t_n+\tau$ give
	\begin{align*}
		\E\!\left[ |X_{t_n+\tau_h}-X_{t_n+\tau}|\mid\mathcal F_{t_n+\tau}\right]
			\le X_{t_n+\tau}\E\!\left[\sup_{0\le v\le h}|R_v-1|\right]
			\le cX_{t_n+\tau}\sqrt h.
	\end{align*}
Finally, the discounted asset price is a martingale, and therefore
	\begin{align*}
    		\E_{t_n,y}\!\left[e^{-r\tau}X_{t_n+\tau}\right]=y\le Y.
	\end{align*}
Combining the three estimates, and using $h\le\sqrt h$ for sufficiently small $h$, yields
	\begin{align*}
		\Bigl|\E\bigl[e^{-r\tau}g_\delta^{[k]}(t_n+\tau,X_{t_n+\tau})\bigr]
			-\E\bigl[e^{-r\tau_h}g_\delta^{[k]}(t_n+\tau_h,X_{t_n+\tau_h})\bigr]\Bigr|
		\le C(1+Y)\sqrt h=: \rho_Y(h),
	\end{align*}
	where $C$ is independent of $n$, $y\le Y$, $\eta$, and the particular $\eta$-optimal stopping time.

Since $\tau_h$ is admissible for the Bermudan problem, \eqref{eq:obs-error} now gives
	\begin{align*}
    		V^{[k]}(t_n,y)\le V_n^{[k],h}(y)+e^{-r\delta}J_{k-1}(h)+\rho_Y(h)+\eta,\qquad y\le Y.
	\end{align*}
Because $\eta>0$ is arbitrary, the discrete and continuous values are uniformly close 
	on every bounded state interval once $J_{k-1}(h)$ is small.

\emph{Step 3: control of large initial states.}
It remains to make the estimate uniform over all $y>0$.  
Lemma~\ref{lem:supermart}(ii) gives
	\begin{align*}
    		V^{[k]}(t_n,y)\le kV^{[1]}(t_n,y).
	\end{align*}
A one-right payoff can be positive only if the asset price falls below $K$ before maturity.  
Since every payoff is bounded by $K$,
	\begin{align*}
    		V^{[1]}(t_n,y)\le K\,   \Pp\!\left(\inf_{0\le u\le T}yR_u\le K\right).
	\end{align*}
The Bermudan recursion is the dynamic-programming recursion of a discrete multiple-stopping problem.  
Unfolding that recursion shows that its total payoff is a sum of at most $k$ put payoffs, 
	each bounded by $K$.  
A positive payoff is possible only if the grid process falls below $K$, 
	which is contained in the event that the continuous path falls below $K$.  
Hence the same tail bound holds for the Bermudan value.  
Thus, for $y\ge Y$,
	\begin{align*}
    		\max\{V^{[k]}(t_n,y),V_n^{[k],h}(y)\}\le kK\,
  		  \Pp\!\left(\inf_{0\le u\le T}R_u\le\frac{K}{Y}\right)=:kK\varpi(Y).
	\end{align*}
Because $R$ has continuous strictly positive paths,
	\begin{align*}
    		\varpi(Y)\longrightarrow0 \qquad\text{as }Y\longrightarrow\infty.
	\end{align*}

Combining the bounded-state estimate with the tail estimate gives
	\begin{align*}
    		J_k(h)\le e^{-r\delta}J_{k-1}(h) +\rho_Y(h)+kK\varpi(Y).
	\end{align*}
For fixed $Y$, let $h\downarrow0$.  
The induction hypothesis gives $J_{k-1}(h)\to0$, and $\rho_Y(h)\to0$.  
Then let $Y\uparrow\infty$ to obtain
	\begin{align*}
   		 J_k(h)\longrightarrow0.
	\end{align*}
This completes the induction.

Finally, the convergence of the delayed residual-right values follows directly from their definitions:
	\begin{align*}
		\bigl|\Psi_0^{[k-1],h}(x)-\Psi_\delta^{[k-1]}(t,x)\bigr|
		\le e^{-r\delta}J_{k-1}(h)\longrightarrow0.
	\end{align*}
Since
	\begin{align*}
    		f_0^{[k],h}=V_0^{[k],h}-\Psi_0^{[k-1],h},\qquad
    		f_\delta^{[k]}=V^{[k]}-\Psi_\delta^{[k-1]},
	\end{align*}
	the convergence of the marginal-right values follows by subtraction.
\end{proof}

The proof of Lemma~\ref{lem:berm-conv} uses only the recursive formulation \eqref{eq:Vm-rec}. 
We can now show that the recursive value is the value of the original multiple-stopping problem.
\begin{lemma}[Direct and recursive formulations coincide]
\label{lem:direct-rec}
For every $m\ge1$ and $(t,x)\in[0,T]\times(0,\infty)$,
	\begin{align*}
		V^{[m]}_{\mathrm{dir}}(t,x)=V^{[m]}(t,x).
	\end{align*}
\end{lemma}
\begin{proof}
We prove
	\begin{align*}
		V_{\mathrm{dir}}^{[m]}(t,x)=V^{[m]}(t,x)
	\end{align*}
	by induction on the number of exercise rights $m$.
For $m=1$, the two definitions are identical.
Assume therefore that
	\begin{align*}
	V_{\mathrm{dir}}^{[m-1]}(s,y)=V^{[m-1]}(s,y)
\end{align*}
	holds for every initial state $(s,y)$.
Recall also that, by \eqref{eq:extension}, both sides are defined to be zero
	when the initial time is later than $T$.

\medskip
\noindent
\emph{Step 1: the upper bound $V_{\mathrm{dir}}^{[m]}\le V^{[m]}$.}
Fix an arbitrary admissible $m$-exercise rule
	$\boldsymbol\tau^{(m)}=(\tau_1,\ldots,\tau_m).$ 
After the first exercise at time $\tau_1$, the holder must wait for the
	refractory period $\delta$ before another exercise is possible.
We therefore put
	$\vartheta:=\tau_1+\delta.$
The event
	$A_2=\{\vartheta\le T-t\}$
	is precisely the event that a second exercise can still be made before maturity.
On $A_2$, all later exercise times satisfy
	$\tau_i\ge \vartheta,\quad i=2,\ldots,m.$
Hence the shifted times
	$\tau_i-\vartheta,\quad i=2,\ldots,m,$ 
	form an admissible $(m-1)$-exercise rule for the problem starting from
	time $t+\vartheta$ and state $X_{t+\vartheta}$.
Indeed, the refractory-period constraints are unchanged by the shift.
For example,
	$\tau_k+\delta\le T-t$
	is equivalent to
	$(\tau_k-\vartheta)+\delta\le T-(t+\vartheta).$
Thus the remaining feasibility conditions are exactly those of the
	$(m-1)$-right problem starting at $(t+\vartheta,X_{t+\vartheta})$.
Condition now on $\mathcal F_{t+\vartheta}$.
Once the state $X_{t+\vartheta}$ is known, the strong Markov property says that
	the future evolution of the asset is governed by fresh Brownian increments,
	independent of the past.
Therefore the conditional value of any particular remaining
	$(m-1)$-exercise rule cannot exceed the optimal $(m-1)$-right value.
Consequently,
	\begin{align*}
		\E\!\left[\sum_{i=2}^m e^{-r\tau_i}g(X_{t+\tau_i})\one_{A_i}
			\Bigm|\mathcal F_{t+\vartheta}\right]
		\le\one_{A_2}e^{-r\vartheta}V_{\mathrm{dir}}^{[m-1]}(t+\vartheta,X_{t+\vartheta}).
	\end{align*}
By the induction hypothesis,
	$V_{\mathrm{dir}}^{[m-1]}=V^{[m-1]}$ on $A_2$.
On $A_2^c$ we have $t+\vartheta>T$, and the value is zero by \eqref{eq:extension}.
Hence
	\begin{align*}
		\E\!\left[\sum_{i=2}^m e^{-r\tau_i} g(X_{t+\tau_i}) \one_{A_i}
			\Bigm|\mathcal F_{t+\vartheta}\right]
				\le e^{-r\vartheta}V^{[m-1]}(t+\vartheta,X_{t+\vartheta}).
	\end{align*}
We next condition one step earlier, at the first exercise time $\tau_1$.
Since $\vartheta=\tau_1+\delta$, the strong Markov property gives
	\begin{align*}
		&\E\!\left[e^{-r\vartheta}V^{[m-1]}(t+\vartheta,X_{t+\vartheta})\Bigm|
			\mathcal F_{t+\tau_1}\right]\\
		&\qquad =e^{-r\tau_1}\E_{t+\tau_1,X_{t+\tau_1}}\!\left[e^{-r\delta}
			V^{[m-1]}(t+\tau_1+\delta,X_{t+\tau_1+\delta})\right]\\
		&\qquad = e^{-r\tau_1} \Psi_\delta^{[m-1]} (t+\tau_1,X_{t+\tau_1}).
	\end{align*}
Thus the expected payoff from the first exercise and all later exercises is
	bounded above by
	\begin{align*}
		\E_{t,x}\!\left[e^{-r\tau_1}\left\{g(X_{t+\tau_1})+\Psi_\delta^{[m-1]}
			(t+\tau_1,X_{t+\tau_1})\right\}\right].
	\end{align*}
By the definition of the stopping reward,
	$g_\delta^{[m]}(s,y)=g(y)+\Psi_\delta^{[m-1]}(s,y),$ 
	this is
	\begin{align*}
		\E_{t,x}\!\left[e^{-r\tau_1}g_\delta^{[m]}(t+\tau_1,X_{t+\tau_1})\right].
	\end{align*}
Since $V^{[m]}$ is the supremum of this quantity over all possible first stopping times,
	\begin{align*}
		\E_{t,x}\!\left[e^{-r\tau_1}g_\delta^{[m]}(t+\tau_1,X_{t+\tau_1})\right]
		\le V^{[m]}(t,x).
	\end{align*}
The original rule $\boldsymbol\tau^{(m)}$ was arbitrary, so taking the
	supremum over all admissible rules gives
	\begin{equation}\label{eq:direct-upper}
		V_{\mathrm{dir}}^{[m]}(t,x)\le V^{[m]}(t,x).
	\end{equation}

\medskip
\noindent
\emph{Step 2: the reverse inequality on an aligned Bermudan grid.}
We next prove
	\begin{align*}
		V_{\mathrm{dir}}^{[m]}(t,x)\ge V^{[m]}(t,x).
	\end{align*}
We first consider times $t$ for which $(T-t)/\delta\in\mathbb Q.$
For such a time, choose the Bermudan grid introduced above,
	\begin{align*}
		h=\frac{\delta}{N},\qquad T-t=Mh,
	\end{align*}
	so that every feasibility deadline is a grid point.
Realize the one-step factors by
	\begin{align*}
		R_i^{(h)}:=\frac{R_{ih}}{R_{(i-1)h}},
	\end{align*}
	so that at the grid times $t_n=t+nh$,
	\begin{align*}
		X_{t_n}=xR_1^{(h)}\cdots R_n^{(h)}.
	\end{align*}
The recursion \eqref{eq:bermrec} is precisely the dynamic-programming recursion 
	for the discrete-time $m$-right stopping problem on this grid.
At a grid point with $k$ rights remaining, it compares two alternatives:
	\begin{enumerate}
		\item exercise immediately and receive $g_n^{[k],h}$;
		\item continue for one more grid step.
	\end{enumerate}
Thus, by backward induction, the following feedback rule is optimal:
	with $k$ rights remaining at grid index $n$, exercise exactly when
	$V_n^{[k],h}=g_n^{[k],h}.$
After exercising, wait exactly $N$ grid steps, corresponding to the refractory
	period $Nh=\delta,$ and then continue with $k-1$ rights.
Because this policy is obtained by choosing at every grid point the branch
	that attains the maximum in the dynamic-programming recursion, its expected payoff is
	exactly $V_0^{[m],h}(x)$. 
Moreover, all its exercise times are grid-valued stopping times, and two
	successive exercises are separated by at least $Nh=\delta.$
Hence this Bermudan rule is also an admissible rule for the original
	continuous-time direct problem.
Therefore
	\begin{align*}
		V_{\mathrm{dir}}^{[m]}(t,x)\ge V_0^{[m],h}(x).
	\end{align*}
Now let $h\downarrow0$ along the aligned grids.
By Lemma~\ref{lem:berm-conv},
	\begin{align*}
		V_0^{[m],h}(x)\longrightarrow V^{[m]}(t,x).
	\end{align*}
Consequently,
	\begin{equation}\label{eq:direct-lower-rational}
		V_{\mathrm{dir}}^{[m]}(t,x)\ge V^{[m]}(t,x)
	\end{equation}
	whenever $(T-t)/\delta\in\mathbb Q$.

\medskip
\noindent
\emph{Step 3: extension to an arbitrary initial time.}
It remains to remove the rationality restriction on $(T-t)/\delta$.
First observe that, for fixed $x$,
	\begin{align*}
		t\longmapsto V_{\mathrm{dir}}^{[m]}(t,x)
	\end{align*}
	is nonincreasing.
	Indeed, if $t_1<t_2$, then the remaining horizon at $t_1$ is longer:
	$T-t_1>T-t_2.$
By time homogeneity, we may realize every relative stopping time as a functional of 
	the Brownian increments after the initial time, so that a rule for the horizon $T-t_2$ 
	induces a rule with the same law of payoffs over the horizon $T-t_1$. 
Such a rule need not be admissible for the longer horizon, however: 
	the separation constraint is imposed only on $\{\tau_i+\delta\le T-t_2\}$, 
	so it may fail on $\{T-t_2<\tau_i+\delta\le T-t_1\}$. 
We therefore modify it. 
Let $\boldsymbol\tau^{(m)}=(\tau_1,\ldots,\tau_m)$ be admissible for the horizon $T-t_2$, 
	with feasibility events $A_i$, and define recursively
	\begin{align*}
		\tau_1':=\tau_1,\qquad
		\tau_{i+1}':=\bigl(\tau_{i+1}\vee(\tau_i'+\delta)\bigr)\wedge(T-t_1),
		\qquad i=1,\ldots,m-1.
	\end{align*}
Each $\tau_i'$ is a stopping time with values in $[0,T-t_1]$, 
	$\tau_1'\le\cdots\le\tau_m'$, and $\tau_{i+1}'\ge\tau_i'+\delta$ on 
	$\{\tau_i'+\delta\le T-t_1\}$; hence $\boldsymbol\tau'^{(m)}$ is admissible 
	for the horizon $T-t_1$. 
Denote its feasibility events by $A_i'$. 
On $A_i$ we have $\tau_k+\delta\le T-t_2$ for $k<i$, hence $\tau_{k+1}\ge\tau_k+\delta$, 
	and induction on $k$ gives $\tau_k'=\tau_k$ for $k\le i$; 
	consequently $A_i\subseteq A_i'$. 
Thus, on each $A_i$, the $i$th exercise and its payoff are unchanged, 
	while all additional terms in the modified payoff are nonnegative. 
Taking expectations and then the supremum over $\boldsymbol\tau^{(m)}$, 
	the value cannot decrease.
Thus
	\begin{align*}
		V_{\mathrm{dir}}^{[m]}(t_1,x)\ge V_{\mathrm{dir}}^{[m]}(t_2,x).
	\end{align*}
If $t$ itself is a feasibility deadline, then
	$(T-t)/\delta$
	is an integer, so Step 2 already applies.
Suppose now that $t$ is not a deadline.
Then $t$ is not the right endpoint of the deadline strip containing it.
Choose a sequence
	$t_n\downarrow t$ 
	within the same strip such that
	$(T-t_n)/\delta\in\mathbb Q$
	for every $n$.
By the time monotonicity just proved,
	\begin{align*}
		V_{\mathrm{dir}}^{[m]}(t,x)\ge V_{\mathrm{dir}}^{[m]}(t_n,x).
	\end{align*}
Step 2 gives
	$V_{\mathrm{dir}}^{[m]}(t_n,x)\ge V^{[m]}(t_n,x).$ 
Finally, Theorem~\ref{th:Vprops}(vii) gives continuity of $V^{[m]}$ in $t$
	within the strip, so
	\begin{align*}
		V^{[m]}(t_n,x)\longrightarrow V^{[m]}(t,x).
	\end{align*}
Hence
	\begin{align*}
		V_{\mathrm{dir}}^{[m]}(t,x)
			\ge V_{\mathrm{dir}}^{[m]}(t_n,x)
			\ge V^{[m]}(t_n,x)\longrightarrow V^{[m]}(t,x),
	\end{align*}
	and therefore
	$V_{\mathrm{dir}}^{[m]}(t,x)\ge V^{[m]}(t,x).$ 
Combining this inequality with \eqref{eq:direct-upper} proves
	$V_{\mathrm{dir}}^{[m]}(t,x)=V^{[m]}(t,x),$
	as required.
\end{proof}

\begin{lemma}[Single-crossing property]
\label{lem:singlecross}
For $m\ge2$ and fixed $t$, define the gap function
	\begin{align*}
		 G_t^{[m]}(x):=f_\delta^{[m]}(t,x)-(K-x)^+.
	\end{align*}
Then $G_t^{[m]}$ is nondecreasing on $(0,K)$. 
Equivalently, for $0<y<x<K$,
	\begin{align*}
		 G_t^{[m]}(x)\ge G_t^{[m]}(y).
	\end{align*}
\end{lemma}
\begin{proof}
The statement will follow once we prove the slope bound
	\begin{equation}\label{eq:slope}
    		f_\delta^{[m]}(t,x)-f_\delta^{[m]}(t,y) \ge -(x-y), \qquad 0<y<x.
	\end{equation}
Indeed, for $0<y<x<K$,
	\begin{align*}
    		G_t^{[m]}(x)-G_t^{[m]}(y)&=f_\delta^{[m]}(t,x)-f_\delta^{[m]}(t,y)+(x-y),
	\end{align*}
	so \eqref{eq:slope} is exactly the assertion that $G_t^{[m]}$ is nondecreasing.

We first reduce the continuous-time problem to the aligned Bermudan problem.  
By Proposition~\ref{prop:stage}, if $t$ lies in a later deadline strip, 
	both $V^{[m]}$ and its delayed residual-right term reduce to the corresponding lower-right level,
	so that $f_\delta^{[m]}(t,\cdot)=f_\delta^{[j]}(t,\cdot)$ for $t\in(d_j,d_{j-1}]$;
	for $j=1$ the slope bound below is immediate because $f_\delta^{[1]}=V^{[1]}$ is $1$-Lipschitz.
Arguing by induction on $m$, it is therefore enough to consider the main strip $t\le d_{m-1}$.
On that strip the marginal-right value is continuous in $t$ by Lemma~\ref{lem:fprops}.  
Hence it suffices to prove \eqref{eq:slope} for the dense set of times 
	for which the aligned Bermudan construction is available, and then pass to the limit.

The key point in the Bermudan argument is a simultaneous induction.  	
For each exercise level $k$ we prove two properties:
	\begin{align*}
    		(\mathrm{SC}_k):\quad f_n^{[k],h}(x)-f_n^{[k],h}(y)
    			&\ge -(x-y), \qquad 0<y<x,
	\end{align*}
	and
	\begin{align*}
    		(\mathrm E_k):\quad x\longmapsto e_n^{[k]}(x) &\text{ is nonincreasing.}
	\end{align*}
The logical order of the induction is
	\begin{align*}
   		 (\mathrm E_{k-1})\quad\Longrightarrow\quad (\mathrm{SC}_k)
    			\quad\Longrightarrow\quad (\mathrm E_k).
	\end{align*}
Thus there is no circularity: monotonicity of the exercise premium 
	at level $k-1$ gives the slope bound at level $k$, 
	and that slope bound is then used to prove monotonicity 
	of the exercise premium at level $k$.

\emph{Base level $k=1$.}
Since $f_n^{[1],h}=V_n^{[1],h}$ and the one-right Bermudan value is one-Lipschitz,
	\begin{align*}
    		f_n^{[1],h}(x)-f_n^{[1],h}(y)\ge -(x-y),
	\end{align*}
	so $(\mathrm{SC}_1)$ holds.
At maturity,
    	$	e_M^{[1]}(x)=(K-x)^+$ 
	is nonincreasing.  
For $n<M$, \eqref{eq:e-zero} gives
	\begin{align*}
    		e_n^{[1]}(x)=0,\qquad x\ge K.
	\end{align*}
On $(0,K)$, use the bracket representation \eqref{eq:ebracket}.  
If $0<y<x<K$, the payoff part changes by exactly $-(x-y)$.  
The continuation value is one-Lipschitz, and
    	$	e^{-rh}\E[R_1^{(h)}]=1,$ 
	so the absolute value of the corresponding continuation-value difference is at most $x-y$.  
Therefore the bracket cannot increase when $x$ increases.  
Taking the positive part preserves this monotonicity, and $(\mathrm E_1)$ follows.

\emph{Induction step.}
Assume that $(\mathrm{SC}_{k-1})$ and $(\mathrm E_{k-1})$ have been proved.  
From \eqref{eq:correction},
	\begin{align*}
    		c_n^{[k-1]}(x)=e^{-rNh}\E\!\left[e_{n+N}^{[k-1]}(xR_1^{(h)}\cdots R_N^{(h)})\right].
	\end{align*}
Because $e_{n+N}^{[k-1]}$ is nonincreasing by $(\mathrm E_{k-1})$ 
	and the multiplicative factors are positive, 
	the correction term $c_n^{[k-1]}$ is also nonincreasing in $x$.
We next prove $(\mathrm{SC}_k)$ by backward induction in the grid index $n$.  
At maturity,
	\begin{align*}
    		f_M^{[k],h}(x)=(K-x)^+,
	\end{align*}
	which has lower slope bound $-1$.  
Assume the slope bound holds at time $n+1$.  
For $0<y<x$, the continuation branch in \eqref{eq:corrected-rec} satisfies
	\begin{align*}
		&e^{-rh}\E\!\left[f_{n+1}^{[k],h}(xR_1^{(h)})-f_{n+1}^{[k],h}(yR_1^{(h)})\right]
			-\bigl(c_n^{[k-1]}(x)-c_n^{[k-1]}(y)\bigr)\\
		&\qquad\ge -(x-y)e^{-rh}\E[R_1^{(h)}]+0 =-(x-y).
	\end{align*}
Here the first inequality uses the induction hypothesis at time $n+1$, 
	while the second contribution is nonnegative because $c_n^{[k-1]}$ is nonincreasing.
The immediate-payoff branch $(K-x)^+$ has the same lower slope bound.  
Finally, for any real numbers $a_1,a_2,b_1,b_2$,
	\begin{align*}
    		\max\{a_1,b_1\}-\max\{a_2,b_2\}\ge \min\{a_1-a_2,b_1-b_2\}.
	\end{align*}
Therefore, if each branch has slope bounded below by $-1$, their maximum has the same property.  
This proves $(\mathrm{SC}_k)$.

It remains to prove $(\mathrm E_k)$.  
Again, \eqref{eq:e-zero} gives
    	$e_n^{[k]}(x)=0,\ x\ge K.$ 
Consider $0<y<x<K$.  
The bracket in \eqref{eq:ebracket} changes from $y$ to $x$ by three contributions.  
The payoff contributes $-(x-y)$. 
By $(\mathrm{SC}_k)$,
	\begin{align*}
    		f_{n+1}^{[k],h}(xR_1^{(h)})-f_{n+1}^{[k],h}(yR_1^{(h)})\ge -(x-y)R_1^{(h)},
	\end{align*}
	so after multiplying by $-e^{-rh}$ and taking expectations, 
	the continuation contribution is at most
	\begin{align*}
    		(x-y)e^{-rh}\E[R_1^{(h)}]=x-y.
	\end{align*}
Finally, because $c_n^{[k-1]}$ is nonincreasing,
	\begin{align*}
    		c_n^{[k-1]}(x)-c_n^{[k-1]}(y)\le0.
	\end{align*}
The sum of the three contributions is therefore nonpositive.  
Hence the bracket in \eqref{eq:ebracket} is nonincreasing on $(0,K)$, and so is its positive part.  
Together with $e_n^{[k]}=0$ on $[K,\infty)$, this proves $(\mathrm E_k)$.

We have now established $(\mathrm{SC}_m)$ on every aligned Bermudan grid.  
Letting $h\downarrow0$ and using Lemma~\ref{lem:berm-conv} gives \eqref{eq:slope} 
	for every aligned time $t$.  
Continuity of $f_\delta^{[m]}$ in $t$ extends the inequality from the dense aligned set to all $t\le d_{m-1}$, 
	and Proposition~\ref{prop:stage} gives the result on the later strips.  
Therefore, for $0<y<x<K$,
	\begin{align*}
  		  G_t^{[m]}(x)-G_t^{[m]}(y)=f_\delta^{[m]}(t,x)-f_\delta^{[m]}(t,y)+(x-y) \ge0,
	\end{align*}
	which is the desired single-crossing property.
\end{proof}

\section{Exercise boundaries and optimal hitting times}\label{sec:boundary}
\subsection{Strict loss from forced waiting}
\begin{lemma}[Strict loss from forced waiting]\label{lem:strict-wait}
Under the standing assumptions $r>0$, $\sigma>0$, $K>0$ and optional
	exercise rights, for every $k\ge1$, $t<T$, $x>0$ and $0<h\le T-t$,
	\begin{equation}\label{eq:strict-wait}
		 V^{[k]}(t,x)>\E_{t,x}\bigl[e^{-rh}V^{[k]}(t+h,X_{t+h})\bigr].
	\end{equation}
Consequently, with the zero extension beyond maturity,
	\begin{equation}\label{eq:strict-W-all}
 		W^{[k]}(t,x):=V^{[k]}(t,x)-\Psi_\delta^{[k]}(t,x)>0,
			 \qquad k\ge1,\quad t<T,\quad x>0.
	\end{equation}

\end{lemma}
\begin{proof}
\emph{Step 1: a common region of immediate exercise.}
Lemma~\ref{lem:supermart}(ii) and value domination give, for $k\ge2$,
	\begin{align*}
		 0\le V^{[k]}(s,y)-g_\delta^{[k]}(s,y)\le V^{[1]}(s,y)-g(y).
	\end{align*}
Thus, for every $k\ge1$ and $0<y\le b^{[1]}(s)$,
	\begin{equation}\label{eq:common-contact-strict}
 		V^{[k]}(s,y)=g(y)+\Psi_\delta^{[k-1]}(s,y),
	\end{equation}
	where $\Psi_\delta^{[0]}=0$. 
This uses only the known one-right boundary,
	not the existence or regularity of a multiple-right boundary.

\emph{Step 2: a strict local loss inside the common region.}
Fix $s<T$, $\varepsilon>0$ with $s+\varepsilon<T$, and $0<\ell<u<b^{[1]}(s)$. 
For $y\in(\ell,u)$, set
	\begin{align*}
 		\theta=\varepsilon\wedge\inf\{v\ge0:X_{s+v}\notin(\ell,u)\}.
	\end{align*}
Path continuity implies $\theta>0$ almost surely and $X_{s+\theta}\in[\ell,u]$. 
The one-right boundary is nondecreasing,
	so \eqref{eq:common-contact-strict} holds both at $(s,y)$ and at
	$(s+\theta,X_{s+\theta})$. 
On this stopped interval $g(X)=K-X$.
It\^o's formula for $e^{-rv}(K-X_{s+v})$, together with the
	optional-sampling inequality for the delayed residual-right value in
	Lemma~\ref{lem:supermart}(i), gives
	\begin{align}
 		V^{[k]}(s,y)-\E_{s,y}\bigl[e^{-r\theta}
          		V^{[k]}(s+\theta,X_{s+\theta})\bigr]
 		\ge rK\,\E_{s,y}\!\left[\int_0^\theta e^{-rv}\dd v\right]>0.
 			\label{eq:strict-local-loss}
	\end{align}
The stochastic integral has mean zero because $X$ is stopped in the
	compact interval $[\ell,u]$ and $\theta\le\varepsilon$.
Only the linear payoff is differentiated; the delayed residual-right value is handled
	by optional sampling, including at deadlines.
	
For every $y>0$, define the deterministic waiting loss
	\begin{align*}
 		A_k(s,y;\varepsilon):=V^{[k]}(s,y)-\E_{s,y}\bigl[e^{-r\varepsilon}
                 V^{[k]}(s+\varepsilon,X_{s+\varepsilon})\bigr].
	\end{align*}
It is nonnegative by Lemma~\ref{lem:supermart}(i).
For $y\in(\ell,u)$, optional sampling between $\theta$ and
	$\varepsilon$ and \eqref{eq:strict-local-loss} imply $A_k(s,y;\varepsilon)>0$.

\emph{Step 3: propagate strictness to every initial state.}
Fix $t<T$, $x>0$ and $0<h\le T-t$. 
Put $a=h/3$, $\varepsilon=h/3$ and $s=t+a$, and choose $0<\ell<u<b^{[1]}(s)$. 
Notice that $s+\varepsilon<T$.
The supermartingale inequalities from $t$ to $s$ and from
	$s+\varepsilon$ to $t+h$, and the Markov property, yield
	\begin{align}
		 V^{[k]}(t,x)-\E_{t,x}\bigl[e^{-rh}V^{[k]}(t+h,X_{t+h})\bigr]
 		\ge\E_{t,x}\bigl[e^{-ra}A_k(s,X_s;\varepsilon)\bigr]>0.
		\label{eq:propagate-strict}
	\end{align}
The final inequality follows because the lognormal law of $X_s$ has a
	strictly positive density on $(0,\infty)$, while $A_k$ is nonnegative
	everywhere and strictly positive on $(\ell,u)$. 
All quantities are bounded by constants depending on $kK$, so the expectations are finite.
This proves \eqref{eq:strict-wait} even when $[t,t+h]$ crosses a deadline.

If $t+\delta\le T$, take $h=\delta$. 
If $t+\delta>T$, then $\Psi_\delta^{[k]}(t,x)=0$ and $V^{[k]}(t,x)\ge V^{[1]}(t,x)>0$,
	the last inequality following from the positive probability of an
	in-the-money payoff at maturity. 
This proves \eqref{eq:strict-W-all}.
\end{proof}
\begin{remark}[Scope of Lemma~\ref{lem:strict-wait}]
\label{rem:strict-wait-scope}
The proof of Lemma~\ref{lem:strict-wait} assumes no inequality between marginal-right values 
	and uses no differentiability of
 $V^{[k]}$ or $\Psi_\delta^{[k-1]}$: only the linear payoff is differentiated, 
 	and the delayed residual-right value is handled by optional sampling.
\end{remark}
\subsection{\texorpdfstring{Incremental values of successive exercise rights}
{Incremental values of successive exercise rights}}\label{subsec:incremental-values}
For $k\ge1$, define the \emph{incremental value of the $k$th right} by
	\begin{align*}
		 D^{[k]}(t,x):=V^{[k]}(t,x)-V^{[k-1]}(t,x),\qquad V^{[0]}\equiv0.
	\end{align*}
Thus $D^{[k]}$ compares adjacent exercise-right levels directly and measures the
	additional value obtained by adding the $k$th exercise right. 
This is distinct from the marginal-right value $f_\delta^{[k]}=V^{[k]}-\Psi_\delta^{[k-1]}$, 
	which subtracts the delayed residual-right value.
Although a difference of supermartingales need not be a supermartingale,
	the particular recursion of this problem provides the required comparison.
The key is to compare the one-step exercise premiums across exercise levels.
\begin{lemma}[Ordering of Bermudan exercise premiums]\label{lem:premium-rights}
On every aligned Bermudan grid of Section~\ref{sec:singlecross}, for $k\ge2$,
	\begin{equation}\label{eq:premium-rights}
		 e_n^{[k]}\ge e_n^{[k-1]},\qquad f_n^{[k],h}\le f_n^{[k-1],h}.
	\end{equation}
The inequalities hold pointwise in the state variable at every grid time.
Writing $D_n^{[k],h}:=V_n^{[k],h}-V_n^{[k-1],h}$ and
	$\mathcal P_h a(x):=e^{-rh}\E[a(xR_h)]$, we also have
	\begin{equation}\label{eq:HD-grid}
 		D_n^{[k],h}-\mathcal P_hD_{n+1}^{[k],h}=e_n^{[k]}-e_n^{[k-1]}\ge0.
	\end{equation}
For $k=1$, the same supermartingale inequality follows from $e_n^{[1]}\ge0$.
\end{lemma}
\begin{proof}
We induct on $k$, using a backward induction in $n$ within each level.
Set $e^{[0]}\equiv0$ and $c^{[0]}\equiv0$.
The base comparison $e_n^{[1]}\ge e_n^{[0]}$ follows directly from the dynamic-programming recursion.
Suppose that $e_n^{[k-1]}\ge e_n^{[k-2]}$ is already known for all $n$.
Positivity of the transition operator and \eqref{eq:correction} give
	\begin{align*}
		 c_n^{[k-1]}=\mathcal P_h^N e_{n+N}^{[k-1]}\ge \mathcal P_h^N e_{n+N}^{[k-2]}=c_n^{[k-2]}.
	\end{align*}
At maturity, $f_M^{[k],h}=f_M^{[k-1],h}=g$.
If $f_{n+1}^{[k],h}\le f_{n+1}^{[k-1],h}$, the corrected recursion gives
	\begin{align*}
 		f_n^{[k],h}&=\max\{g,\mathcal P_hf_{n+1}^{[k],h}-c_n^{[k-1]}\}\\
 			&\le\max\{g,\mathcal P_hf_{n+1}^{[k-1],h}-c_n^{[k-2]}\}=f_n^{[k-1],h}.
	\end{align*}
Backward induction proves this ordering at every time.
Now use the bracket identity \eqref{eq:ebracket}:
	\begin{align*}
 		e_n^{[k]}&=\bigl(g-\mathcal P_hf_{n+1}^{[k],h}+c_n^{[k-1]}\bigr)^+\\
 			&\ge\bigl(g-\mathcal P_hf_{n+1}^{[k-1],h}+c_n^{[k-2]}\bigr)^+ =e_n^{[k-1]},\qquad n<M.
	\end{align*}
At $n=M$ both premiums equal $g$, and beyond maturity both are zero.
This completes the induction. Its order is
	\begin{align*}
		 e^{[k-1]}\ge e^{[k-2]} \ \Longrightarrow\ c^{[k-1]}\ge c^{[k-2]}
 			\ \Longrightarrow\ f^{[k],h}\le f^{[k-1],h} \ \Longrightarrow\ e^{[k]}\ge e^{[k-1]}.
	\end{align*}
Thus no ordering at level $k$ is assumed in its own proof.
Subtracting the definitions of $e^{[k]}$ and $e^{[k-1]}$ proves
\eqref{eq:HD-grid}.
\end{proof}
\begin{theorem}[Supermartingale property of the incremental values]\label{th:HD}
Under the standing model assumptions, for every finite $k\ge1$,
	$0\le t\le s\le T$ and $x>0$,
	\begin{equation}\label{eq:HD-deterministic}
 		D^{[k]}(t,x)\ge\E_{t,x}\bigl[e^{-r(s-t)}D^{[k]}(s,X_s)\bigr].
	\end{equation}
With zero extension beyond $T$, the discounted process associated with the
	incremental value of the $k$th right is a bounded strong supermartingale:
	for bounded stopping times $0\le\rho\le\eta$,
	\begin{equation}\label{eq:HD-strong}
		 \E_{t,x}\bigl[e^{-r\eta}D^{[k]}(t+\eta,X_{t+\eta}) \mid\mathcal F_{t+\rho}\bigr]
 			\le e^{-r\rho}D^{[k]}(t+\rho,X_{t+\rho}).
	\end{equation}

For $k\ge2$ we further have
	\begin{equation}\label{eq:f-rights-cont}
 		f_\delta^{[k]}(t,x)\le f_\delta^{[k-1]}(t,x), \qquad 0\le t\le T,\quad x>0,
	\end{equation}
	and
	\begin{equation}\label{eq:W-rights-cont}
 		W^{[k]}(t,x)\ge W^{[k-1]}(t,x),\qquad 0\le t\le T,\quad x>0.
	\end{equation}

\end{theorem}
\begin{proof}
\emph{Step 1: deterministic times from an aligned initial time.}
First suppose $(T-t)/\delta\in\mathbb Q$ and choose the aligned grids of
	Lemma~\ref{lem:berm-conv}. Fix $s\in[t,T]$, and let $s_h$ be the first grid
	time at or after $s$. Iterating \eqref{eq:HD-grid} yields
	\begin{align*}
 		D_0^{[k],h}(x)\ge \E_{t,x}\bigl[e^{-r(s_h-t)}D_{n_h}^{[k],h}(X_{s_h})\bigr],
 			\qquad s_h=t+n_hh.
	\end{align*}
The proof of Lemma~\ref{lem:berm-conv} gives
	\begin{align*}
 		\sup_{n,y>0}|D_n^{[k],h}(y)-D^{[k]}(t_n,y)| \le J_k(h)+J_{k-1}(h)\longrightarrow0.
	\end{align*}
If $s$ is a deadline or $T$, it is a grid point and $s_h=s$.
Otherwise $s_h$ eventually lies in the same open strip as $s$.
Joint continuity on that strip and path continuity imply
	$D^{[k]}(s_h,X_{s_h})\to D^{[k]}(s,X_s)$ almost surely.
The bound $0\le D^{[k]}\le kK$, together with the analogous Bermudan bound,
	allows dominated convergence. 
This proves \eqref{eq:HD-deterministic} from every aligned initial time.

\emph{Step 2: arbitrary initial times.}
Every deadline is an aligned time. If $t$ is not a deadline and $t<s$,
	choose aligned $t_j\downarrow t$, with $t_j<s$ in the same strip as $t$.
Step 1 gives, using time homogeneity,
	\begin{align*}
		 D^{[k]}(t_j,x)\ge \E\bigl[e^{-r(s-t_j)}D^{[k]}(s,xR_{s-t_j})\bigr].
	\end{align*}
Stripwise continuity gives convergence on the left. 
On the right, spatial continuity at the fixed time $s$, continuity of $R$, and boundedness
	give convergence by dominated convergence. 
This proves \eqref{eq:HD-deterministic} for all $t\le s$; $s=t$ is immediate.
The same limiting argument applied to \eqref{eq:premium-rights}, 
	using convergence of the delayed residual-right values, proves \eqref{eq:f-rights-cont}.
For $s>T$, the deterministic inequality follows from the zero extension
	and nonnegativity.

\emph{Step 3: bounded stopping times, including the deadlines.}
Fix $(t,x)$ and write
	\begin{align*}
		 Y_u=e^{-ru}D^{[k]}(t+u,X_{t+u}).
	\end{align*}
The Markov property and the deterministic inequalities make $Y$ a
	bounded supermartingale at deterministic times. 
To justify optional sampling without assuming right continuity at a deadline, 
	choose finite deterministic grids on $[0,B]$, 
	where $B$ bounds $\eta$, with mesh tending
	to zero and containing every deadline offset $d_j-t$ in this interval,
	as well as $T-t$ when applicable. 
Round $\rho$ and $\eta$ upward on these grids to $\rho_n\le\eta_n$. 
These are finite-valued stopping times.
For every event $A$ in $\mathcal F_{t+\rho}$, which is contained in
	$\mathcal F_{t+\rho_n}$, finite-grid optional sampling gives
	\begin{align*}
 		\E_{t,x}[\one_A Y_{\eta_n}]\le\E_{t,x}[\one_A Y_{\rho_n}].
	\end{align*}
At a deadline the rounding leaves the time unchanged; elsewhere the
	value process is right-continuous along each path. 
Consequently $Y_{\rho_n}\to Y_\rho$ and $Y_{\eta_n}\to Y_\eta$ almost surely.
Bounded convergence now proves \eqref{eq:HD-strong}.

Finally, take $s=t+\delta$ in the deterministic inequality, with the
	zero extension when necessary. 
Linearity gives
	\begin{align*}
 		D^{[k]}(t,x)-\E_{t,x}[e^{-r\delta}D^{[k]}(t+\delta,X_{t+\delta})]
 		=W^{[k]}(t,x)-W^{[k-1]}(t,x)\ge0.
	\end{align*}
This proves \eqref{eq:W-rights-cont} for all $x>0$.
\end{proof}
\begin{remark}[Scope of Theorem~\ref{th:HD}]
\label{rem:HD-conclusion}
The proof of Theorem~\ref{th:HD} uses only the aligned Bermudan scheme
	(Lemmas~\ref{lem:berm-conv} and~\ref{lem:premium-rights}) and stripwise continuity;
	it uses no boundary ordering, no time monotonicity of a multiple-right boundary, no classical regularity of the
	delayed residual-right values, and no premium representation.
	This is why Theorem~\ref{th:HD} can be used below to prove the boundary hierarchy.
\end{remark}
\subsection{Weak regularity of the delayed residual-right values}\label{sec:weak-R}
We next obtain time regularity without a boundary regularity hypothesis.
Write
	\begin{align*}
 		\mathcal P_s\phi(x):=\E[e^{-rs}\phi(xR_s)]
	\end{align*}
	for the discounted Black--Scholes semigroup. 
In this subsection let
	\begin{align*}
 		\mathcal D_k&:=\{T-j\delta:1\le j\le k-1\}\cap[0,T),\\
 		\mathcal E_k&:=\{T-j\delta:1\le j\le k\}\cap[0,T),\qquad
 			Q_k:=rK\sum_{j=0}^{k-1}e^{-rj\delta}\le krK.
	\end{align*}
The sets $\mathcal D_k$ and $\mathcal E_k$ are the possible deadlines in $[0,T)$
of $V^{[k]}$ and $\Psi_\delta^{[k]}$, respectively; zero is included when it is a deadline.
\begin{lemma}[A small-step bound away from deadline nodes]\label{lem:premium-small-step}
On an aligned grid $h=\delta/N$, $T=t_*+Mh$, the exercise premiums
	of \eqref{eq:e-h} satisfy, for $0\le n<M$ with
	$n\notin\{M-jN:0\le j\le k-1\}$,
	\begin{equation}\label{eq:premium-small-step}
		 0\le e_n^{[k]}(x) \le K(1-e^{-rh})\sum_{j=0}^{k-1}e^{-rj\delta}
 			\le Q_kh,\qquad x>0.
	\end{equation}
Only indices belonging to the grid are excluded.
\end{lemma}
\begin{proof}
By Jensen's inequality and $\E[e^{-rh}R_h]=1$,
	\begin{align*}
 		\mathcal P_hg(x)&=\E[(Ke^{-rh}-e^{-rh}xR_h)^+]\ge(Ke^{-rh}-x)^+.
	\end{align*}
Hence $g-\mathcal P_hg\le K(1-e^{-rh})=:a_h$.
Since $f_{n+1}^{[k],h}\ge g$, \eqref{eq:correction} and \eqref{eq:ebracket} give, for $n<M$,
	\begin{align*}
		 e_n^{[k]} =\bigl(g-\mathcal P_h f_{n+1}^{[k],h}
             		+\mathcal P_\delta e_{n+N}^{[k-1]}\bigr)^+
 			\le a_h+\mathcal P_\delta e_{n+N}^{[k-1]}.
	\end{align*}
Induction on $k$ proves the result, since $\mathcal P_\delta1=e^{-r\delta}$.
If $n+N$ is a deadline node of level $k-1$, then $n$ is precisely
	an excluded node of level $k$; for $n+N>M$ the premium is zero.
Finally $1-e^{-rh}\le rh$. The exclusion is essential: $e_M^{[k]}=g$,
	and premiums at other deadline nodes may also be of order one.
\end{proof}
\begin{theorem}[Weak delayed-value regularity]\label{th:R-weak}
Under the standing model assumptions, for every finite $k\ge1$ there is
	a bounded measurable function $a^{[k]}$ on $(0,T)\times(0,\infty)$ such that
	\begin{equation}\label{eq:weak-density}
		 0\le a^{[k]}\le Q_k,\qquad -(\partial_t+\cL-r)V^{[k]}=a^{[k]}
	\end{equation}
	where the bounds hold almost everywhere and the equation holds in the sense of
	distributions on $\bigl((0,T)\setminus\mathcal D_k\bigr)\times(0,\infty)$.
On each open component of $(0,T)\setminus\mathcal E_k$,
	all spatial derivatives of $\Psi_\delta^{[k]}$ are jointly continuous
	and locally bounded, and its weak time derivative is bounded.
For $t<T-\delta$, set
	\begin{equation}\label{eq:Gamma-weak}
		 \mathcal B_k(t,x):=\mathcal P_\delta[a^{[k]}(t+\delta,\cdot)](x),
		 \qquad (r-\partial_t-\cL)\Psi_\delta^{[k]}=\mathcal B_k .
	\end{equation}
The last equality is distributional on each open component, and
	$0\le\mathcal B_k\le e^{-r\delta}Q_k$ almost everywhere.
For $t>T-\delta$, both $\Psi_\delta^{[k]}$ and its weak derivatives vanish.
With $\mu=r-\sigma^2/2$ and
	\begin{align*}
		 C_\delta:=e^{-r\delta}
		 	\left(r+\frac{|\mu|\sqrt{2/\pi}}{\sigma\sqrt\delta}+\frac1\delta\right),
	\end{align*}
	the quantitative estimate
	\begin{equation}\label{eq:Psi-time-Lipschitz}
		 |\Psi_\delta^{[k]}(t,x)-\Psi_\delta^{[k]}(s,x)| \le \bigl(kKC_\delta+e^{-r\delta}Q_k\bigr)|t-s|
	\end{equation}
	holds for every $x>0$ and $s,t$ in the same open component.
No regularity across a deadline is asserted.
\end{theorem}
\begin{proof}
The proof is given in Appendix~\ref{app:weak-regularity-proof}.
\end{proof}
The weak theorem leaves open only the continuity of the weak time derivative.
The one-right case can be identified explicitly, and after the exercise boundary has been constructed
	we close the higher-level regularity recursively by a localized level-set argument; see
Theorem~\ref{th:finite-level-regularity} below.

\begin{proposition}[Regularity of the one-right delayed residual-right value]\label{prop:R-one}
Under the standing model assumptions, $\Psi_\delta^{[1]}$ is $C^{1,2}$
	on $t<T-\delta$, with continuous, locally bounded derivatives, and
	\begin{equation}\label{eq:Gamma-one}
 		(r-\partial_t-\cL)\Psi_\delta^{[1]}(t,x)
			 =rK e^{-r\delta}\Pp_{t,x}\{X_{t+\delta}\le b^{[1]}(t+\delta)\}.
	\end{equation}
For $t>T-\delta$ it is identically zero. 
Thus the classical regularity of the delayed residual-right value holds at level $1$ without an additional assumption. Moreover, for every $k\ge1$,
	$\Psi_\delta^{[k]}=\Psi_\delta^{[1]}$ on
	$(T-2\delta,T-\delta)\cap[0,T]$.
\end{proposition}
\begin{proof}
Use the discounted semigroup introduced in Section~\ref{sec:weak-R}.
Applying $\mathcal P_\delta$ to the classical one-right premium formula gives
	\begin{equation}\label{eq:Psi-one-semigroup}
		 \Psi_\delta^{[1]}(t,x)=\mathcal P_{T-t}g(x)+rK\int_{t+\delta}^T
 		\mathcal P_{u-t}\one_{(0,b^{[1]}(u)]}(x)\dd u.
	\end{equation}
Every semigroup time in this expression is at least $\delta>0$.
The lognormal kernel and its time and first two space derivatives have
	integrable bounds, uniform on compact sets of $x>0$ and on semigroup
	times in $[\delta,T]$. 
The payoff and the indicators are bounded.
Consequently the kernels may be differentiated under the integral.
Continuity of $b^{[1]}$ also makes the lower-limit term continuous.
Since $(\partial_t+\cL-r)\mathcal P_{u-t}=0$ for fixed $u$,
	only the moving lower limit contributes, yielding \eqref{eq:Gamma-one}.
This proves the asserted joint continuity and local bounds; no derivative
	of the boundary is used. 
The identity for higher levels on the final pre-deadline strip follows 
	from stagewise reduction at time $t+\delta$.
No regularity across $t=T-\delta$ is asserted.
\end{proof}

The proposition supplies the base case for the higher-level induction developed after the
	boundary and optimal-hitting results.  
No classical regularity assumption is imposed at higher levels.
\begin{lemma}[Weak superharmonicity of the obstacle]
\label{lem:obstacle-gen}
Under the standing model assumptions, on every open time strip on which the weak identity of
	Theorem~\ref{th:R-weak} applies, and for $x<K$,
	\begin{align}
 		(\partial_t+\cL-r)g_\delta^{[m]}=-rK-\mathcal B_{m-1}\le-rK<0
		\label{eq:obstacle-generator}
	\end{align}
	in the sense of distributions.  
After Theorem~\ref{th:finite-level-regularity},
	$\mathcal B_{m-1}$ has the continuous representative $\Gamma_\delta^{[m-1]}$, 
	and the same identity holds classically on the interiors of the open deadline strips.
\end{lemma}
\begin{proof}
For $x<K$ the payoff is affine and
	\begin{align*}
		(\partial_t+\cL-r)(K-x)=-rK.
	\end{align*}
Theorem~\ref{th:R-weak} gives, distributionally on each open delayed deadline strip,
	\begin{align*}
		 (\partial_t+\cL-r)\Psi_\delta^{[m-1]}=-\mathcal B_{m-1},
			 \qquad \mathcal B_{m-1}\ge0.
	\end{align*}
Adding the two identities proves the weak statement.  
The classical statement follows from Theorem~\ref{th:finite-level-regularity}.
\end{proof}

\subsection{Existence of a stopping boundary}
Define the contact set at time $t$ by
	\begin{align*}
		 \cS_t^{[m]}:=\{x>0:V^{[m]}(t,x)=g_\delta^{[m]}(t,x)\}
			=\{x>0:f_\delta^{[m]}(t,x)=(K-x)^+\}.
	\end{align*}
\begin{theorem}[Existence and uniqueness of the exercise boundary]
\label{th:boundary-m}
Let $m\ge2$ and $t\in[0,T)$. Under the standing model assumptions, the quantity
	\begin{equation}\label{eq:bm-def}
 		b^{[m]}(t):=\sup\cS_t^{[m]}
	\end{equation}
	is well defined and satisfies
	\begin{equation}\label{eq:bm-bounds}
 		0<b^{[1]}(t)\le b^{[m]}(t)<K.
	\end{equation}
Moreover,
	\begin{equation}\label{eq:Sm-interval}
 		\cS_t^{[m]}=(0,b^{[m]}(t)],\qquad \cC_t^{[m]}=(b^{[m]}(t),\infty).
	\end{equation}
For notational convenience, set $b^{[m]}(T):=K$; this is a terminal convention, consistent with 
	the terminal payoff and with the one-right boundary limit. 
The associated boundary-hitting time is
	\begin{equation}\label{eq:tau-m-star}
		 \tau_t^{[m]*}:=\inf\{u\in[0,T-t]:X_{t+u}\le b^{[m]}(t+u)\},
			 \qquad \inf\varnothing:=T-t.
	\end{equation}
Its optimality is proved in Theorem~\ref{th:optimal-first} below.
\end{theorem}
\begin{proof}
We argue by induction on the number of rights.

\emph{Step 0: reduce later strips to lower-right problems.}
For $m=1$ the result is the classical one-right boundary theorem, Theorem~\ref{th:one-right}.  
Suppose $m\ge2$.  
If $t>d_{m-1}$, then $t$ belongs to a strip $(d_j,d_{j-1}]$ for some $1\le j\le m-1$.  
Proposition~\ref{prop:stage} and \eqref{eq:obstacle-strip} give
	\begin{align*}
    		V^{[m]}(t,\cdot)=V^{[j]}(t,\cdot),\qquad 
    		g_\delta^{[m]}(t,\cdot)=g_\delta^{[j]}(t,\cdot).
	\end{align*}
Thus
	\begin{align*}
    		\cS_t^{[m]}=\cS_t^{[j]},
	\end{align*}
	and the assertion follows from the induction hypothesis at level $j$.  
It therefore remains only to consider the main strip $t\le d_{m-1}.$ 

\emph{Step 1: the stopping set is nonempty and contains the one-right stopping set.}
Define the excess of the value over the stopping reward by
	\begin{align*}
    		\Xi^{[m]}(t,x):=V^{[m]}(t,x)-g_\delta^{[m]}(t,x).
	\end{align*}
By Lemma~\ref{lem:supermart}(ii),
	\begin{align*}
    		V^{[m]}(t,x)\le V^{[1]}(t,x)+\Psi_\delta^{[m-1]}(t,x).
	\end{align*}
Subtracting
	\begin{align*}
   		g_\delta^{[m]}(t,x)=(K-x)^++\Psi_\delta^{[m-1]}(t,x)
	\end{align*}
	gives
	\begin{equation}\label{eq:HmH1}
    		\Xi^{[m]}(t,x) \le V^{[1]}(t,x)-(K-x)^+ =:\Xi^{[1]}(t,x).
	\end{equation}
If $x\le b^{[1]}(t)$, then the one-right problem is in its stopping region, so $\Xi^{[1]}(t,x)=0$.  
On the other hand, the value always dominates the stopping reward, hence $\Xi^{[m]}\ge0$.  
Therefore \eqref{eq:HmH1} forces
	\begin{align*}
   		 \Xi^{[m]}(t,x)=0.
	\end{align*}
Consequently,
	\begin{align*}
  		  (0,b^{[1]}(t)]=\cS_t^{[1]} \subseteq\cS_t^{[m]}.
	\end{align*}
Thus the stopping set is nonempty and
	\begin{align*}
    		\sup\cS_t^{[m]}\ge b^{[1]}(t)>0.
	\end{align*}

\emph{Step 2: stopping is impossible at or above the strike.}
For $x\ge K$ the immediate put payoff is zero, and therefore
    	$	g_\delta^{[m]}(t,x)=\Psi_\delta^{[m-1]}(t,x).$ 
Thus, to show that immediate stopping is not optimal, 
	it is enough to show that keeping the remaining rights has strictly positive value.  
Since $V^{[m]}\ge V^{[m-1]}$, it suffices to prove
	\begin{equation}\label{eq:W-positive}
    		W^{[m-1]}(t,x) :=V^{[m-1]}(t,x)-\Psi_\delta^{[m-1]}(t,x)>0.
	\end{equation}
Indeed, once \eqref{eq:W-positive} holds,
	\begin{align*}
    		V^{[m]}(t,x)\ge V^{[m-1]}(t,x) >\Psi_\delta^{[m-1]}(t,x)=g_\delta^{[m]}(t,x),
	\end{align*}
	so $(t,x)$ lies in the continuation region.

Lemma~\ref{lem:strict-wait}, applied with $k=m-1$, gives
\eqref{eq:W-positive} for every $m\ge2$ and every $t<T$, $x>0$.
This does not use Theorem~\ref{th:HD}.
We conclude that
	\begin{align*}
    		\cS_t^{[m]}\cap[K,\infty)=\varnothing.
	\end{align*}
Continuity of $V^{[m]}-g_\delta^{[m]}$ in $x$, and its strict positivity at $K$,
exclude an entire neighborhood of $K$ from the contact set. Together with Step 1, this proves
	\begin{align*}
    		0<b^{[1]}(t)\le \sup\cS_t^{[m]}<K.
	\end{align*}

\emph{Step 3: the single-crossing property turns the stopping set into one interval.}
On $(0,K)$, recall the gap function $G_t^{[m]}$ from Lemma~\ref{lem:singlecross}.
By the value-dominates-reward inequality, $G_t^{[m]}(x)\ge 0.$ 
Lemma~\ref{lem:singlecross} says that $G_t^{[m]}$ is nondecreasing in $x$, 
	and Lemma~\ref{lem:fprops} gives its continuity.  
Step 1 shows that it vanishes on the nonempty interval $(0,b^{[1]}(t)]$, while Step 2 gives
	\begin{align*}
    		G_t^{[m]}(K)=f_\delta^{[m]}(t,K) \ge W^{[m-1]}(t,K)>0.
	\end{align*}
Let
	\begin{align*}
    		b^{[m]}(t):=\sup\{x\in(0,K):G_t^{[m]}(x)=0\}.
	\end{align*}
Because $G_t^{[m]}$ is continuous, the supremum is itself a zero.  
Because it is nonnegative and nondecreasing, if it is zero at some point 
	then it must also be zero at every smaller point.  
Hence its zero set is exactly
	\begin{align*}
    		(0,b^{[m]}(t)].
	\end{align*}
For $x>b^{[m]}(t)$ it is strictly positive.  
Together with the exclusion of $[K,\infty)$ this gives
	\begin{align*}
    		\cS_t^{[m]}=(0,b^{[m]}(t)],\qquad 
    		\cC_t^{[m]}=(b^{[m]}(t),\infty).
	\end{align*}
This proves both the existence and uniqueness of the threshold boundary.

\emph{Step 4: the natural candidate for the optimal rule.}
Once the stopping set has the lower-threshold form, the first entry time into the stopping set is exactly
	\begin{align*}
   		 \tau_t^{[m]*} =\inf\{u\in[0,T-t]:X_{t+u}\le b^{[m]}(t+u)\}.
	\end{align*}
Theorem~\ref{th:optimal-first} below proves that this candidate hitting time is indeed optimal.
\end{proof}
\begin{remark}[Logical structure]\label{rem:HD-scope}
The proof of Theorem~\ref{th:boundary-m} rests on Lemmas~\ref{lem:singlecross}
	(single crossing) and~\ref{lem:strict-wait} (strict waiting loss); it does not use Theorem~\ref{th:HD}.
Theorem~\ref{th:HD}, whose comparison comes from the ordered Bermudan exercise premiums of
	Lemma~\ref{lem:premium-rights} rather than from subtracting two supermartingale inequalities,
	is used for the adjacent-level boundary ordering in Corollary~\ref{cor:hierarchy}.
The stripwise regularity needed for smooth fit, the QVI and the premium representation
	is proved below from the standing model assumptions; only the verification theorem
	imposes regularity conditions, and these concern an arbitrary candidate pair.
\end{remark}
\begin{corollary}[Deadline strips, jumps, and a sufficient monotonicity condition]
\label{cor:strip-boundary}
Under the assumptions of Theorem~\ref{th:boundary-m}, for $1\le j\le m-1$,
	\begin{equation}\label{eq:boundary-strips}
 		b^{[m]}(t)=b^{[j]}(t),\quad t\in(d_j,d_{j-1}],\ t<T,\qquad
 		b^{[m]}(d_j)=b^{[j+1]}(d_j).
	\end{equation}
Hence on the final strip $(d_1,T)$ the boundary is $b^{[1]}$ and
	\begin{align*}
 		\lim_{t\uparrow T}b^{[m]}(t)=K.
	\end{align*}
Whenever the right limit of $b^{[j]}$ at $d_j$ exists,
	\begin{equation}\label{eq:boundary-jump}
 		b^{[m]}(d_j)-\lim_{s\downarrow d_j}b^{[m]}(s)
 						=b^{[j+1]}(d_j)-\lim_{s\downarrow d_j}b^{[j]}(s).
	\end{equation}
If, in addition, $b^{[j]}$ is right-continuous at $d_j$, the right-hand side is nonnegative, 
	since $b^{[j]}(d_j)\le b^{[j+1]}(d_j)$ by Lemma~\ref{lem:deadline}(iii) and \eqref{eq:Sm-interval}; 
	so the boundary can only jump downward when a deadline is crossed from left to right. 
Finally, if $t\mapsto f_\delta^{[m]}(t,x)$ is nonincreasing on $[0,d_{m-1}]$ 
	for every $x$, then $b^{[m]}$ is nondecreasing there.
\end{corollary}
\begin{proof}
As in Step~0 of the proof of Theorem~\ref{th:boundary-m}, the stagewise identities for $V^{[m]}$ 
	in Proposition~\ref{prop:stage} hold also for the corresponding stopping rewards, 
	so the contact sets  and their upper endpoints agree on each strip. 
The identities at $d_j$ follow in the same way from the deadline values. 
This proves \eqref{eq:boundary-strips} and then \eqref{eq:boundary-jump}. 
For the last statement, if $t_1<t_2$ and $x\le b^{[m]}(t_1)$, then
	$G_{t_1}^{[m]}(x)=0$. 
Since $f_\delta^{[m]}(t,x)$ is assumed nonincreasing in $t$ 
	while $(K-x)^+$ is time independent, $G_{t_2}^{[m]}(x)\le0$.
 Since the value dominates the stopping reward, $G_{t_2}^{[m]}(x)\ge0$, 
 	hence it is zero and $x\le b^{[m]}(t_2)$.
\end{proof}
\begin{remark}[The time-monotonicity gap]\label{rem:monotonicity-gap}
The sufficient condition in the last part of Corollary~\ref{cor:strip-boundary} 
	is not known to hold in the present model. 
It does not follow from the time monotonicity in Theorem~\ref{th:Vprops}: 
	$f_\delta^{[m]}=V^{[m]}-\Psi_\delta^{[m-1]}$ is the difference of two functions 
	that are both nonincreasing in $t$, so its own monotonicity in $t$ is not automatic, 
	even for $m=2$. 
The numerical calculations reported in Section~\ref{sec:numerics} show 
	no violation of boundary monotonicity on any open deadline strip, 
	including for $m=3$; the visible downward jumps occur only when a feasibility deadline is crossed. 
A complete analytical proof of stripwise time monotonicity of $b^{[m]}$ is therefore left open. 
We do not use this time-monotonicity property in the threshold, regularity, smooth-fit, QVI, premium,
	or Volterra results below.  
It remains an independent qualitative question about the shape of the boundary.
\end{remark}
\begin{corollary}[Upper semicontinuity of the boundary]
\label{cor:boundary-regularity}
Under the assumptions of Theorem~\ref{th:boundary-m}, $b^{[m]}$ is upper semicontinuous on
	$[0,d_{m-1}]$.  
On every later open strip $(d_j,d_{j-1})$, its regularity reduces to that of
	$b^{[j]}$ through \eqref{eq:boundary-strips}; deadlines may carry right-hand jumps as
	in \eqref{eq:boundary-jump}.  
Local Lipschitz continuity on the interiors of all open strips is
	proved later in Theorem~\ref{th:finite-level-regularity}.
\end{corollary}
\begin{proof}
On $[0,d_{m-1}]\times(0,\infty)$ both $V^{[m]}$ and $g_\delta^{[m]}$ are continuous, 
	so the contact set is closed.  
If $t_n\to t$ and $x_n=b^{[m]}(t_n)$, the bounds in \eqref{eq:bm-bounds} keep $x_n$ in a
	compact subinterval of $(0,K]$.  
Every convergent subsequence $x_{n_k}\to L$ therefore satisfies
	$(t,L)$ in the closed contact set and hence $L\le b^{[m]}(t)$.  
Thus
	\begin{align*}
 		\limsup_{s\to t}b^{[m]}(s)\le b^{[m]}(t).
	\end{align*}
The stripwise and deadline statements are Corollary~\ref{cor:strip-boundary}.
\end{proof}

\begin{corollary}[Hierarchy of exercise boundaries]
\label{cor:hierarchy}
Under the standing model assumptions, for every $t\in[0,T)$,
	\begin{align*}
		 b^{[m-1]}(t)\le b^{[m]}(t),\qquad \cS_t^{[m-1]}\subseteq\cS_t^{[m]}.
	\end{align*}
The conclusion holds for every finite $m$. At the deadlines $t=d_j$,
$1\le j\le m-1$, the inclusion also follows directly from the transmission identities: 
	for $j=m-1$ this is Lemma~\ref{lem:deadline}(iii), and for $j<m-1$ both levels coincide 
	at $d_j$ by \eqref{eq:dvalue}.
\end{corollary}
\begin{proof}
For $t>T-\delta$, every level coincides with the one-right problem, so the claim is trivial. 
Let $t\le T-\delta$ and take $x\in\cS_t^{[m-1]}$. 
Then
	\begin{equation}\label{eq:touch-mminus1}
 		V^{[m-1]}(t,x)=(K-x)^++\E_{t,x}[e^{-r\delta}V^{[m-2]}(t+\delta,X_{t+\delta})].
	\end{equation}
For $\varepsilon>0$, let $\tau_\varepsilon$ be $\varepsilon$-optimal 
	for the $m$-right recursive problem. 
Then
	\begin{align*}
	 V^{[m]}(t,x)
 		&\le \E_{t,x}[e^{-r\tau_\varepsilon}g_\delta^{[m]}
			(t+\tau_\varepsilon,X_{t+\tau_\varepsilon})]+\varepsilon.
	\end{align*}
Using the same $\tau_\varepsilon$ as a first exercise time in the $(m-1)$-right problem gives
	\begin{align*}
 		\E_{t,x}[e^{-r\tau_\varepsilon}g_\delta^{[m]}(t+\tau_\varepsilon,X_{t+\tau_\varepsilon})]
		\le V^{[m-1]}(t,x)
 		+\E_{t,x}[e^{-r(\tau_\varepsilon+\delta)}D^{[m-1]}
			(t+\tau_\varepsilon+\delta,X_{t+\tau_\varepsilon+\delta})].
	\end{align*}
By the optional-sampling inequality in Theorem~\ref{th:HD}, the last expectation is at most
	\begin{align*}
 		\E_{t,x}[e^{-r\delta}D^{[m-1]}(t+\delta,X_{t+\delta})].
	\end{align*}
Together with \eqref{eq:touch-mminus1} and $D^{[m-1]}=V^{[m-1]}-V^{[m-2]}$, 
	this upper bound equals
	\begin{align*}
		 (K-x)^++\E_{t,x}[e^{-r\delta}V^{[m-1]}(t+\delta,X_{t+\delta})]=g_\delta^{[m]}(t,x).
	\end{align*}
Hence $V^{[m]}(t,x)\le g_\delta^{[m]}(t,x)+\varepsilon$. 
Letting $\varepsilon\downarrow0$ and using $V^{[m]}\ge g_\delta^{[m]}$ yields equality, 
	so $x\in\cS_t^{[m]}$. 
The boundary inequality follows from the interval representation.
\end{proof}

\subsection{Optimality of the hitting rule}
\begin{theorem}[Optimal first exercise time]
\label{th:optimal-first}
Under the assumptions of Theorem~\ref{th:boundary-m}, for every $t\in[0,T)$ 
	the hitting time \eqref{eq:tau-m-star} is optimal for the recursive $m$-right problem:
	\begin{equation}\label{eq:optimal-first-value}
 		V^{[m]}(t,x)=\E_{t,x}\!\left[e^{-r\tau_t^{[m]*}}
		 g_\delta^{[m]}(t+\tau_t^{[m]*},X_{t+\tau_t^{[m]*}})\right].
	\end{equation}
\end{theorem}
\begin{proof}
We prove the result by induction on the number of rights.  
For $m=1$ this is the classical one-right result, Theorem~\ref{th:one-right}.  
Assume the assertion is known for all levels up to $m-1$.

If $t>d_{m-1}$, Proposition~\ref{prop:stage} reduces the $m$-right problem 
	to a lower-right problem on the strip containing $t$, and the induction hypothesis applies.  
We may therefore assume
    	$	t\le d_{m-1}.
	$ 
If $x\le b^{[m]}(t)$, then the initial state is already in the stopping set, 
	so $\tau_t^{[m]*}=0$ and the assertion is immediate.  
We henceforth suppose
   	$	 x>b^{[m]}(t).
	$ 

\emph{Step 1: martingale property up to the first contact or the deadline.}
At the first feasibility deadline, Lemma~\ref{lem:deadline}(i) gives
	\begin{equation}\label{eq:deadline-max}
    		V^{[m]}(d_{m-1},y)=\max\{g_\delta^{[m]}(d_{m-1},y),V^{[m-1]}(d_{m-1},y)\}.
	\end{equation}
Define
	\begin{align*}
  		 \widehat\tau :=\tau_t^{[m]*}\wedge(d_{m-1}-t).
	\end{align*}
On the strip $[t,d_{m-1}]$, the $m$-right recursive problem is an ordinary one-stopping problem 
	with continuous stopping reward in the interior 
	and terminal reward \eqref{eq:deadline-max}.  
Its discounted value process is the Snell envelope of the discounted reward;
	see \citet[Chapter~I, Section~2]{PeskirShiryaev2006}.
The reward process is continuous on $[0,d_{m-1}-t)$ and can only jump upward at the terminal time
	$d_{m-1}-t$, because $V^{[m]}(d_{m-1},\cdot)\ge g_\delta^{[m]}(d_{m-1},\cdot)$; in particular it is
	upper semicontinuous from the left, which is the regularity under which the first contact time is optimal.

A Snell envelope is in general a supermartingale.  
In its Doob--Meyer decomposition, the increasing compensator can grow only 
	when the value is in contact with the stopping reward.  
Before the first contact, therefore, the compensator is constant and the Snell envelope is a martingale.  
Since $\widehat\tau$ stops either at the first contact or at the deadline, optional sampling gives
	\begin{equation}\label{eq:mart-until-hit}
    		V^{[m]}(t,x)=\E_{t,x}\!\left[e^{-r\widehat\tau}
        	V^{[m]}(t+\widehat\tau,X_{t+\widehat\tau})\right].
	\end{equation}

\emph{Step 2: identify the value at $\widehat\tau$.}
There are two possibilities.
First suppose that the boundary is reached strictly before the deadline:
	\begin{align*}
    		\tau_t^{[m]*}<d_{m-1}-t.
	\end{align*}
The contact set is closed and the asset path is continuous.  Hence at the hitting time
	\begin{align*}
   		 V^{[m]}(t+\tau_t^{[m]*},X_{t+\tau_t^{[m]*}})
    			=g_\delta^{[m]}(t+\tau_t^{[m]*},X_{t+\tau_t^{[m]*}}).
	\end{align*}
Thus the hitting rule exercises optimally on this event.

Now suppose that the deadline is reached before any contact, and put
	\begin{align*}
    		Y:=X_{d_{m-1}}.
	\end{align*}
If $Y\le b^{[m]}(d_{m-1}),$ 
	then the state is in the stopping set at the deadline, so value matching again gives
	\begin{align*}
    		V^{[m]}(d_{m-1},Y)=g_\delta^{[m]}(d_{m-1},Y).
	\end{align*}

The only remaining case is
    	$Y>b^{[m]}(d_{m-1}).$ 
Then the state is not in the $m$-right stopping set.  
In the maximum in \eqref{eq:deadline-max}, the stopping-reward branch is therefore not selected, and
	\begin{align*}
   		 V^{[m]}(d_{m-1},Y)=V^{[m-1]}(d_{m-1},Y).
	\end{align*}
Lemma~\ref{lem:deadline}(iii) says
	\begin{align*}
    		\cS_{d_{m-1}}^{[m-1]}\subseteq
    		\cS_{d_{m-1}}^{[m]}.
	\end{align*}
Hence $Y\notin\cS_{d_{m-1}}^{[m]}$ implies $Y\notin\cS_{d_{m-1}}^{[m-1]}$, or equivalently
    	$	Y>b^{[m-1]}(d_{m-1}).$ 
Thus the lower-level hitting rule does not exercise immediately.  
More explicitly, upper semicontinuity of $b^{[m-1]}$ and continuity of $X$ imply 
	that the level-$(m-1)$ boundary-hitting time 
	started from $(d_{m-1},Y)$ is strictly positive almost surely.

Immediately to the right of $d_{m-1}$, Proposition~\ref{prop:stage} 
	and \eqref{eq:boundary-strips} show that the stopping rewards 
	and boundaries of levels $m$ and $m-1$ coincide.  
By the induction hypothesis, the first hitting time of this common boundary realizes exactly
	\begin{align*}
    		V^{[m-1]}(d_{m-1},Y).
	\end{align*}
Therefore, on the event that the deadline is reached first and $Y>b^{[m]}(d_{m-1})$, 
	continuing with the same boundary-hitting rule is optimal.

\emph{Step 3: paste the two parts together.}
Equation \eqref{eq:mart-until-hit} gives the value up to the first hit or the deadline.  
On an interior hit, the hitting rule realizes the stopping reward immediately.  
At the deadline, either it stops there by value matching or, 
	on the noncontact branch, the strong Markov property restarts the problem 
	as the optimal $(m-1)$-right problem just described.  
Splitting the expectation in \eqref{eq:mart-until-hit} according to these events 
	and using the tower property therefore yields
	\begin{align*}
    		V^{[m]}(t,x)=\E_{t,x}\!\left[e^{-r\tau_t^{[m]*}}
        		g_\delta^{[m]}(t+\tau_t^{[m]*},X_{t+\tau_t^{[m]*}})\right],
	\end{align*}
	which is \eqref{eq:optimal-first-value}.
\end{proof}
\begin{corollary}[Optimal sequence of exercise times]
\label{cor:sequence}
Starting from $(t,x,m)$, define recursively
	\begin{align*}
 		\tau_1^{(m)*}&:=\tau_t^{[m]*}(x),\\
		\tau_{k+1}^{(m)*}&:=\tau_k^{(m)*}+\delta
 			+\tau_{t+\tau_k^{(m)*}+\delta}^{[m-k]*}
		 \bigl(X_{t+\tau_k^{(m)*}+\delta}\bigr),
			 \qquad k=1,\ldots,m-1,
	\end{align*}
	whenever the next right remains feasible. 
This sequence is optimal. 
Thus each successive exercise is the first hitting time of the boundary associated 
	with the number of rights remaining after the refractory period.
\end{corollary}
\begin{proof}
By Lemma~\ref{lem:direct-rec}, the recursive value equals the value of the direct problem 
	\eqref{eq:Vm-direct}, so it suffices to show that the sequence attains $V^{[m]}(t,x)$. 
Apply Theorem~\ref{th:optimal-first} to the first exercise. 
After the refractory period, the strong Markov property restarts the problem with one fewer right. 
Iterating this argument proves optimality of the recursively defined sequence.
\end{proof}

\subsection{Interior and stripwise boundary regularity}
\label{sec:strip-regularity}
The next theorem is the local regularity input used to close the classical delayed-value regularity.
It is a one-dimensional, model-specific version of the positive-level-set method of
	\citet{DeAngelisStabile2019}.  
The formulation below is tailored to the constant-coefficient log-price
	process and allows the terminal trace at an interior truncation time to have a positive measure-valued curvature component.
The proof below is self-contained.
\begin{theorem}[Localized level-set Lipschitz criterion]
\label{th:localized-levelset}
Let $I\subset\R$ be an open interval, $\mu\in\R$, $\sigma>0$,
 	\begin{align*}
  		 Y_s^z:=z+\mu s+\sigma B_s,\qquad
   		\widetilde\cL:=\frac{\sigma^2}{2}\partial_{zz}+\mu\partial_z .
 	\end{align*}
Let $w:I\times\R\to[0,\infty)$ be continuous and suppose that, for every $t\in I$,
 	\begin{align*}
   		\{z\in\R:\ w(t,z)=0\}=(-\infty,\beta(t)]
 	\end{align*}
 	for some $\beta(t)\in\R$.
Put $\mathcal C:=\{(t,z)\in I\times\R:\ w(t,z)>0\}=\{(t,z):\ z>\beta(t)\}$, an open set.
Assume that for every compact interval $J\subset I$ there exist
 	\begin{align*}
   		t_-<\inf J\le\sup J<t_+<t_2,\qquad [t_-,t_2]\subset I,\qquad L<R<R_*,
 	\end{align*}
 	an open set $\mathcal N\subset I\times(-\infty,R_*)$ containing $[t_-,t_2]\times[L,R]$,
 	and a function $M:\mathcal N\to\R$ such that the following five conditions hold.
 	\begin{enumerate}[label={\rm(\arabic*)}]
  		\item \emph{Band.} $L<\beta(t)<R$ for $t\in[t_-,t_2]$, and $w(t,\cdot)$ is nondecreasing on
   			$(-\infty,R_*]$ for each $t\in[t_-,t_2]$.
 		 \item \emph{Regularity in the continuation region.}
  			 $M$ and $M_z$ are continuous on $\mathcal N$, the functions $w$ and $w_z$ belong to
  			 $C^{1,2}(\mathcal C\cap\mathcal N)$, and
   			\begin{align}
     				(\partial_t+\widetilde\cL)w=-M,\qquad (\partial_t+\widetilde\cL)w_z=-M_z
     				 \qquad\text{in }\mathcal C\cap\mathcal N. \label{eq:LL-pde}
   			\end{align}
  	\item \emph{Running reward.} There are constants $m_0,m_1<\infty$ and $\alpha>0$ such that, 
		on $[t_-,t_2]\times[L,R]$,
  		 \begin{align}
   			  |M|\le m_0,\qquad |M(t,z)-M(s,z)|\le m_1|t-s|,\qquad M_z\ge\alpha. \label{eq:LL-M}
  		 \end{align}
 	 \item \emph{Local dynamic-programming principle.}
  		 For $s\in[t_-,t_2)$ and $z\in(L,R)$ put $Y=Y^z$,
  		 \begin{align*}
     			\eta_s^z:=\inf\{u\ge0:\ u\ge t_2-s\ \text{ or }\ Y_u\notin(L,R)\},
   		\end{align*}
  		 and, for every stopping time $\tau$ of the filtration generated by $B$ with values in $[0,\infty]$,
   		\begin{align*}
     			\mathcal J_s^z(\tau):=\E\!\left[\int_0^{\tau\wedge\eta}M(s+u,Y_u)\dd u
        		+\one_{\{\eta\le\tau\}}\,w(s+\eta,Y_\eta)\right],\qquad \eta:=\eta_s^z .
   		\end{align*}
   		Then
   		\begin{align}
    			 w(s,z)=\sup_\tau\mathcal J_s^z(\tau), \label{eq:LL-dpp}
   		\end{align}
  		 and the supremum is attained by $\tau_s^z:=\inf\{u\ge0:\ w(s+u,Y_u)=0\}\wedge\eta_s^z$.
  	\item \emph{Terminal trace.}
   		Put $H:=w(t_2,\cdot)$. There exist an open interval $(L_1,R_1)\supset[L,R]$,
   		a finite nonnegative Radon measure $\nu$ on $(L_1,R_1)$ and $q\in L^\infty(L_1,R_1)$ such that
   		\begin{align}
     			\widetilde\cL H=\nu+q\qquad\text{in }\mathcal D'(L_1,R_1). \label{eq:LL-H}
   		\end{align}
 	\end{enumerate}
Then $\beta\in C^{0,1}_{\rm loc}(I)$.
\end{theorem}
\begin{remark}[On the hypotheses]
Condition~(4) is the dynamic-programming principle of the \emph{excess-value problem}:
 	stopping at a time $\tau<\eta$ earns nothing beyond the running reward $M$ collected up to $\tau$,
	 whereas continuing up to the exit time $\eta$ earns the boundary value $w$ at the exit point
 	(on the lateral face $z=L$ this value is $0$, on the face $z=R$ it is $w(\cdot,R)>0$,
	 and on the time face it is $H$).
Only one-sided information on the terminal trace is used at each stage of the proof in Appendix~\ref{app:localized-levelset-proof}:
 	$\nu\ge0$ gives $\widetilde\cL H\ge-\|q\|_\infty$, which controls one direction of the time-difference quotient;
 	the finiteness of $\nu$, combined with Gaussian smoothing over a time interval of positive length,
	 controls the other direction.
Neither monotonicity nor Lipschitz continuity of $H$ is needed beyond the continuity of $w$.
The criterion is verified for $w=w_k$ in the proof of Theorem~\ref{th:finite-level-regularity}.
\end{remark}
\begin{proof}
The proof is given in Appendix~\ref{app:localized-levelset-proof}.
\end{proof}
\begin{theorem}[Stripwise local Lipschitz regularity and closure of delayed-value regularity]
\label{th:finite-level-regularity}
Under the standing Black--Scholes assumptions, for every finite $k\ge1$ the value
	$V^{[k]}$ is $C^{1,2}$ in the interior of its continuation region on each open deadline strip.
Whenever $d_{k-1}:=T-(k-1)\delta>0$, the genuine $k$-right boundary satisfies
	\begin{align}
		 b^{[k]}\in C_{\rm loc}^{0,1}\bigl((0,d_{k-1})\bigr).
		\label{eq:finite-b-lip}
	\end{align}
On every later open strip the corresponding conclusion is inherited from the lower-right problem by
	stagewise reduction.  
On every open deadline strip the weak density of Theorem~\ref{th:R-weak} is therefore
	identified almost everywhere by
	\begin{align}
 		a^{[k]}(t,x)=\bigl(rK+\Gamma_\delta^{[k-1]}(t,x)\bigr) \one_{\{x<b^{[k]}(t)\}}.
		 \label{eq:ak-explicit-global}
	\end{align}

Moreover, for every finite $k$, $\Psi_\delta^{[k]}$ is $C^{1,2}$ on every open delayed deadline strip.
For $t<T-\delta$ away from delayed deadlines,
	\begin{align}
		 \Gamma_\delta^{[k]}(t,x)&:=(r-\partial_t-\cL)\Psi_\delta^{[k]}(t,x)\nonumber\\
			 &=\E_{t,x}\!\left[e^{-r\delta}
 			\bigl(rK+\Gamma_\delta^{[k-1]}(t+\delta,X_{t+\delta})\bigr)
 			\one_{\{X_{t+\delta}\le b^{[k]}(t+\delta)\}}\right],
 			\label{eq:Gamma-general-rec}
	\end{align}
	with $\Gamma_\delta^{[0]}\equiv0$; 
	for $t>T-\delta$ we set $\Gamma_\delta^{[k]}(t,\cdot)\equiv0$.
On each nontrivial open delayed strip, $x\mapsto\Gamma_\delta^{[k]}(t,x)$ is strictly decreasing, 
	and in log-price coordinates it is, together with its spatial derivative, locally Lipschitz in time,
	uniformly in the spatial variable on compact time intervals.
\end{theorem}
\begin{proof}[Proof sketch]
The argument is by induction on the number of exercise rights.  The one-right case follows from the
classical regularity of the American-put boundary together with Proposition~\ref{prop:R-one}.
Assume the conclusions through level $k-1$.  The induction hypothesis gives the required spatial
monotonicity and local time control of the delayed source $\Gamma_\delta^{[k-1]}$.
These estimates verify the hypotheses of Theorem~\ref{th:localized-levelset} for the excess value at
level $k$, and hence yield local Lipschitz continuity of $b^{[k]}$ on the genuine $k$-right strip.
The weak source of Theorem~\ref{th:R-weak} can then be identified with the stopping-region source,
and the fixed-lag Black--Scholes semigroup upgrades the delayed value to $C^{1,2}$ regularity and gives
recursion~\eqref{eq:Gamma-general-rec}.  Stagewise reduction transfers the conclusion to the later
strips.  The strict spatial decrease and the stated time regularity of $\Gamma_\delta^{[k]}$ follow from
the Gaussian kernel representation and the induction hypothesis.
The complete induction, including the quantitative estimates needed to verify the localized criterion,
is given in Appendix~\ref{app:finite-level-proof}.
\end{proof}
\begin{remark}[Initial time and deadlines]
\label{rem:initial-time}
The model is time-homogeneous, since the recursion \eqref{eq:Vm-rec} depends on $(t,T)$ only
	through the remaining horizon $T-t$.
Let $V^{[k]}(t,x;T)$ and $b^{[k]}(t;T)$ denote the value and the boundary of the problem
	with maturity $T$.
Then, for every $a>0$,
	\begin{align*}
 		V^{[k]}(t,x;T)=V^{[k]}(t+a,x;T+a),\qquad b^{[k]}(t;T)=b^{[k]}(t+a;T+a),
	\end{align*}
	and the deadlines of the problem with maturity $T+a$ are $d_j+a$.
Hence the initial time $t=0$ of the original problem corresponds to the interior time $a$ of the extended
	problem, and every statement proved for interior times in Theorems~\ref{th:R-weak}
	and~\ref{th:finite-level-regularity} holds at $t=0$ as well, with one-sided time derivatives.
In particular, $b^{[k]}\in C^{0,1}_{\rm loc}\bigl([0,d_{k-1})\bigr)$ whenever $d_{k-1}>0$, and the mild
	representation \eqref{eq:V-weak-mild} holds at $t=0$, also when $0$ is itself a deadline.

Moreover, for $1\le j\le m-1$ the deadline $d_j$ belongs to the main strip $[0,d_{j-1})$ of level $j$,
	so $b^{[j]}$ is continuous at $d_j$ (right-continuous if $d_j=0$).
Consequently the right-continuity hypothesis in Corollary~\ref{cor:strip-boundary} is always satisfied, and
	\begin{align*}
 		b^{[m]}(d_j)-\lim_{s\downarrow d_j}b^{[m]}(s)=b^{[j+1]}(d_j)-b^{[j]}(d_j)\ge0,
 		\qquad 1\le j\le m-1,
	\end{align*}
	so that the exercise boundary can only jump downward when a deadline is crossed from left to right.
\end{remark}

\begin{corollary}[Classical obstacle generator]
\label{cor:obstacle-classical}
Under the standing model assumptions, on the interior of every open deadline strip and for $x<K$,
	\begin{align*}
		 (\partial_t+\cL-r)g_\delta^{[m]}=-rK-\Gamma_\delta^{[m-1]}<0.
	\end{align*}
\end{corollary}
\begin{proof}
Combine Lemma~\ref{lem:obstacle-gen} with the $C^{1,2}$ regularity of $\Psi_\delta^{[m-1]}$ and the
	continuity of $\Gamma_\delta^{[m-1]}$ established in Theorem~\ref{th:finite-level-regularity}.
\end{proof}

\subsection{Smooth fit away from deadlines}
The stripwise local Lipschitz regularity obtained above is sufficient for the boundary-hitting argument.
\begin{lemma}[Vanishing hitting time at a locally Lipschitz boundary]
\label{lem:tauzero}
Fix $t<T$ away from the deadlines and put $b=b^{[m]}(t)$. 
 For $x\ge b$ let
	\begin{align*}
 		\tau_x:=\inf\{u\in[0,T-t]:X_{t+u}^{t,x}\le b^{[m]}(t+u)\}.
	\end{align*}
Then, as $x\downarrow b$,
	\begin{align*}
		\tau_x\downarrow0\quad\Pp\text{-a.s.},\qquad R_{\tau_x}\to1\quad\Pp\text{-a.s.}
	\end{align*}
\end{lemma}
\begin{proof}
Write $\beta(s)=\log b^{[m]}(s)$ and $\mu=r-\sigma^2/2$.  By
Theorem~\ref{th:finite-level-regularity} (and Remark~\ref{rem:initial-time} when $t=0$), for some $L<\infty$ and $\rho>0$ contained 
	in the same open strip,
	\begin{align*}
		\beta(t+u)\ge\beta(t)-Lu, \qquad 0\le u\le\rho.
	\end{align*}
Let $x>b$ and set $a_x:=\log(x/b)\downarrow0$.  
A sufficient condition for exercise by time $u$ is
	\begin{align*}
 		a_x+\sigma B_u+(\mu+L)u\le0.
	\end{align*}
For every $\varepsilon\in(0,\rho]$, Brownian oscillation with drift gives
	\begin{align*}
 		\inf_{0<u\le\varepsilon}\{\sigma B_u+(\mu+L)u\}<0\quad\Pp\text{-a.s.}
	\end{align*}
Hence for almost every path and all $x$ sufficiently close to $b$, 
	the boundary is hit before $\varepsilon$.  
Since $\tau_x$ is nondecreasing in $x$, letting $x\downarrow b$ and then
	$\varepsilon\downarrow0$ gives $\tau_x\downarrow0$.  
Path continuity yields $R_{\tau_x}\to1$.
\end{proof}
\begin{theorem}[Smooth fit]
\label{th:smoothfit-m}
Under the standing model assumptions, let $t\in[0,T)$ be different from every deadline.
Then $x\mapsto V^{[m]}(t,x)$ is differentiable at $b=b^{[m]}(t)$ and
	\begin{equation}\label{eq:smoothfit-m}
		 V_x^{[m]}(t,b) =-1+\left.\frac{\partial}{\partial x}
 		\E_{t,x}[e^{-r\delta}V^{[m-1]}(t+\delta,X_{t+\delta})]\right|_{x=b}.
	\end{equation}
\end{theorem}
\begin{proof}
Write $\Psi:=\Psi_\delta^{[m-1]}$.  
Since $b<K$, the stopping reward is locally
	$K-x+\Psi(t,x)$ and its spatial derivative is $-1+\Psi_x(t,x)$; the derivative is jointly continuous
	near $(t,b)$ by the fixed-lag lognormal smoothing in Theorem~\ref{th:R-weak}.
On the stopping side,
	\begin{align*}
 		V_x^{[m]}(t,b-)=-1+\Psi_x(t,b).
	\end{align*}
Convexity of $V^{[m]}(t,\cdot)$ gives the reverse one-sided ordering, so it remains only to prove
	\begin{align*}
 		V_x^{[m]}(t,b+)\le-1+\Psi_x(t,b).
	\end{align*}

For $x>b$, let $\tau_x$ be the optimal boundary-hitting time. 
Using the same stopping time as an admissible rule from the smaller initial state $b$ gives
	\begin{align*}
		 \frac{V^{[m]}(t,x)-V^{[m]}(t,b)}{x-b} \le\E\!\left[e^{-r\tau_x}R_{\tau_x}
 		\frac{g_\delta^{[m]}(t+\tau_x,xR_{\tau_x})-g_\delta^{[m]}(t+\tau_x,bR_{\tau_x})}
   		   {(x-b)R_{\tau_x}}\right].
	\end{align*}
Lemma~\ref{lem:tauzero} gives $\tau_x\downarrow0$ and $R_{\tau_x}\to1$ almost surely. 
The stopping reward is $m$-Lipschitz in space, so the integrand is dominated by
	$m\sup_{0\le u\le T}e^{-ru}R_u$, which is integrable.  
Joint continuity of the spatial derivative of the
	stopping reward and dominated convergence therefore give
	\begin{align*}
 		\limsup_{x\downarrow b}\frac{V^{[m]}(t,x)-V^{[m]}(t,b)}{x-b} \le -1+\Psi_x(t,b).
	\end{align*}
Combining this with the left derivative and convexity proves smooth fit.
\end{proof}

\section{QVI, free-boundary problem, and integral representations}
\label{sec:qvi}\label{sec:eep}
\subsection{QVI and free-boundary problem}
Let
	\begin{align*}
		 \mathcal D_m:=\{d_j=T-j\delta:j=1,\ldots,m-1\}\cap[0,T)
	\end{align*}
	be the deadline set. 
This agrees with Section~\ref{sec:weak-R}, including $0$ when it is a deadline.
Away from $\mathcal D_m$, the obstacle is continuous in time 
	and the usual infinitesimal dynamic-programming argument applies.
\begin{theorem}[Quasi-variational inequality]
\label{th:qvi}
Under the standing Black--Scholes assumptions, away from the deadline set $\mathcal D_m$,
	the classical complementarity relations
	\begin{align*}
 		\begin{cases}
			 V^{[m]}-g_\delta^{[m]}\ge0,\\[1mm]
			 (\partial_t+\cL-r)V^{[m]}\le0,\\[1mm]
			 (V^{[m]}-g_\delta^{[m]})((\partial_t+\cL-r)V^{[m]})=0
		 \end{cases}
	\end{align*}
hold in the interiors of the continuation and stopping regions.  More precisely,
	\begin{align*}
		 (\partial_t+\cL-r)V^{[m]}&=0
			&&\text{on the continuation interior},\\
		 (\partial_t+\cL-r)V^{[m]}&=-rK-\Gamma_\delta^{[m-1]}<0
 			&&\text{on the stopping interior}.
	\end{align*}
No differentiability across the free boundary is asserted.  
The terminal condition is $V^{[m]}(T,x)=(K-x)^+$, 
	and at a deadline the stripwise transmission conditions of
	Proposition~\ref{prop:stage} replace ordinary time-continuous matching.
\end{theorem}
\begin{proof}
Obstacle domination gives the first inequality.  
The continuation identity and the required $C^{1,2}$ regularity are part of Theorem~\ref{th:finite-level-regularity}. 
On the stopping interior, Theorem~\ref{th:boundary-m} gives $x<K$ and
	\begin{align*}
		 V^{[m]}=K-x+\Psi_\delta^{[m-1]}.
	\end{align*}
Theorem~\ref{th:finite-level-regularity} therefore yields
	\begin{align*}
		 (\partial_t+\cL-r)V^{[m]}=-rK-\Gamma_\delta^{[m-1]}<0.
	\end{align*}
The complementarity relation follows on the two open regions, and the terminal and deadline statements
	follow from the contract definition and Proposition~\ref{prop:stage}.
\end{proof}

The QVI becomes a free-boundary system after inserting the threshold structure.
\begin{theorem}[Free-boundary problem]
\label{th:fbp}
Under the standing Black--Scholes assumptions, away from $\mathcal D_m$ the pair
$(V^{[m]},b^{[m]})$ satisfies
	\begin{align}
		 (\partial_t+\cL-r)V^{[m]}(t,x)&=0,
 		&&x>b^{[m]}(t),\label{eq:fbp-cont}\\
		 V^{[m]}(t,x)&=g_\delta^{[m]}(t,x),
 		&&0<x\le b^{[m]}(t),\label{eq:fbp-stop}\\
 		V^{[m]}(t,b^{[m]}(t))&=g_\delta^{[m]}(t,b^{[m]}(t)),\notag\\
 		V_x^{[m]}(t,b^{[m]}(t))&=\partial_xg_\delta^{[m]}(t,b^{[m]}(t)),\label{eq:fbp-sf}
	\end{align}
	where \eqref{eq:fbp-sf} holds at every time away from the deadlines by
Theorem~\ref{th:smoothfit-m}. 
In addition,
	\begin{align*}
 		V^{[m]}(T,x)&=(K-x)^+,\\
 			\lim_{t\uparrow T}b^{[m]}(t)&=K,\\
 			\lim_{x\to\infty}V^{[m]}(t,x)&=0.
	\end{align*}
At every deadline $d_j$ the matching conditions are instead
	\begin{align}
 		V^{[m]}(d_j,x)
 			&=\max\{g_\delta^{[m]}(d_j,x),V^{[j]}(d_j,x)\},\label{eq:deadline-condition}\\
 		V^{[m]}(d_j+,x)&=V^{[j]}(d_j,x).\label{eq:deadline-right}
	\end{align}
Smooth fit is not imposed at $d_j$.
\end{theorem}
\begin{proof}
Equations \eqref{eq:fbp-cont}--\eqref{eq:fbp-stop} are the two active branches of Theorem~\ref{th:qvi}; 
	value matching follows from the definition of the contact boundary 
	and smooth fit from Theorem~\ref{th:smoothfit-m}. 
The terminal conditions follow from the one-right final strip and the one-right boundary limit. 
Since each put payoff is bounded by $K$ and the probability of reaching the in-the-money region 
	before maturity vanishes as the initial asset price tends to infinity, 
	the value tends to zero at spatial infinity. 
Finally, \eqref{eq:deadline-condition} is Lemma~\ref{lem:deadline}(i), 
	while \eqref{eq:deadline-right} is Proposition~\ref{prop:stage}. 
The max structure can create a spatial kink at a deadline, so smooth fit need not hold there.
\end{proof}
\subsection{Verification}
The next theorem gives a sufficient characterization on the main strip.  It is deliberately stated
with a monotone candidate boundary because bounded variation of the candidate curve is used by the
local-time-space formula of \citet{Peskir2005}.
\begin{theorem}[Verification theorem]
\label{th:verification}
Let $m\ge2$ and suppose $V^{[m-1]}$ is known. 
Let $c:[0,d_{m-1}]\to(0,K)$ be continuous and nondecreasing. 
Suppose a continuous function
	$v:[0,d_{m-1}]\times(0,\infty)\to\R$ satisfies $0\le v\le mK$ and:
	\begin{enumerate}[label={\rm(\roman*)}]
	\item at $d_{m-1}$,
		\begin{equation}\label{eq:ver-terminal}
 			v(d_{m-1},x)=\max\{g_\delta^{[m]}(d_{m-1},x),V^{[m-1]}(d_{m-1},x)\};
		\end{equation}
	\item $v\in C^{1,2}$ on $\{x>c(t),\,t<d_{m-1}\}$, 
		its derivatives extend continuously to the boundary from the continuation side
		locally in time on $[0,d_{m-1})$,
		and $v(t,\cdot)\in C^1((0,\infty))$ for every $0\le t<d_{m-1}$;
	\item $v\ge g_\delta^{[m]}$ everywhere on the main strip and $v=g_\delta^{[m]}$ for $0<x\le c(t)$;
	\item $(\partial_t+\cL-r)v=0$ for $x>c(t)$ and $0\le t<d_{m-1}$.
	\end{enumerate}
At $d_{m-1}$, continuity and \eqref{eq:ver-terminal} are required,
but no derivative regularity is imposed: the terminal maximum may have a spatial kink.
Then
	\begin{equation}\label{eq:ver-value}
		 v(t,x)=V^{[m]}(t,x),\qquad 0\le t\le d_{m-1},
	\end{equation}
	and
	\begin{align*}
 		c(t)=b^{[m]}(t),\qquad 0<t<d_{m-1}.
	\end{align*}
Since $b^{[m]}$ is continuous at $0$ by Remark~\ref{rem:initial-time}, equality also holds at $t=0$.
\end{theorem}
\begin{proof}
Fix an initial state $(t,x)$ with $t<d_{m-1}$.  
The proof has three parts: first we show that the candidate $v$ dominates the true value, 
	then we prove the reverse inequality, 
	and finally we identify the candidate curve $c$ with the true exercise boundary.

\emph{Step 1: the discounted candidate value is a supermartingale.}
By Lemma~\ref{lem:obstacle-gen}, on the candidate stopping region $x\le c(t)<K$,
	\begin{align*}
  		  (\partial_t+\cL-r)g_\delta^{[m]}\le-rK<0.
	\end{align*}
On that region condition (iii) gives $v=g_\delta^{[m]}$.  
On the candidate continuation region $x>c(t)$, condition (iv) gives
	\begin{align*}
    		(\partial_t+\cL-r)v=0.
	\end{align*}

Apply the local-time-space change-of-variable formula to
	\begin{align*}
    		e^{-ru}v(t+u,X_{t+u})
	\end{align*}
first up to time $d_{m-1}-\varepsilon$, where $0<\varepsilon<d_{m-1}-t$.
The curve $c$ is continuous and nondecreasing, hence of bounded variation.  
Condition (ii) implies that $v_x$ is continuous across the curve; 
	therefore the coefficient of the boundary local-time term is zero.  
The drift is zero on the continuation side and nonpositive on the stopping side.  
Thus, after localization, the discounted candidate is a supermartingale.

For completeness, one may localize further at the first exit of $X$ from $[1/N,N]$.  
On this compact set all coefficients appearing in the stochastic integral are bounded, 
	so the stochastic integral has mean zero.  
Letting $N\to\infty$ preserves the supermartingale inequality because $0\le v\le mK$.  
Letting $\varepsilon\downarrow0$ then preserves this inequality by continuity of $v$
and bounded convergence. Thus the discounted candidate is a supermartingale up to $d_{m-1}$,
without requiring derivatives at that endpoint.

Let $\tau$ be any admissible stopping time and set
	\begin{align*}
    		\widehat\tau:=\tau\wedge(d_{m-1}-t).
	\end{align*}
If $\tau\le d_{m-1}-t$, condition (iii) gives
	\begin{align*}
    		g_\delta^{[m]}(t+\tau,X_{t+\tau})\le v(t+\tau,X_{t+\tau}).
	\end{align*}
If $\tau>d_{m-1}-t$, then after the deadline only $m-1$ effective rights remain.  
Conditional on $\mathcal F_{d_{m-1}}$, the continuation payoff is therefore bounded by the lower-level value:
	\begin{align*}
		&\E\!\left[e^{-r\tau}g_\delta^{[m]}(t+\tau,X_{t+\tau})\mid\mathcal F_{d_{m-1}}\right]\\
		&\qquad\le e^{-r(d_{m-1}-t)}V^{[m-1]}(d_{m-1},X_{d_{m-1}})\\
		&\qquad\le e^{-r(d_{m-1}-t)}v(d_{m-1},X_{d_{m-1}}),
	\end{align*}
	where the last inequality follows from the terminal condition (i).  
Optional sampling of the candidate supermartingale at $\widehat\tau$ now yields
	\begin{align*}
    		\E_{t,x}\!\left[e^{-r\tau}g_\delta^{[m]}(t+\tau,X_{t+\tau})\right]\le v(t,x).
	\end{align*}
Taking the supremum over $\tau$ gives
	\begin{align*}
    		V^{[m]}(t,x)\le v(t,x).
	\end{align*}

\emph{Step 2: the candidate value cannot exceed the true value.}
Define the first time the process reaches the candidate stopping region, capped at the first deadline:
	\begin{align*}
   		 \tau_c:=\inf\{u\in[0,d_{m-1}-t):X_{t+u}\le c(t+u)\}\wedge(d_{m-1}-t).
	\end{align*}
Before $\tau_c$ the process remains in the candidate continuation region, 
	where the drift of the discounted candidate is zero.  
The same localization argument as above therefore shows that 
	the stopped discounted candidate is a martingale, and
	\begin{align*}
    		v(t,x)=\E_{t,x}\!\left[e^{-r\tau_c}v(t+\tau_c,X_{t+\tau_c})\right].
	\end{align*}
If $\tau_c<d_{m-1}-t$, then the process hits the candidate boundary and condition (iii) gives
	\begin{align*}
    		v(t+\tau_c,X_{t+\tau_c})=g_\delta^{[m]}(t+\tau_c,X_{t+\tau_c})
   		 \le V^{[m]}(t+\tau_c,X_{t+\tau_c}).
	\end{align*}
If $\tau_c=d_{m-1}-t$, condition (i) agrees exactly with the true deadline value 
	by Lemma~\ref{lem:deadline}(i), so again
	\begin{align*}
   	 	v(t+\tau_c,X_{t+\tau_c})  \le V^{[m]}(t+\tau_c,X_{t+\tau_c}).
	\end{align*}
Finally, Lemma~\ref{lem:supermart}(i) says that the discounted true value is a strong supermartingale.  
Hence
	\begin{align*}
    		v(t,x)\le\E_{t,x}\!\left[ e^{-r\tau_c}V^{[m]}(t+\tau_c,X_{t+\tau_c})
    			\right]\le V^{[m]}(t,x).
	\end{align*}
Together with Step 1, this proves \eqref{eq:ver-value}.

\emph{Step 3: identify the candidate boundary.}
Since $v=V^{[m]}$ and condition (iii) gives $v=g_\delta^{[m]}$ for $x\le c(t)$, 
	every point below $c(t)$ belongs to the true stopping set.  
Therefore
	\begin{align*}
		c(t)\le b^{[m]}(t).
	\end{align*}
Suppose, for a contradiction, that the inequality is strict at some interior time $t_0$:
	\begin{align*}
		c(t_0)<b^{[m]}(t_0).
	\end{align*}
Choose $x_0\in(c(t_0),b^{[m]}(t_0)).$ 
Since the true stopping set is the interval $(0,b^{[m]}(t_0)]$ and $v=V^{[m]}$, we have
	\begin{align*}
   		 v(t_0,x)=g_\delta^{[m]}(t_0,x)
	\end{align*}
	for all $x$ in a small spatial neighborhood of $x_0$.  
Hence the first and second spatial derivatives of $v$ and $g_\delta^{[m]}$ agree at $(t_0,x_0)$.

Moreover, the function $v-g_\delta^{[m]}$ 
	is nonnegative everywhere and vanishes at the interior point $(t_0,x_0)$.  
Because $x_0>c(t_0)$ and $c$ is continuous, $(t_0,x_0)$ lies in the interior of the candidate continuation region, 
	where $v$ is $C^{1,2}$.  
Theorem~\ref{th:finite-level-regularity} gives the same time differentiability for the stopping reward.  
Thus $(t_0,x_0)$ is a differentiable local minimum of $v-g_\delta^{[m]}$, and
	\begin{align*}
    		\partial_t(v-g_\delta^{[m]})(t_0,x_0)=0.
	\end{align*}
Combining the equality of the time and space derivatives with condition (iv) gives
	\begin{align*}
    		0=(\partial_t+\cL-r)v(t_0,x_0)=(\partial_t+\cL-r)g_\delta^{[m]}(t_0,x_0)\le-rK<0,
	\end{align*}
	a contradiction.  
Therefore
	\begin{align*}
    		c(t)=b^{[m]}(t),\qquad 0<t<d_{m-1}.
	\end{align*}
Since $c$ is continuous and $b^{[m]}$ is continuous at $0$ by Remark~\ref{rem:initial-time}, the equality extends to $t=0$.
\end{proof}

\subsection{Early-exercise premium representation}
The weak Duhamel representation of Theorem~\ref{th:R-weak}, together with the explicit source density
identified in Theorem~\ref{th:finite-level-regularity}, gives the early-exercise premium directly.
No boundary monotonicity, continuation-side derivative extension, or local-time-space formula is needed.
Recall that on every open strip
	\begin{align}
 		a^{[m]}(t,x)=\bigl(rK+\Gamma_\delta^{[m-1]}(t,x)\bigr)
 		\one_{\{x<b^{[m]}(t)\}} \quad\text{a.e.}
		 \label{eq:am-explicit-eep}
	\end{align}
	with $\Gamma_\delta^{[0]}\equiv0$.
\begin{theorem}[Early-exercise premium]
\label{th:eep}
Under the standing Black--Scholes assumptions, for every finite $m\ge1$,
	\begin{align}
		 V^{[m]}(t,x)&=P^E(t,x)+\int_t^T e^{-r(u-t)} \E_{t,x}\!\left[(rK+\Gamma_\delta^{[m-1]}(u,X_u))
			 \one_{\{X_u\le b^{[m]}(u)\}}\right]\dd u\nonumber\\
			 &\quad+\sum_{j=1}^{m-1}\one_{\{t\le d_j\}}e^{-r(d_j-t)}
				 \E_{t,x}[\Delta^{[j+1]}(X_{d_j})],
 			\label{eq:eep-m}
	\end{align}
	where
	\begin{align*}
		P^E(t,x):=\E_{t,x}[e^{-r(T-t)}(K-X_T)^+].
	\end{align*}
\end{theorem}
\begin{proof}
Theorem~\ref{th:R-weak} gives
	\begin{align*}
 		V^{[m]}(t,x)&=\mathcal P_{T-t}g(x)+\int_t^T\mathcal P_{u-t}[a^{[m]}(u,\cdot)](x)\dd u\\
			 &\quad+\sum_{\substack{d\in\mathcal D_m\\t\le d}}
 				\mathcal P_{d-t}[\Delta_m(d,\cdot)](x).
	\end{align*}
The first term is $P^E(t,x)$.  
Substituting \eqref{eq:am-explicit-eep} in the integral gives the premium
	term in \eqref{eq:eep-m}; replacing $<$ by $\le$ does not alter the expectation because the lognormal
	law has no atom at the boundary.  
At $d_j$, stagewise reduction gives
	$\Delta_m(d_j,\cdot)=\Delta^{[j+1]}$, so the atomic terms are exactly the deadline sum in
	\eqref{eq:eep-m}.  
This proves the formula for $0<t<T$; the endpoint $t=0$ follows from
	time homogeneity (Remark~\ref{rem:initial-time}), also when $0$ is itself a deadline.
\end{proof}

The jump size itself has a useful stopping-reward representation. Recall 
	$W^{[j]}=V^{[j]}-\Psi_\delta^{[j]}\ge0$. 
Lemma~\ref{lem:deadline}(ii) gives
	\begin{equation}\label{eq:Delta-explicit}
 		\Delta^{[j+1]}(x)=\bigl(g_\delta^{[j+1]}(d_j,x)-V^{[j]}(d_j,x)\bigr)^+
 			=\bigl(K-x-W^{[j]}(d_j,x)\bigr)^+,
	\end{equation}
	and its support is contained in $\cS^{[j+1]}_{d_j}$, which equals $(0,b^{[j+1]}(d_j)]$ 
	under the assumptions of Theorem~\ref{th:boundary-m}.

The waiting loss also has a premium representation under the standing assumptions.
\begin{remark}[A premium representation of the waiting loss]
\label{rem:W-regularity}
Fix $2\le j\le m-1$ and $t\le T-\delta$.
Applying Theorem~\ref{th:eep} over $[t,t+\delta]$ at level $j$ gives
	\begin{align*}
		W^{[j]}(t,x)
		&=\E_{t,x}\!\int_0^\delta e^{-ru}\bigl(rK+\Gamma_\delta^{[j-1]}(t+u,X_{t+u})\bigr)
			\one_{\{X_{t+u}\le b^{[j]}(t+u)\}}\dd u\\
		&\quad+\sum_{i:\,t\le d_i<t+\delta}e^{-r(d_i-t)}\E_{t,x}\bigl[\Delta^{[i+1]}(X_{d_i})\bigr],
	\end{align*}
	where the sum runs over the deadlines $d_i$, $i\le j-1$, of level $j$. 
Since $\Gamma_\delta^{[j-1]}\ge0$, $\Delta^{[i+1]}\ge0$, and $b^{[j]}\ge b^{[1]}$ on $[t,t+\delta]$ 
	(by \eqref{eq:bm-bounds} and \eqref{eq:boundary-strips}), we obtain
	\begin{align*}
		W^{[j]}(t,x)\ge W^{[1]}(t,x)=\Theta_\delta(t,x)>0\qquad\text{for every }x>0.
	\end{align*}
Theorem~\ref{th:HD} also gives directly
	$W^{[j]}\ge W^{[1]}$ under the standing model assumptions, and
Lemma~\ref{lem:residual} identifies $W^{[1]}=\Theta_\delta>0$.
The one-boundary theorem does not depend on this remark.
\end{remark}
\subsection{Recursive Volterra boundary equations}
For the integral-equation formulation let $\Phi$ denote the standard normal distribution function 
	and define, for $s>0$ (we avoid the customary symbol $d_2$, 
	which would clash with the deadline $d_2=T-2\delta$),
	\begin{equation}\label{eq:d2}
 		\zeta(x,z;s):=\frac{\log(x/z)+(r-\tfrac12\sigma^2)s}{\sigma\sqrt{s}},
 		\qquad
 		\Pp_{t,x}(X_{t+s}\le z)=\Phi(-\zeta(x,z;s)).
	\end{equation}
The lognormal transition density is
	\begin{align*}
 		p(s;x,y)=\frac{1}{y\sigma\sqrt{2\pi s}}
 		\exp\!\left[-\frac{(\log(y/x)-(r-\tfrac12\sigma^2)s)^2}{2\sigma^2s}\right].
	\end{align*}
\begin{theorem}[Volterra equation for $b^{[m]}$]
\label{th:volterra}
Under the standing Black--Scholes assumptions, let $t<T-\delta$ and put $b(t)=b^{[m]}(t)$. 
Then
	\begin{align}
 	K-b(t)+\Psi_\delta^{[m-1]}(t,b(t))&=P^E(t,b(t))+rK\int_t^T e^{-r(u-t)}
 		\Phi\!\bigl(-\zeta(b(t),b^{[m]}(u);u-t)\bigr)\dd u\nonumber\\
 		&\quad+\int_t^{T-\delta}e^{-r(u-t)}
 			\int_0^{b^{[m]}(u)}\Gamma_\delta^{[m-1]}(u,y)
			 p(u-t;b(t),y)\dd y\dd u\nonumber\\
		 		 &\quad+\sum_{j=1}^{m-1}\one_{\{t<d_j\}}e^{-r(d_j-t)}
			 \int_0^{b^{[j+1]}(d_j)}\Delta^{[j+1]}(y)
			 p(d_j-t;b(t),y)\dd y\nonumber\\
		 &\quad+\sum_{j=1}^{m-1}\one_{\{t=d_j\}}\Delta^{[j+1]}(b(t)).
		 \label{eq:volterra}
	\end{align}
When $t$ belongs to a later strip, the identity reduces to the corresponding lower-level equation 
	through Proposition~\ref{prop:stage}.
\end{theorem}
\begin{proof}
Insert $x=b(t)$ into \eqref{eq:eep-m}. 
Value matching and $b(t)<K$ give
	\begin{align*}
 		V^{[m]}(t,b(t))=K-b(t)+\Psi_\delta^{[m-1]}(t,b(t)).
	\end{align*}
For the $rK$ part, Fubini's theorem and \eqref{eq:d2} give
	\begin{align*}
		 \E_{t,b(t)}\!\left[\int_t^T e^{-r(u-t)}rK\one_{\{X_u\le b^{[m]}(u)\}}\dd u\right]
 		=rK\int_t^T e^{-r(u-t)}\Phi(-\zeta(b(t),b^{[m]}(u);u-t))\dd u.
	\end{align*}
For the $\Gamma$ part, $\Gamma_\delta^{[m-1]}$ vanishes after $T-\delta$ 
	and the transition density gives
	\begin{align*}
		 &\E_{t,b(t)}\!\left[\int_t^T e^{-r(u-t)}\Gamma_\delta^{[m-1]}(u,X_u)
			\one_{\{X_u\le b^{[m]}(u)\}}\dd u\right]\\
 		&\quad=\int_t^{T-\delta}e^{-r(u-t)}\int_0^{b^{[m]}(u)}
		 \Gamma_\delta^{[m-1]}(u,y)p(u-t;b(t),y)\dd y\dd u.
	\end{align*}
Finally, for $t<d_j$ the support property following \eqref{eq:Delta-explicit} converts each deadline 
	expectation into the integral against $p(d_j-t;b(t),\cdot)$ in \eqref{eq:volterra}, 
	while for $t=d_j$ the expectation is simply $\Delta^{[j+1]}(b(t))$. 
Combining the three pieces proves the claim.
\end{proof}

The preceding equation still contains $\Psi_\delta^{[m-1]}$ and $\Gamma_\delta^{[m-1]}$. 
Both can be reduced recursively to lower-level boundaries.
\begin{lemma}[Premium representation of the delayed residual-right value]
\label{lem:Psi-eep}
For $t\le T-\delta$,
	\begin{align}
		 \Psi_\delta^{[m-1]}(t,x) &=P^E(t,x)+rK\int_{t+\delta}^T e^{-r(u-t)}
 			\Phi\!\bigl(-\zeta(x,b^{[m-1]}(u);u-t)\bigr)\dd u\nonumber\\
		 &\quad+\int_{t+\delta}^{T-\delta}e^{-r(u-t)}
			\int_0^{b^{[m-1]}(u)}\Gamma_\delta^{[m-2]}(u,y)p(u-t;x,y)\dd y\dd u\nonumber\\
 		&\quad+\sum_{j=1}^{m-2}\one_{\{t\le d_{j+1}\}}e^{-r(d_j-t)}
			 \int_0^{b^{[j+1]}(d_j)}\Delta^{[j+1]}(y)p(d_j-t;x,y)\dd y.
		 \label{eq:Psi-eep}
	\end{align}
For $m=2$ the sum is empty.
\end{lemma}
\begin{proof}
Apply Theorem~\ref{th:eep} at level $m-1$ with starting state
	$(t+\delta,X_{t+\delta})$, multiply by $e^{-r\delta}$, and take $\E_{t,x}$. 
The European part collapses to $P^E(t,x)$ because the discounted European value is a martingale. 
For the $rK$ part, the tower property changes the conditional exercise probability after 
	$t+\delta$ into the one-time probability from $(t,x)$ at time $u$, 
	giving the first integral. 
Fubini, the tower property and the lognormal transition density give the $\Gamma$ integral. 
For the deadline terms, observe
	$\one_{\{t+\delta\le d_j\}}=\one_{\{t\le d_{j+1}\}}$ and
	$e^{-r\delta}e^{-r(d_j-t-\delta)}=e^{-r(d_j-t)}$. 
These are exactly the terms in \eqref{eq:Psi-eep}.
\end{proof}
\begin{corollary}[Recursive representation of $\Gamma$]
\label{cor:Gamma-rec}
Under the standing Black--Scholes assumptions, for every finite $k\ge1$ and every $t<T-\delta$ away
from delayed deadlines,
	\begin{equation}\label{eq:Gamma-rec}
		 \Gamma_\delta^{[k]}(t,x)=\E_{t,x}\!\left[e^{-r\delta}
			 \bigl(rK+\Gamma_\delta^{[k-1]}(t+\delta,X_{t+\delta})\bigr)
			 \one_{\{X_{t+\delta}\le b^{[k]}(t+\delta)\}}\right].
	\end{equation}
Moreover,
	\begin{align*}
 		0\le\Gamma_\delta^{[k]}\le krK
	\end{align*}
	on every open delayed deadline strip.
\end{corollary}
\begin{proof}
The representation is \eqref{eq:Gamma-general-rec}.  
The bound follows inductively from
	$\Gamma_\delta^{[0]}=0$ and $\mathcal P_\delta1=e^{-r\delta}\le1$.
\end{proof}

\begin{remark}[Recursive closure of the boundary equations]
\label{cor:volterra-rec}
Substituting \eqref{eq:Psi-eep} and \eqref{eq:Gamma-rec} into \eqref{eq:volterra} 
	gives a Volterra-type equation for $b^{[m]}$ coupled only to lower-level boundaries 
	and lower-level $\Gamma$ terms. 
Iterating \eqref{eq:Gamma-rec} down to $\Gamma_\delta^{[0]}=0$ formally 
	closes the system in the boundary family $\{b^{[k]}\}_{k=1}^m$, 
	with the deadline jumps carried by \eqref{eq:Delta-explicit}.
\end{remark}

\section{Random termination}\label{sec:termination}
We now retain the same asset dynamics and contractual maturity $T$ but allow the contract to disappear 
	at an independent random time $\bar T$. 
At current time $t$ the contract is known to be alive, 
	and exercise decisions are adapted to the asset-price filtration; 
	the realization of $\bar T$ cannot be anticipated.

Random-horizon optimal stopping has a long history.
In the best-choice setting, Presman and Sonin~\cite{PresmanSonin1972}
	studied the problem with a random number of objects, one of the early
	formulations of optimal stopping with a random horizon.
Samuel-Cahn~\cite{SamuelCahn1996} later gave a general treatment of
	independent random horizons and showed how survival probabilities can be
	incorporated into an equivalent discounted stopping problem.
Here we adapt this idea to the multiple-exercise American put with a
	deterministic refractory period.

Assume $\Pp(\bar T>s)>0$ for every $0\le s<T$. 
Define the conditional survival probability
	\begin{align*}
		 \pi_t(u):=\Pp(\bar T\ge t+u\mid\bar T>t),\qquad 0\le t<T,\quad u\ge0.
	\end{align*}
At maturity we set $\pi_T(0):=1$ as a terminal convention;
	this does not require conditioning on $\{\bar T>T\}$.
We assume $\bar T$ is independent of the Brownian motion. 
The following elementary separation is the basis of the model.

\begin{lemma}[Separation of survival risk]
\label{lem:survival}
For $0\le t<T$, every asset-filtration stopping time $\tau\in\cT_{[0,T-t]}$ 
	and every bounded asset-measurable payoff $Y_\tau$,
	\begin{equation}\label{eq:survival-sep}
		 \E_{t,x}\!\left[e^{-r\tau}Y_\tau\one_{\{\tau\le\bar T-t\}}\mid\bar T>t\right]
		 =\E_{t,x}[e^{-r\tau}\pi_t(\tau)Y_\tau].
	\end{equation}
\end{lemma}
\begin{proof}
Under the conditional probability $\Pp(\cdot\mid\bar T>t)$, 
	the Brownian sigma-field remains independent of the sigma-field generated by $\bar T$. 
Since $\tau$ and $Y_\tau$ are measurable with respect to the former,
	\begin{align*}
 		\E[\one_{\{\bar T\ge t+\tau\}}\mid\sigma(B),\bar T>t]=\pi_t(\tau).
	\end{align*}
Multiplication by $e^{-r\tau}Y_\tau$ and the tower property yield \eqref{eq:survival-sep}. 
The final conditioning on $\bar T>t$ disappears because the remaining integrand is 
	Brownian-measurable and independent of that event.
\end{proof}

For one right,
	\begin{align*}
 		\bar V^{[1]}(t,x) =\sup_{\tau\in\cT_{[0,T-t]}} \E_{t,x}[e^{-r\tau}\pi_t(\tau)(K-X_{t+\tau})^+].
	\end{align*}
A continuous intensity model
	\begin{align*}
		 \pi_t(u)=\exp\!\left(-\int_t^{t+u}\lambda_s\dd s\right)
	\end{align*}
	has the multiplicative property
	\begin{equation}\label{eq:survival-chain}
 		\pi_t(s+u)=\pi_t(s)\pi_{t+s}(u),\qquad s,u\ge0,\quad s+u\le T-t.
	\end{equation}
More generally, \eqref{eq:survival-chain} holds if the law of $\bar T$ has no atoms in $(0,T)$.
For an intermediate time $t<t+s<T$, the absence of an atom gives
	$\Pp(\bar T\ge t+s)=\Pp(\bar T>t+s)$, which proves the factorization.
The cases $s=0$ and $t+s=T$ follow from $\pi_t(0)=1$ and the terminal convention
	$\pi_T(0)=1$, respectively; no division by $\Pp(\bar T>T)$ is required.
\subsection{Multiple rights under termination risk}
For the same admissible exercise sequences as in Section~\ref{sec:multiple}, define first the
	\emph{direct termination-risk value}. 
For $0\le t<T$,
	\begin{equation}\label{eq:barVm-primal}
		\begin{aligned}
 		\bar V_{\mathrm{dir}}^{[m]}(t,x)&:=\sup_{\boldsymbol\tau^{(m)}}
			 \E_{t,x}\!\left[\sum_{i=1}^m e^{-r\tau_i}(K-X_{t+\tau_i})^+
 			  \one_{A_i}\one_{\{t+\tau_i\le\bar T\}}\,\middle|\,\bar T>t\right]\\
 			&=\sup_{\boldsymbol\tau^{(m)}}
				 \E_{t,x}\!\left[\sum_{i=1}^m e^{-r\tau_i}\pi_t(\tau_i)
  				 (K-X_{t+\tau_i})^+\one_{A_i}\right].
		\end{aligned}
	\end{equation}
The second equality follows by applying Lemma~\ref{lem:survival} term by term.
We put $\bar V_{\mathrm{dir}}^{[0]}\equiv0$ and use the terminal convention
	$\bar V_{\mathrm{dir}}^{[k]}(T,x)=g(x)$.

We now define a recursive value separately. Put $\bar V^{[0]}\equiv0$, set
	$\bar V^{[k]}(s,x)=0$ for $s>T$, and, recursively in $m$, define
	\begin{align}
 		\bar\Psi_\delta^{[m-1]}(s,y)
			&:=\begin{cases}
 			\E_{s,y}[e^{-r\delta}\pi_s(\delta)\bar V^{[m-1]}(s+\delta,X_{s+\delta})], &s+\delta\le T,\\
 			0,    &s+\delta>T,
 			\end{cases}\label{eq:barPsi}\\
 		\bar g_\delta^{[m]}(s,y)
 		&:=(K-y)^++\bar\Psi_\delta^{[m-1]}(s,y),\notag\\
 		\bar V^{[m]}(t,x)
		&:=\sup_{\tau\in\cT_{[0,T-t]}} \E_{t,x}[e^{-r\tau}\pi_t(\tau)
 			\bar g_\delta^{[m]}(t+\tau,X_{t+\tau})].\label{eq:barVm-rec}
	\end{align}
For every $k\ge1$, this recursion gives $\bar V^{[k]}(T,x)=g(x)$.
For $m=1$, the direct and recursive definitions coincide with the one-right value above.

\begin{proposition}[One-sided recursive reduction with termination]
\label{prop:termination-recursion}
Assume that the survival probabilities satisfy the multiplicative relation \eqref{eq:survival-chain}. 
Then, for every $m\ge1$,
	\begin{equation}\label{eq:bar-direct-upper}
 		\bar V_{\mathrm{dir}}^{[m]}(t,x)\le \bar V^{[m]}(t,x).
	\end{equation}
For $m=1$ equality holds.
\end{proposition}
\begin{proof}
We argue by induction on $m$. The case $m=1$ is immediate.
Assume $\bar V_{\mathrm{dir}}^{[m-1]}\le\bar V^{[m-1]}$.
Fix an admissible $m$-exercise rule
	$\boldsymbol\tau^{(m)}=(\tau_1,\ldots,\tau_m)$ and put
	$\vartheta:=\tau_1+\delta$.
On $A_2=\{\vartheta\le T-t\}$, the shifted later exercise times form an admissible
	$(m-1)$-exercise rule for the problem started from
	$(t+\vartheta,X_{t+\vartheta})$.
Moreover, for every later exercise time, the multiplicative relation gives
	\begin{align*}
 		e^{-r\tau_i}\pi_t(\tau_i)=e^{-r\vartheta}\pi_t(\vartheta)
   			e^{-r(\tau_i-\vartheta)}\pi_{t+\vartheta}(\tau_i-\vartheta),
			 \qquad i\ge2,
	\end{align*}
	on the event on which that exercise is feasible.
Conditioning on $\mathcal F_{t+\vartheta}$, the expected contribution of all later
	exercises is therefore bounded above by
	\begin{align*}
		 \one_{A_2}e^{-r\vartheta}\pi_t(\vartheta)
			 \bar V_{\mathrm{dir}}^{[m-1]}(t+\vartheta,X_{t+\vartheta}) \le
		 \one_{A_2}e^{-r\vartheta}\pi_t(\vartheta)
			 \bar V^{[m-1]}(t+\vartheta,X_{t+\vartheta}).
	\end{align*}
Conditioning next on $\mathcal F_{t+\tau_1}$ and using \eqref{eq:survival-chain} once more gives
	\begin{align*}
		 \E\!\left[\one_{A_2}e^{-r\vartheta}\pi_t(\vartheta)
			 \bar V^{[m-1]}(t+\vartheta,X_{t+\vartheta})
			 \middle|\mathcal F_{t+\tau_1}\right]
		=e^{-r\tau_1}\pi_t(\tau_1)
			 \bar\Psi_\delta^{[m-1]}(t+\tau_1,X_{t+\tau_1}).
	\end{align*}
Hence the expected payoff of the fixed direct rule is at most
	\begin{align*}
		 \E_{t,x}\!\left[e^{-r\tau_1}\pi_t(\tau_1)\bar g_\delta^{[m]}(t+\tau_1,X_{t+\tau_1})\right]
 		\le \bar V^{[m]}(t,x).
	\end{align*}
Taking the supremum over all admissible direct rules proves
\eqref{eq:bar-direct-upper}.
\end{proof}
\begin{remark}[Scope of Proposition~\ref{prop:termination-recursion}]
\label{rem:termination-scope}
For a general survival law satisfying \eqref{eq:survival-chain},
	Proposition~\ref{prop:termination-recursion} proves the direction
	$\bar V_{\mathrm{dir}}^{[m]}\le\bar V^{[m]}$ and identifies the correct survival factors
	in the recursion. 
The reverse inequality requires a construction of nearly optimal
	continuation rules after a random first exercise time. 
The elementary proof of Lemma~\ref{lem:direct-rec} obtains this 
	through deadline-aligned Bermudan approximation,
	stripwise time continuity, and time homogeneity. 
Since a general $\pi_t$ makes the problem time-inhomogeneous, 
	we do not repeat that argument here.
For Poisson termination the time homogeneity is restored after absorbing the intensity
 into the discount rate, and equality is proved in Proposition~\ref{prop:poisson-rec} below.
\end{remark}

For the following discussion $\bar V^{[m]}$ always denotes the recursive value
	\eqref{eq:barVm-rec}. 
The deterministic feasibility deadlines themselves are unchanged by termination risk. 
The algebraic part of the proof of Proposition~\ref{prop:stage} therefore
	carries over to the recursive values and gives
	\begin{align*}
 		\bar V^{[m]}(t,x)&=\bar V^{[q_m(t)]}(t,x),\\
 		\bar V^{[m]}(t,x)&=\bar V^{[j]}(t,x),
			 &&t\in(d_j,d_{j-1}],\quad m\ge j,\\
 		\bar V^{[m]}(d_j,x)&=\bar V^{[j+1]}(d_j,x),
 			&&m\ge j+1.
	\end{align*}
The direct value has the same algebraic degeneration by feasibility.
A right-hand deadline limit, however, requires stripwise time regularity of the
	termination-risk value and is not asserted here for a general survival law.

The survival-adjusted marginal-right value
	\begin{align*}
 		\bar f_\delta^{[m]}(t,x):=\bar V^{[m]}(t,x)-\bar\Psi_\delta^{[m-1]}(t,x)
	\end{align*}
	satisfies the exact contact identity
	\begin{align*}
 		\bar V^{[m]}-\bar g_\delta^{[m]}=\bar f_\delta^{[m]}-(K-x)^+.
	\end{align*}
We expect that, under corresponding regularity and dynamic-programming assumptions,
	the comparison of the marginal-right values and the single-crossing argument of Sections~\ref{sec:singlecross}--\ref{sec:boundary} 
	yields a one-sided boundary $\bar b^{[m]}$.
In the slope estimates the identity $\E[e^{-rh}R_1^{(h)}]=1$ is replaced by the corresponding
	inequality after multiplication by the survival factor. 
We do not carry out those details here,
	and statements below involving $\bar b^{[k]}$ are made whenever these one-sided boundaries exist.
In the model without termination, the supermartingale property of the incremental values is a theorem
	(Theorem~\ref{th:HD}), and it yields the adjacent-level hierarchy of Corollary~\ref{cor:hierarchy}.
Its survival-adjusted analogue, namely the corresponding supermartingale property for the
	incremental value $\bar D^{[k]}:=\bar V^{[k]}-\bar V^{[k-1]}$ of the $k$th right,
	is not proved here.
Accordingly, for $m\ge3$, any adjacent-level boundary hierarchy
	$\bar b^{[m-1]}\le\bar b^{[m]}$ under termination risk requires an additional
	supermartingale argument for the survival-adjusted incremental values.
\subsection{Poisson termination}
The exponential case is especially transparent, because it amounts to raising the discount rate 
	above the risk-neutral drift of the asset. 
The recursive reduction can then be proved in full by the methods of 
	Sections~\ref{sec:multiple} and~\ref{sec:singlecross}.
\begin{lemma}[Discount rates above the risk-neutral drift]
\label{lem:rho-discount}
Let $\bar r\ge r$. Replace every discount factor $e^{-ru}$ in \eqref{eq:Vm-direct}, 
	\eqref{eq:Psi} and \eqref{eq:Vm-rec} by $e^{-\bar ru}$, keeping the asset dynamics 
	$\dd X_s=rX_s\dd s+\sigma X_s\dd B_s$ unchanged, and denote the resulting direct value, 
	delayed residual-right value and recursive value by 
	$V^{[m]}_{\bar r,\mathrm{dir}}$, $\Psi^{[m-1]}_{\bar r,\delta}$ and $V^{[m]}_{\bar r}$. 
Then Proposition~\ref{prop:stage}, Theorem~\ref{th:Vprops}, Lemma~\ref{lem:supermart} 
	and Lemma~\ref{lem:berm-conv} hold for these quantities. The bounds in
	Theorem~\ref{th:Vprops}(i) and (v) are unchanged, while the stripwise H\"older
	constant in Theorem~\ref{th:Vprops}(vi) may depend on $\bar r$. Moreover,
	\begin{equation}\label{eq:rho-direct-rec}
		V^{[m]}_{\bar r,\mathrm{dir}}(t,x)=V^{[m]}_{\bar r}(t,x),
			\qquad (t,x)\in[0,T]\times(0,\infty),\quad m\ge1.
	\end{equation}
\end{lemma}
\begin{proof}
In the proofs of the cited results, the martingale identity \eqref{eq:martfactor} 
	is used only through the following consequence: for every bounded stopping time 
	$\tau$ of $(\mathcal F_{t+u})_{u\ge0}$,
	\begin{equation}\label{eq:rho-supermart}
		\E\bigl[e^{-\bar r\tau}R_\tau\bigr]\le1,
			\qquad\text{equivalently}\qquad
		\E_{t,x}\bigl[e^{-\bar r\tau}X_{t+\tau}\bigr]\le x.
	\end{equation}
This inequality holds for $\bar r\ge r$, because 
	$e^{-\bar ru}R_u=e^{-(\bar r-r)u}\,e^{-ru}R_u$, where $(e^{-ru}R_u)_{u\ge0}$ 
	is a positive martingale and $0<e^{-(\bar r-r)u}\le1$; 
	optional sampling gives $\E[e^{-\bar r\tau}R_\tau]\le\E[e^{-r\tau}R_\tau]=1$.
All other ingredients are unaffected by the change of discount rate: 
	the strong Markov property, the multiplicative representation $X^{t,x}_{t+u}=xR_u$, 
	the estimate $\E[\sup_{0\le u\le h}|R_u-1|]\le c\sqrt h$, 
	the zero extension \eqref{eq:extension}, the position of the deadlines, 
	and the bound $e^{-\bar rs}\le1$. 
We go through the proofs in order.

\emph{(a) Proposition~\ref{prop:stage}.}
The proof uses only the recursive structure of the stopping rewards 
	and the position of the deadlines, and is unchanged.

\emph{(b) Theorem~\ref{th:Vprops}.}
Assertions (i)--(iv) and (viii) use only $e^{-\bar r\delta}\le1$, 
	pathwise comparison of $x_1R$ and $x_2R$, and convexity; they are unchanged. 
In \eqref{eq:gmLip}, the second term becomes
	\begin{align*}
		e^{-\bar r\delta}(m-1)\E\bigl[|y_1-y_2|R_\delta\bigr]
			=e^{-(\bar r-r)\delta}(m-1)|y_1-y_2|\le(m-1)|y_1-y_2|,
	\end{align*}
	so the stopping reward is still $m$-Lipschitz, and \eqref{eq:rho-supermart} 
	replaces \eqref{eq:martfactor} in the passage from the stopping reward to 
	$V^{[m]}_{\bar r}$. This gives (v). 
In the proof of (vi), the H\"older estimate for the delayed term becomes
	\begin{align*}
		e^{-\bar r\delta}C_{m-1}\E[1+xR_\delta]\sqrt h\le C_{m-1}(1+x)\sqrt h,
	\end{align*}
	and the remaining terms involve only the spatial Lipschitz constant and the asset dynamics. 
Assertion (vii) follows from (v) and (vi) as before, 
	and so does the right-limit identity \eqref{eq:dright}.

\emph{(c) Lemma~\ref{lem:supermart}.}
Both parts use only the dynamic-programming principle, the strong Markov property 
	and the definition of $\Psi^{[k]}_{\bar r,\delta}$; 
	no martingale identity is involved.

\emph{(d) Lemma~\ref{lem:berm-conv}.}
Define the Bermudan recursion \eqref{eq:Psi-h}--\eqref{eq:bermrec} with $e^{-rh}$ and 
	$e^{-rNh}$ replaced by $e^{-\bar rh}$ and $e^{-\bar rNh}$. 
Step~1 is unchanged, with $e^{-\bar r\delta}J_{k-1}(h)$ on the right of \eqref{eq:obs-error}. 
In Step~2 the three error terms are estimated by $1-e^{-\bar rh}\le\bar rh$, 
	the stripwise H\"older estimate from (b), the $k$-Lipschitz property of the stopping reward, 
	and \eqref{eq:rho-supermart} in place of the identity 
	$\E_{t_n,y}[e^{-r\tau}X_{t_n+\tau}]=y$. 
This yields the same bound $C(1+Y)\sqrt h$, with a constant $C$ that now depends also on $\bar r$. 
Step~3 uses Lemma~\ref{lem:supermart}(ii), available by (c), 
	together with the pathwise tail argument for the Bermudan value, 
	which does not involve the discount rate.

\emph{(e) Lemma~\ref{lem:direct-rec}.}
Step~1 uses the strong Markov property at $t+\vartheta$ and at $t+\tau_1$, 
	and the definition of $\Psi^{[m-1]}_{\bar r,\delta}$; it is unchanged. 
Step~2 uses that the modified recursion \eqref{eq:bermrec} is the dynamic-programming 
	recursion of the discrete-time $m$-right problem with one-step discount factor 
	$e^{-\bar rh}$, together with (d). 
Step~3 uses the time homogeneity of the problem, which persists because the discount factor 
	$e^{-\bar ru}$ depends only on the elapsed time $u$, 
	and stripwise continuity in time, which is (b)(vii). 
This proves \eqref{eq:rho-direct-rec}.
\end{proof}
\begin{proposition}[Recursive reduction under Poisson termination]
\label{prop:poisson-rec}
Suppose the contract disappears at the first event of a Poisson process of intensity
	$\lambda>0$, independent of the asset, and put $\bar r:=r+\lambda$.
Then, for every $m\ge1$,
	\begin{equation}\label{eq:poisson-primal-rec}
		 \bar V_{\mathrm{dir}}^{[m]}(t,x)=\bar V^{[m]}(t,x)=V^{[m]}_{\bar r}(t,x),
		 \qquad (t,x)\in[0,T]\times(0,\infty).
	\end{equation}
Thus the direct termination-risk problem and the recursive problem coincide in the Poisson case. 
In addition, $\bar V^{[m]}$ has the properties listed in
	Theorem~\ref{th:Vprops} and Lemma~\ref{lem:supermart}, with $r$ replaced by
	$\bar r$ in the discount factors and with the qualifications of Lemma~\ref{lem:rho-discount}.
\end{proposition}
\begin{proof}
Memorylessness gives $\pi_t(u)=e^{-\lambda u}$, hence
$e^{-ru}\pi_t(u)=e^{-\bar r u}$ for all $t$ and $u$.
Therefore every term of the direct representation \eqref{eq:barVm-primal} is the
	corresponding term of \eqref{eq:Vm-direct} with discount rate $\bar r$, so that
	\begin{align*}
 		\bar V_{\mathrm{dir}}^{[m]}=V^{[m]}_{\bar r,\mathrm{dir}}.
	\end{align*}
Likewise, \eqref{eq:barPsi} becomes
	\begin{align*}
 		\bar\Psi_\delta^{[m-1]}(s,y)
			=\E_{s,y}\bigl[e^{-\bar r\delta}\bar V^{[m-1]}(s+\delta,X_{s+\delta})\bigr]
	\end{align*}
	for $s+\delta\le T$, and \eqref{eq:barVm-rec} is exactly the recursion
	\eqref{eq:Vm-rec} with discount rate $\bar r$.
Induction on $m$, starting from $\bar V^{[0]}=V^{[0]}_{\bar r}\equiv0$, gives
	$\bar V^{[m]}=V^{[m]}_{\bar r}$.
Lemma~\ref{lem:rho-discount} gives
	$V^{[m]}_{\bar r,\mathrm{dir}}=V^{[m]}_{\bar r}$, proving
	\eqref{eq:poisson-primal-rec}.
The remaining assertions also follow from Lemma~\ref{lem:rho-discount}.
\end{proof}
\begin{theorem}[Effective discount rate under Poisson termination]
\label{th:poisson}
Suppose the contract disappears at the first event of a Poisson process of intensity 
	$\lambda>0$, independent of the asset. 
Then
	\begin{equation}\label{eq:pi-poisson}
 		\pi_t(u)=e^{-\lambda u}.
	\end{equation}
For every $m\ge1$, the termination-risk problem is therefore identical, 
	at the level of its objective function and dynamic program, 
	to the fixed-maturity multiple-stopping problem 
	with effective discounting $e^{-(r+\lambda)u}$, that is,
	\begin{align*}
		 r\longmapsto r+\lambda
	\end{align*}
	in the stopping functional, while the risk-neutral asset dynamics remain
$\dd X_s=rX_s\dd s+\sigma X_s\dd B_s$. In particular, the continuation operator in the QVI is
	\begin{align*}
		 \partial_t+\cL-(r+\lambda),
	\end{align*}
	where $\cL$ still contains the drift $rx$.
\end{theorem}
\begin{proof}
Memorylessness gives \eqref{eq:pi-poisson}, hence $e^{-ru}\pi_t(u)=e^{-(r+\lambda)u}$. 
In the direct representation \eqref{eq:barVm-primal}, every exercise term is therefore exactly 
	the corresponding fixed-maturity term with discount exponent $r+\lambda$. 
The same factorization in \eqref{eq:barPsi} gives $e^{-(r+\lambda)\delta}$ across the refractory period. 
By Proposition~\ref{prop:poisson-rec}, the direct and recursive representations coincide, 
	and the value recursion is \eqref{eq:Vm-rec} with the higher discount rate $r+\lambda$. 
The law of $X$ has not been changed, so the generator retains risk-neutral drift $r$; 
	only the killing/discount term changes from $-rV$ to $-(r+\lambda)V$.
\end{proof}

We stress that this identification concerns the objective functional only. 
Because the drift of $X$ is still $r$, the martingale identity \eqref{eq:martfactor} becomes 
	$\E[e^{-(r+\lambda)u}R_u]=e^{-\lambda u}<1$, so the results of 
	Sections~\ref{sec:multiple}--\ref{sec:eep} do not transfer verbatim; 
	the parts needed for the recursive reduction do, by Lemma~\ref{lem:rho-discount}. 
For instance, in the stopping region the drift of the discounted put payoff becomes
	\begin{align*}
		\bigl(\partial_t+\cL-(r+\lambda)\bigr)(K-x)=-(r+\lambda)K+\lambda x
	\end{align*}
	instead of $-rK$, which modifies the premium terms of Section~\ref{sec:eep} accordingly.
\begin{remark}[Poisson termination as a dividend yield]
\label{rem:poisson-dividend}
Since the asset has drift $r=\bar r-\lambda$ while payoffs are discounted at $\bar r$, 
	the Poisson problem coincides with the multiple-exercise put on an asset paying 
	a continuous dividend yield $\lambda$ in a market with interest rate $\bar r$. 
In particular, the one-right problem is the classical finite-horizon American put 
	with a continuous dividend yield, whose exercise boundary is nondecreasing and continuous, 
	with limit $\min\{K,\bar rK/\lambda\}=K$ at maturity. 
Throughout, \eqref{eq:rho-supermart} replaces \eqref{eq:martfactor}, and for $x\le K$
	\begin{align*}
		\bigl(\partial_t+\cL-\bar r\bigr)(K-x)=-\bar rK+\lambda x\le-rK<0.
	\end{align*}
This suggests that the single-crossing and boundary results of 
	Sections~\ref{sec:singlecross} and~\ref{sec:boundary} can be transferred 
	to the Poisson case; we do not carry out the details here.
\end{remark}
\subsection{Comparative results}
For a general survival law the following optional-sampling statements require the usual
	measurable dynamic-programming machinery for the time-inhomogeneous recursive problem.
We state that requirement explicitly.
\begin{assumption}[Dynamic programming under general survival]
\label{ass:termination-dpp}
For every finite $k$, the recursive stopping problem \eqref{eq:barVm-rec} has a Borel
	value function and satisfies the dynamic-programming principle at bounded stopping times;
	in particular, after a bounded stopping time an $\varepsilon$-optimal continuation rule
	can be selected measurably from the current time and state.
\end{assumption}
In the Poisson case this assumption is not additional: it follows from
	Proposition~\ref{prop:poisson-rec} and Lemma~\ref{lem:rho-discount}.
We first record the survival-adjusted analogue of Lemma~\ref{lem:supermart} for the
	recursive termination-risk problem.
\begin{lemma}[Survival-adjusted supermartingale bounds]
\label{lem:bar-supermart}
Assume \eqref{eq:survival-chain} and Assumption~\ref{ass:termination-dpp}.
Fix $(t,x)$ with $t<T$. 
For $k\ge1$ put
	\begin{align*}
 		\bar Z_u^{[k]}&:=e^{-ru}\pi_t(u)\bar V^{[k]}(t+u,X_{t+u}),\\
 		\bar Y_u^{[k]}&:=e^{-ru}\pi_t(u)\bar\Psi_\delta^{[k]}(t+u,X_{t+u}),
	\end{align*}
	for $0\le u\le T-t$, and $\bar Z_u^{[k]}:=\bar Y_u^{[k]}:=0$ for $u>T-t$.
\begin{enumerate}[label={\rm(\roman*)}]
 	\item $\bar Z^{[k]}$ and $\bar Y^{[k]}$ are strong supermartingales with respect to
 			$(\mathcal F_{t+u})_{u\ge0}$: for bounded stopping times $0\le\rho\le\eta$,
 			\begin{align*}
  				\E_{t,x}[\bar Z_\eta^{[k]}\mid\mathcal F_{t+\rho}]\le\bar Z_\rho^{[k]},
				\qquad 
  				\E_{t,x}[\bar Y_\eta^{[k]}\mid\mathcal F_{t+\rho}]\le\bar Y_\rho^{[k]}.
 			\end{align*}
 	\item For $m\ge2$,
 		\begin{align*}
  			\bar V^{[m]}\le\bar V^{[1]}+\bar\Psi_\delta^{[m-1]}.
 		\end{align*}
\end{enumerate}
\end{lemma}
\begin{proof}
(i) Fix bounded stopping times $\rho\le\eta$.
On $\{\rho>T-t\}$ both sides vanish, so work on $\{\rho\le T-t\}$.
By Assumption~\ref{ass:termination-dpp}, conditional on $\mathcal F_{t+\rho}$
	we may wait until $\eta\wedge(T-t)$ and, on $\{\eta\le T-t\}$, follow an
	$\varepsilon$-optimal rule for the problem started at
	$(t+\eta,X_{t+\eta})$, whose relative exercise time we denote by $\varsigma$.
By \eqref{eq:survival-chain}, the combined discount and survival factor satisfies
	\begin{align*}
 		e^{-r(\eta-\rho+\varsigma)}\pi_{t+\rho}(\eta-\rho+\varsigma)
 		 =e^{-r(\eta-\rho)}\pi_{t+\rho}(\eta-\rho)e^{-r\varsigma}\pi_{t+\eta}(\varsigma).
	\end{align*}
Letting $\varepsilon\downarrow0$ gives
	\begin{align*}
 		\E_{t,x}\!\left[e^{-r(\eta-\rho)}\pi_{t+\rho}(\eta-\rho)
 			\bar V^{[k]}(t+\eta,X_{t+\eta})\one_{\{\eta\le T-t\}}
 			\middle|\mathcal F_{t+\rho}\right]
 		\le \bar V^{[k]}(t+\rho,X_{t+\rho}).
	\end{align*}
Multiplying by $e^{-r\rho}\pi_t(\rho)$ and using
	$\pi_t(\rho)\pi_{t+\rho}(\eta-\rho)=\pi_t(\eta)$ proves the inequality for
	$\bar Z^{[k]}$.

For $\bar Y^{[k]}$, if $\tau+\delta\le T-t$, the strong Markov property,
	\eqref{eq:barPsi}, and \eqref{eq:survival-chain} give
	\begin{align*}
		 \bar Y_\tau^{[k]}&=\E[\bar Z_{\tau+\delta}^{[k]}\mid\mathcal F_{t+\tau}].
	\end{align*}
If $\tau+\delta>T-t$, both sides vanish by the zero extension.
The tower property and the supermartingale inequality for $\bar Z^{[k]}$ then imply
	\begin{align*}
 		\E[\bar Y_\eta^{[k]}\mid\mathcal F_{t+\rho}]\le \bar Y_\rho^{[k]}.
	\end{align*}

(ii) For every $\tau\in\cT_{[0,T-t]}$, part (i) with $\rho=0$ gives
	\begin{align*}
 		\E_{t,x}[e^{-r\tau}\pi_t(\tau) \bar g_\delta^{[m]}(t+\tau,X_{t+\tau})]
 		&=\E_{t,x}[e^{-r\tau}\pi_t(\tau)(K-X_{t+\tau})^+]
  			 +\E_{t,x}[\bar Y_\tau^{[m-1]}]\\
 		&\le \bar V^{[1]}(t,x)+\bar\Psi_\delta^{[m-1]}(t,x).
	\end{align*}
Taking the supremum in \eqref{eq:barVm-rec} proves (ii).
\end{proof}
The contact sets in the next lemma are those of the recursive termination-risk problem.
For Poisson termination, Proposition~\ref{prop:poisson-rec} identifies them with the
	contact sets of the original direct problem.
\begin{lemma}[One-right stopping-set inclusion under termination]
\label{lem:barS1barSm}
Under the assumptions of Lemma~\ref{lem:bar-supermart}, for every $m\ge2$,
	\begin{align*}
 		\bar{\mathcal S}_t^{[1]}\subseteq \bar{\mathcal S}_t^{[m]},
	\end{align*}
	where $\bar{\mathcal S}_t^{[k]}$ denotes the contact set of the recursive
	$k$-right termination-risk problem.
Consequently, whenever the corresponding contact sets are one-sided with boundaries
	$\bar b^{[k]}$, one has $\bar b^{[m]}(t)\ge\bar b^{[1]}(t)$.
\end{lemma}
\begin{proof}
By Lemma~\ref{lem:bar-supermart}(ii),
	\begin{align*}
 		\bar V^{[m]}(t,x)\le\bar V^{[1]}(t,x)+\bar\Psi_\delta^{[m-1]}(t,x).
	\end{align*}
Subtracting the $m$-right stopping reward gives
	\begin{align*}
 		\bar V^{[m]}-\bar g_\delta^{[m]}\le \bar V^{[1]}-(K-x)^+.
	\end{align*}
If $x$ belongs to the one-right contact set, the right-hand side is zero,
	while obstacle domination makes the left-hand side nonnegative.
Thus $x$ also belongs to the $m$-right contact set.
\end{proof}

Because $0\le\pi_t(u)\le1$, both the direct termination-risk value and the
	recursive termination-risk value are dominated by the corresponding value without termination.
\begin{lemma}[Value comparison]
\label{lem:termination-value-compare}
For every $m\ge1$,
	\begin{align*}
 		\bar V_{\mathrm{dir}}^{[m]}(t,x)\le V^{[m]}(t,x),\qquad
 		\bar V^{[m]}(t,x)\le V^{[m]}(t,x).
	\end{align*}
\end{lemma}
\begin{proof}
For the direct value, every term in \eqref{eq:barVm-primal} is bounded above by the
	corresponding term in \eqref{eq:Vm-direct}; hence
	$\bar V_{\mathrm{dir}}^{[m]}\le V_{\mathrm{dir}}^{[m]}=V^{[m]}$ by
Lemma~\ref{lem:direct-rec}.

For the recursive value, argue by induction on $m$.
The case $m=1$ follows immediately from $\pi_t(\tau)\le1$.
If $\bar V^{[m-1]}\le V^{[m-1]}$, then
	\begin{align*}
		 \bar\Psi_\delta^{[m-1]}(s,y) &\le \E_{s,y}[e^{-r\delta}V^{[m-1]}(s+\delta,X_{s+\delta})]
 	=\Psi_\delta^{[m-1]}(s,y),
	\end{align*}
	where both sides are zero when $s+\delta>T$.
Thus $\bar g_\delta^{[m]}\le g_\delta^{[m]}$, and since the stopping rewards are
	nonnegative and $\pi_t(\tau)\le1$,
	\begin{align*}
		 \E_{t,x}[e^{-r\tau}\pi_t(\tau)\bar g_\delta^{[m]}(t+\tau,X_{t+\tau})]
		 \le
		 \E_{t,x}[e^{-r\tau}g_\delta^{[m]}(t+\tau,X_{t+\tau})].
	\end{align*}
Taking suprema proves $\bar V^{[m]}\le V^{[m]}$.
\end{proof}

For one right the direct and recursive termination values coincide and the stopping reward is the
	same as without termination. 
Hence, whenever the one-sided boundaries exist,
	\begin{align*}
 		\bar b^{[1]}(t)\ge b^{[1]}(t).
	\end{align*}
Under Assumption~\ref{ass:termination-dpp}, Lemma~\ref{lem:barS1barSm} therefore gives
	for the recursive termination problem
	\begin{align*}
 		\bar b^{[m]}(t)\ge\bar b^{[1]}(t)\ge b^{[1]}(t).
	\end{align*}
On the final strip $(T-\delta,T]$, the recursive stagewise reduction gives
	\begin{align*}
		\bar b^{[m]}(t)=\bar b^{[1]}(t)\ge b^{[1]}(t)=b^{[m]}(t).
	\end{align*}
For $m\ge2$, however, the same-level comparison
	\begin{equation}\label{eq:open-random-boundary}
 		\bar b^{[m]}(t)\stackrel{?}{\ge}b^{[m]}(t),\qquad t\le T-\delta,
	\end{equation}
	is not proved in general. 
The reason is that termination lowers both the value of waiting and the
	continuation value embedded in the stopping reward; value domination alone does not order the
	marginal-right values $\bar f_\delta^{[m]}$ and $f_\delta^{[m]}$.
In the Poisson case the recursive and direct values coincide, so
	\eqref{eq:open-random-boundary} becomes the nontrivial comparative-statics question of how the
	multiple-exercise boundary responds when the discount rate in the objective is raised while the
	risk-neutral drift of the asset remains fixed.

\section{Numerical illustration}\label{sec:numerics}
We solve the QVI numerically by finite differences and projected successive over-relaxation (SOR), 
	while the refractory-period expectation is evaluated by Gauss--Hermite quadrature. 
Table~\ref{tab:grid} reports a representative grid-refinement study. 
Parameters are
	\begin{align*}
 	K=100,\qquad T=1,\qquad r=0.05,\qquad \sigma=0.30,\qquad \delta=0.10,
	\end{align*}
	with the spatial domain truncated at $x_{\max}=300$ and up to three exercise rights. 
The Gauss--Hermite rule uses 256 nodes. In every run $\delta/\Delta t$ is an integer, 
	so all feasibility deadlines are time-grid points, in line with the aligned Bermudan scheme of 	
	Section~\ref{sec:singlecross}.

\begin{center}
\begin{minipage}{\textwidth}
\centering
\small
	\captionof{table}{Grid refinement for the multiple-exercise American put. 
		Boundary values are obtained from the midpoint between the last stopping grid point 
		and the first continuation grid point.}
	\label{tab:grid}
\resizebox{\textwidth}{!}{%
\begin{tabular}{rrrrrrrrrr}
\toprule
$N_t$ & $N_x$ & $\Delta t$ & $\Delta x$ & $V^{[1]}(0,K)$ & $V^{[2]}(0,K)$ & $V^{[3]}(0,K)$ & $b^{[1]}(0)$ & $b^{[2]}(0)$ & $b^{[3]}(0)$\\
\midrule
400  & 600  & 0.002500 & 0.500000 & 9.863938 & 19.249703 & 28.124411 & 69.250000 & 71.750000 & 73.750000\\
800  & 1000 & 0.001250 & 0.300000 & 9.867241 & 19.253155 & 28.127412 & 69.150000 & 71.550000 & 73.350000\\
1200 & 1600 & 0.000833 & 0.187500 & 9.868143 & 19.254027 & 28.128066 & 69.281250 & 71.531250 & 73.406250\\
1600 & 2200 & 0.000625 & 0.136364 & 9.868607 & 19.254488 & 28.128448 & 69.204545 & 71.522727 & 73.431818\\
2000 & 2600 & 0.000500 & 0.115385 & 9.868892 & 19.254787 & 28.128712 & 69.173077 & 71.480769 & 73.326923\\
\bottomrule
\end{tabular}}
\end{minipage}
\end{center}

The table illustrates three features consistent with the analytical development: 
	the value increases with the number of exercise rights; the computed initial boundaries satisfy
	$b^{[1]}(0)<b^{[2]}(0)<b^{[3]}(0)$; and all observed boundaries remain below $K$. 
The boundary estimates are resolved only up to the spatial mesh size, 
	which explains their small non-monotone variation across refinements.
\begin{figure}[htbp]
\centering
\includegraphics[width=0.65\textwidth]{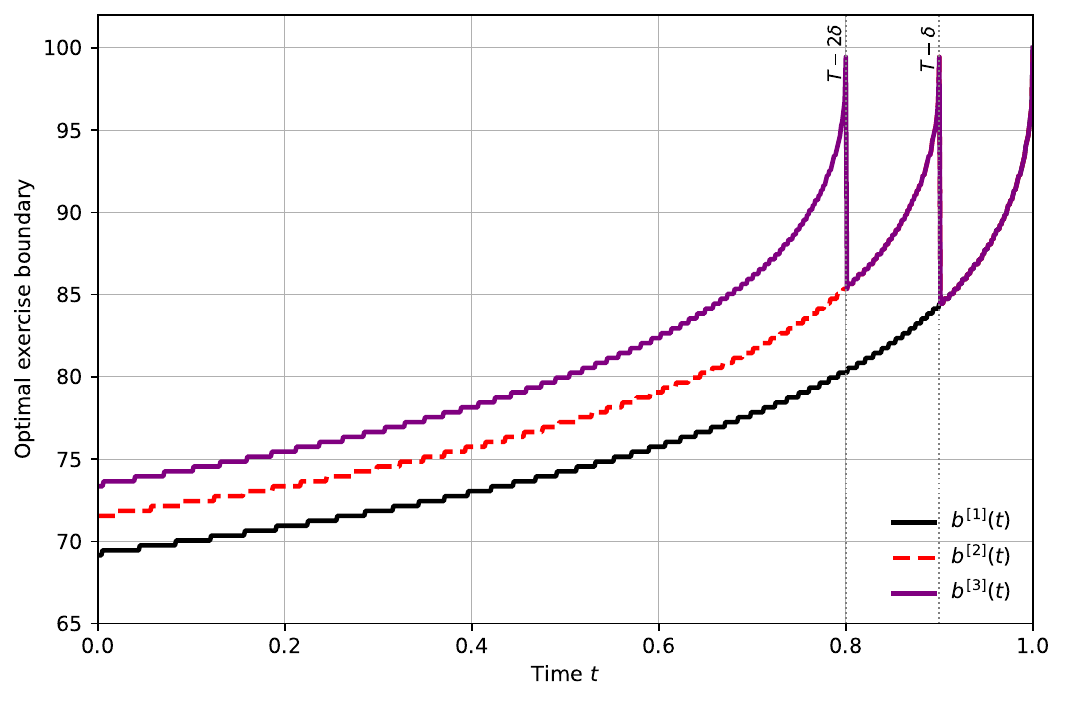}
	\caption{Computed optimal exercise boundaries for one, two, and three exercise rights. 
	The parameters are $K=100$, $T=1$, $r=0.05$, $\sigma=0.30$, and $\delta=0.10$; 
	the curves use $N_t=800$, $N_x=1000$, and 256 Gauss--Hermite nodes, 
	matching the corresponding row of Table~\ref{tab:grid}. 
	The vertical dotted lines mark the feasibility deadlines $T-2\delta$ and $T-\delta$. 
	The observed ordering is consistent with the boundary hierarchy proved under the standing model
	assumptions, while the $m=2$ and $m=3$ curves 
	display right-hand downward jumps at the deadlines.}
\label{fig:boundaries123}
\end{figure}
Figure~\ref{fig:boundaries123} visualizes the full time profiles of the computed boundaries. 
In particular, the jumps at $T-2\delta$ and $T-\delta$ are clearly visible. 
They reflect the fact that immediately after a deadline one fewer future exercise 
	can be accommodated within the remaining horizon.
On every open deadline strip, the computed boundaries for $m=1,2,3$ are nondecreasing in time
	up to the spatial mesh size; the only departures from monotonicity are the downward jumps at the
	feasibility deadlines.
These computations support, but do not replace, the still-open general proof of 
	stripwise time monotonicity discussed in Remark~\ref{rem:monotonicity-gap}.

\section*{Acknowledgments}
The author thanks Aiko Kurushima for fruitful discussions on multiple stopping problems.
During the preparation of this work, the author used ChatGPT (OpenAI) to assist with
	English-language editing, numerical calculations underlying the figures,
	and auxiliary checks of the exposition and mathematical consistency.
In particular, the proof strategy and detailed argument for the local Lipschitz continuity of the
	exercise boundaries within each open deadline strip in Section~\ref{sec:strip-regularity}
	were developed through iterative exchanges with ChatGPT.
The author independently reviewed and verified all mathematical arguments and
	takes full responsibility for their correctness and for the final content of the manuscript.
\appendix
\section{Weak regularity and the Duhamel representation}
\label{app:weak-regularity-proof}
\subsection*{Proof of Theorem~\ref{th:R-weak}}
\begin{proof}
\emph{Step 1: the bounded weak density.}
Use log-price $z=\log x$ and put
	\begin{align*}
		 \widetilde V^{[k]}(t,z)&:=V^{[k]}(t,e^z),\\
		 \widetilde A&:=\tfrac12\sigma^2\partial_{zz}+\mu\partial_z-r.
	\end{align*}
Let $\widetilde{\mathcal P}_s\phi(z):=\E[e^{-rs}\phi(z+\mu s+\sigma B_s)]$ denote the discounted semigroup
	in log-price, so that $\widetilde{\mathcal P}_s[\phi\circ\exp]=(\mathcal P_s\phi)\circ\exp$, and let
	$\widetilde{\mathcal P}_s^{\,*}$ and $\widetilde A^*$ denote the adjoints of $\widetilde{\mathcal P}_s$ and
	$\widetilde A$ with respect to Lebesgue measure $\dd z$.
On an aligned grid, interpolate $e_n^{[k]}(e^z)/h$ constantly on
	$[t_n,t_{n+1})$, setting it to zero on cells starting at deadline nodes.
Denote the result by $\widetilde a_h$. 
Lemma~\ref{lem:premium-small-step} gives $0\le\widetilde a_h\le Q_k$. 
A subsequence therefore has a weak-star limit $\widetilde a$ in $L^\infty$.
For a smooth compactly supported test function $\varphi$ whose time
	support is compactly contained in one open deadline strip and is away from
	the initial and terminal endpoints of the chosen aligned grid, summation by parts
	and the adjoint discounted semigroup give
	\begin{align*}
		 \sum_n\int_{\mathbb R}e_n^{[k]}(e^z)\varphi(t_n,z)\dd z
		 =\sum_n\int_{\mathbb R}\widetilde V_n^{[k],h}(z)
		 \bigl(\varphi(t_n,z)-\widetilde{\mathcal P}_h^{\,*}\varphi(t_{n-1},z)\bigr)\dd z .
	\end{align*}
The bracket divided by $h$ converges in the test-function $L^1$ norm
	to $\partial_t\varphi-\widetilde A^*\varphi$.
The uniform node convergence $J_k(h)\to0$ established in the proof of
	Lemma~\ref{lem:berm-conv}, stripwise continuity, and
	$0\le V^{[k],h}\le kK$ allow passage to the limit. 
The Gaussian tails control the spatial tails introduced by the adjoint semigroup.
Moreover, because $\widetilde a_h=e_n^{[k]}(e^z)/h$ on each non-deadline cell,
$0\le e_n^{[k]}\le Q_kh$, and $\varphi$ is smooth with compact support, the difference between
	the left-hand Riemann sum and $\iint\widetilde a_h\varphi$ is $O(h)$.
Hence the left side has the same limit as $\iint\widetilde a_h\varphi$.
Consequently
	\begin{align*}
 		\iint\widetilde a\varphi\,\dd t\dd z=\iint\widetilde V^{[k]}
      		 (\partial_t\varphi-\widetilde A^*\varphi)\,\dd t\dd z .
	\end{align*}
This proves \eqref{eq:weak-density} with
	$a^{[k]}(t,x)=\widetilde a(t,\log x)$.
The distributional identity makes the density unique almost everywhere
	away from the deadlines; its values at the deadlines are immaterial.
For arbitrary $T/\delta$, choose an aligned starting time $t_*$
	before any given compact interior time interval, with
	$(T-t_*)/\delta\in\mathbb Q$, and refine that grid.
These intervals cover $(0,T)$, and uniqueness makes the local densities consistent.

\emph{Step 2: retaining the deadline atoms.}
For $d\in\mathcal D_k$, define
	\begin{align*}
 		\Delta_k(d,x):=V^{[k]}(d,x)-V^{[k]}(d+,x)\ge0.
	\end{align*}
This is the jump already described by the stagewise reduction and the
	deadline transmission condition. 
At the corresponding grid node, uniform Bermudan convergence, the right limit, 
	spatial Lipschitz continuity and $R_h\to1$ give
	\begin{align*}
		 e_d^{[k],h}(x) :=V^{[k],h}(d,x)-\mathcal P_h[V^{[k],h}(d+h,\cdot)](x)
			 \longrightarrow\Delta_k(d,x).
	\end{align*}
These bounded terms converge also against compactly supported tests.
The full distributional source on $(0,T)$ therefore consists of
$a^{[k]}(t,x)\dd t$ and the finitely many atoms
$\Delta_k(d,x)\delta_d(\dd t)$ with $d\in\mathcal D_k\cap(0,T)$. 
The corresponding mild representation can be read off directly from the Bermudan recursion, 
	so that no uniqueness theorem for the backward equation is needed. 
Since $V_M^{[k],h}=g$ and $e_{n'}^{[k]}=V_{n'}^{[k],h}-\mathcal P_hV_{n'+1}^{[k],h}$ for $n'<M$, 
	telescoping and the semigroup property $\mathcal P_h^{\,j}=\mathcal P_{jh}$ 
	of the exact one-step transition give the discrete Duhamel identity
	\begin{equation}\label{eq:discrete-duhamel}
		V_n^{[k],h}(x)=\mathcal P_{T-t_n}g(x)
			+\sum_{n'\in I_n}h\,\mathcal P_{t_{n'}-t_n}[a_h(t_{n'},\cdot)](x)
			+\sum_{\substack{d\in\mathcal D_k\\t_n\le d}}\mathcal P_{d-t_n}[e_d^{[k],h}](x),
	\end{equation}
	where $I_n$ is the set of non-deadline indices $n'$ with $n\le n'<M$ 
	and $a_h(s,x):=\widetilde a_h(s,\log x)$; 
	recall that $h\,a_h(t_{n'},\cdot)=e_{n'}^{[k]}$ for non-deadline indices $n'$.
Fix $t\in(t_*,T)$ and let $t_{n_h}$ be the first grid time at or after $t$; 
	if $t$ is a deadline, then $t_{n_h}=t$. 
We pass to the limit in \eqref{eq:discrete-duhamel} along the subsequence of Step~1, term by term.
The left side converges to $V^{[k]}(t,x)$, by $J_k(h)\to0$ and stripwise continuity. 
The first term on the right converges to $\mathcal P_{T-t}g(x)$. 
In the last sum, the set of deadlines $d\ge t_{n_h}$ coincides with the set of deadlines $d\ge t$, 
	because every deadline is a grid time. 
The functions $e_d^{[k],h}$ take values in $[0,kK]$ and are equi-Lipschitz:
	indeed, $V^{[k],h}$ is $k$-Lipschitz and $\mathcal P_h$ preserves the same Lipschitz constant because
	$e^{-rh}\E[R_h]=1$. Hence the Lipschitz constants of $e_d^{[k],h}$ are bounded uniformly in $h$.
	Since $\Delta_k(d,\cdot)$ is continuous, the pointwise convergence
	$e_d^{[k],h}\to\Delta_k(d,\cdot)$ shown above is therefore locally uniform; 
	since $\Delta_k(d,\cdot)$ is continuous and $R_{d-t_{n_h}}\to R_{d-t}$ almost surely, 
	dominated convergence gives 
	$\mathcal P_{d-t_{n_h}}[e_d^{[k],h}](x)\to\mathcal P_{d-t}[\Delta_k(d,\cdot)](x)$.
In the middle sum, the term with $n'=n_h$, when present, is at most $Q_kh$ and can be discarded. 
In log-price, the remaining terms equal $\iint K_h(u,z')\,\widetilde a_h(u,z')\dd z'\dd u$, 
	where, for $u\ge t_{n_h+1}$, $K_h(u,\cdot)$ is the discounted Gaussian kernel of 
	$\widetilde{\mathcal P}_{\lfloor u\rfloor_h-t_{n_h}}$ started at $\log x$, 
	$\lfloor u\rfloor_h$ denotes the grid time at or before $u$, 
	and $K_h(u,\cdot):=0$ for $u<t_{n_h+1}$. 
Let $K(u,\cdot)$ be the discounted Gaussian kernel of $\widetilde{\mathcal P}_{u-t}$ started at $\log x$ 
	for $u>t$. 
Every $K_h(u,\cdot)$ and $K(u,\cdot)$ has total mass at most one, 
	and $K_h(u,\cdot)\to K(u,\cdot)$ in $L^1(\dd z')$ for every $u>t$, 
	because the Gaussian laws depend continuously on a strictly positive time parameter. 
Dominated convergence in $u$ therefore gives $K_h\to K$ in $L^1(\dd u\,\dd z')$. 
Since $K\in L^1((t,T)\times\mathbb R)$, it is an admissible test function for the weak-star
	convergence. Together with $|\widetilde a_h|\le Q_k$ and
	$\widetilde a_h\overset{*}{\rightharpoonup}\widetilde a$ in $L^\infty$,
	\begin{align*}
		\iint K_h\widetilde a_h-\iint K\widetilde a
			=\iint(K_h-K)\,\widetilde a_h+\iint K\,(\widetilde a_h-\widetilde a)\longrightarrow0.
	\end{align*}
Hence, for $t_*<t<T$,
	\begin{equation}\label{eq:V-weak-mild}
		 V^{[k]}(t,x)={}\mathcal P_{T-t}g(x) +\int_t^T\mathcal P_{u-t}[a^{[k]}(u,\cdot)](x)\dd u
			 +\sum_{\substack{d\in\mathcal D_k\\t\le d}}
        			\mathcal P_{d-t}[\Delta_k(d,\cdot)](x).
	\end{equation}
Since $t_*$ can be chosen arbitrarily close to $0$ and $a^{[k]}$ is determined almost everywhere 
	by Step~1, \eqref{eq:V-weak-mild} holds for $0<t<T$.
At $t=d$, the left-continuous value includes the atom at $d$; the right limit excludes it. 
This explains the condition $t\le d$ in the sum.

\emph{Step 3: the delayed representation and weak differentiation.}
Apply $\mathcal P_\delta$ to \eqref{eq:V-weak-mild} at $t+\delta$.
For $0<t<T-\delta$ this gives
	\begin{equation}\label{eq:Psi-weak-mild}
 		\Psi_\delta^{[k]}(t,x)={}\mathcal P_{T-t}g(x)
       			+\int_{t+\delta}^T\mathcal P_{u-t}[a^{[k]}(u,\cdot)](x)\dd u
 		 +\sum_{\substack{d\in\mathcal D_k\\t+\delta\le d}}
        		\mathcal P_{d-t}[\Delta_k(d,\cdot)](x).
	\end{equation}
On an open delayed deadline strip the set of atoms in this expression is fixed. 
Every semigroup time is at least $\delta$.
Gaussian derivative bounds therefore justify differentiation of the kernels under the integral. 
The moving lower endpoint contributes
	$-\mathcal P_\delta[a^{[k]}(t+\delta,\cdot)]$ as a weak derivative;
	since $a^{[k]}$ is only measurable, this step asserts an almost-everywhere
	identity, not a continuous classical derivative. 
All fixed-endpoint terms solve the homogeneous backward equation, 
	proving \eqref{eq:Gamma-weak} and its bounds.

\emph{Step 4: the time-Lipschitz bound.}
The Gaussian kernel, with a standard normal variable $Z$, gives
	\begin{align*}
		 \|\widetilde A\widetilde{\mathcal P}_\delta\phi\|_\infty
			 \le C_\delta\|\phi\|_\infty; \qquad \E|Z|=\sqrt{2/\pi},\quad \E|Z^2-1|\le2 .
	\end{align*}
Use this estimate with $\phi(z)=V^{[k]}(t+\delta,e^z)$ and $V^{[k]}\le kK$. 
Equation~\eqref{eq:Gamma-weak} implies
	\begin{align*}
 		|\partial_t\Psi_\delta^{[k]}|\le kKC_\delta+e^{-r\delta}Q_k
		 \quad\hbox{almost everywhere}.
	\end{align*}
By Fubini's theorem and the one-dimensional Sobolev characterization,
	the bounded weak time derivative implies that, for almost every fixed $x$,
	$t\mapsto\Psi_\delta^{[k]}(t,x)$ has an absolutely continuous representative on each compact
	subinterval of the strip, with the same derivative bound. Stripwise continuity identifies this
	representative with $\Psi_\delta^{[k]}$, yielding \eqref{eq:Psi-time-Lipschitz} for almost every $x$.
	Spatial continuity at the two endpoint times then extends the estimate to every $x>0$. 
Finally, the spatial derivatives can be obtained directly from the fixed-lag lognormal transition kernel in
	\begin{align*}
		\Psi_\delta^{[k]}(t,\cdot)=\mathcal P_\delta V^{[k]}(t+\delta,\cdot).
	\end{align*}
Since this transition kernel is smooth in the spatial variable, 
	derivatives of any order may be taken under the integral sign. 
The boundedness and stripwise continuity of \(V^{[k]}\) provide the domination 
	needed for this differentiation. 
Dominated convergence then yields joint continuity of the resulting derivatives in \((t,x)\), 
	together with local bounds on each compact strip.
\end{proof}

\section{Boundary regularity and finite-level closure}
\label{app:boundary-regularity}
\subsection{Proof of Theorem~\ref{th:localized-levelset}}
\label{app:localized-levelset-proof}
\begin{proof}
Fix a compact interval $J\subset I$ and the data of the statement.
Identities and inequalities between random variables hold almost surely, and events are identified up to
 	$\Pp$-null sets.
Write $\widetilde P_vf(y):=\E f(Y_v^y)=\int p_v(y,y')f(y')\dd y'$, where
 	$p_v(y,y')=(\sigma\sqrt{2\pi v})^{-1}\exp\bigl(-(y'-y-\mu v)^2/(2\sigma^2v)\bigr)$;
 	in particular $p_v\le(\sigma\sqrt{2\pi v})^{-1}$.

\medskip\noindent\emph{Step 0: constants, notation and elementary events.}
Put
	 \begin{align*}
   		\eta_J:=t_+-\sup J>0,\qquad s_*:=\tfrac12(t_2-t_+)>0,\qquad
   		h_0:=\min\{\inf J-t_-,\ s_*\}>0 .
	 \end{align*}
By (1) and (2), the compact segment $[t_-,t_2]\times\{R\}$ lies in $\mathcal C\cap\mathcal N$.
Hence $\phi(s):=w(s,R)$ is of class $C^1$ on $[t_-,t_2]$, and
 	\begin{align*}
   		\lambda:=\max_{[t_-,t_2]}|\phi'|<\infty,\qquad \phi_0:=\min_{[t_-,t_2]}\phi>0 .
 	\end{align*}
Since $L<\beta(t)$ we have $w(t,L)=0$ for $t\in[t_-,t_2]$; therefore $H:=w(t_2,\cdot)$ satisfies
 	\begin{equation}\label{eq:LL-HLR}
  		 H\ge0,\qquad H(L)=0,\qquad H(R)=\phi(t_2).
 	\end{equation}
We set $\nu_{\rm tot}:=\nu((L_1,R_1))$ and $q_\infty:=\|q\|_{L^\infty}$.

Fix $t\in J$ and $z\in(L,R)$ with $(t,z)\in\mathcal C$, that is, $z>\beta(t)$.
Write $Y:=Y^z$, $a:=t_2-t$, $r_+:=t_+-t\ (\ge\eta_J)$, and
 	\begin{align*}
  		\tau_0&:=\inf\{u\ge0:\ Y_u\le\beta(t+u)\},\\
  		\eta&:=\eta_t^z=\inf\{u\ge0:\ u\ge a\ \text{or}\ Y_u\notin(L,R)\},\qquad \tau^*:=\tau_0\wedge\eta,\\
 		 \rho&:=\inf\{u\ge0:\ Y_u=R\}\wedge r_+,\qquad \theta:=\tau_0\wedge\rho,
	 \end{align*}
 and
 \begin{align*}
  A_R:=\{\rho<r_+,\ \rho<\tau_0\},\qquad A_T:=\{\rho=r_+<\tau_0\}.
 \end{align*}
Thus $\{\rho<\tau_0\}=A_R\cup A_T$, and $Y_\rho=R$ on $A_R$.
Since $w$ is continuous and $w(t,z)>0$, we have $\tau_0>0$; also $\rho>0$.
Hence $\theta>0$ and
 	\begin{align*}
   		\Xi:=\E[\theta]+\Pp(A_R)>0 .
 	\end{align*}
The following elementary facts are used repeatedly.
 \begin{enumerate}[label={\rm(E\arabic*)}]
  	\item $\Pp(\tau^*\ge r_+)\le\E[\theta]/\eta_J$.
   		Indeed, if $\tau^*\ge r_+$ then $\eta\ge r_+$, so $Y$ does not reach $R$ before $r_+$;
   		hence $\rho=r_+\le\tau_0$ and $\theta=r_+$.
   		Therefore $\{\tau^*\ge r_+\}\subset\{\theta\ge r_+\}$, and Markov's inequality gives the claim.
  	\item $\Pp(A_T)\le\E[\theta]/\eta_J$, because $\theta=r_+\ge\eta_J$ on $A_T$.
 	 \item $\tau^*=\theta$ on $\{\tau_0\le\rho\}$. Indeed on this event $\tau_0\le\rho\le r_+\le a$,
  	 	$\tau_0\le\rho\le\inf\{u:Y_u=R\}$, and $\tau_0\le\inf\{u:Y_u=L\}$ 
		because $L<\beta$; hence $\tau_0\le\eta$.
   		Since $\tau^*\le\eta\le a\le t_2-t_-$ always, we obtain
   		$\E[\tau^*]\le\E[\theta]+(t_2-t_-)\,\Pp(\rho<\tau_0)$.
  	\item Let $G_R:=\{\tau^*=\eta<a,\ Y_\eta=R\}$ be the event of exit through the lateral face $z=R$
   		before the time face. 
		Then $G_R\subset\{\rho<\tau_0\}$.
  	 	Indeed $\tau_0>\eta$ on $G_R$, because $Y_\eta=R>\beta(t+\eta)$; and $\rho\le\min\{\eta,r_+\}<\tau_0$.
 \end{enumerate}
Consequently,
 	\begin{equation}\label{eq:LL-E-sum}
  		 \E[\tau^*]+\Pp(G_R)+\Pp(\tau^*\ge r_+)\le C_0\,\Xi,\qquad
  		C_0:=(t_2-t_-+2)(1+\eta_J^{-1}).
 	\end{equation}
To see \eqref{eq:LL-E-sum}, note that
 	$\Pp(\rho<\tau_0)\le\Pp(A_R)+\E[\theta]/\eta_J\le(1+\eta_J^{-1})\Xi$ by (E2);
 	hence $\E[\tau^*]\le\Xi\{1+(t_2-t_-)(1+\eta_J^{-1})\}$ by (E3), 
	$\Pp(G_R)\le(1+\eta_J^{-1})\Xi$ by (E4),
 	and $\Pp(\tau^*\ge r_+)\le\eta_J^{-1}\Xi$ by (E1).

\medskip\noindent\emph{Step 1: a lower bound for $w_z$.}
Put $u:=w_z$. By (1), $u\ge0$ on $\mathcal C\cap\mathcal N$.

\emph{(1a) Positivity on the face $z=R$.}
Let $s\in[\inf J,t_+]$.
By continuity of $M_z$ and \eqref{eq:LL-M} there is $\varepsilon\in(0,t_2-t_+)$ such that
 	$Q:=[s,s+\varepsilon]\times[R-\varepsilon,R+\varepsilon]\subset\mathcal C\cap\mathcal N$ 
	and $M_z\ge\alpha/2$ on $Q$.
Let $\vartheta:=\inf\{v\ge0:\ |Y_v^R-R|=\varepsilon\}\wedge\varepsilon$.
By (2), It\^o's formula for $u(s+v,Y^R_v)$ up to $\vartheta$ (the stochastic integral is a martingale,
 	since $u_z$ is bounded on $Q$) and $u\ge0$ on $Q$,
 	\begin{align*}
   		u(s,R)=\E\bigl[u(s+\vartheta,Y^R_\vartheta)\bigr]+\E\int_0^\vartheta M_z(s+v,Y^R_v)\dd v
   		\ge\tfrac\alpha2\,\E[\vartheta]>0 .
 	\end{align*}
Since $u$ is continuous on the compact segment $[\inf J,t_+]\times\{R\}$,
 	\begin{align*}
   		\kappa_J:=\min_{s\in[\inf J,t_+]}u(s,R)>0 .
 	\end{align*}

\emph{(1b) The main estimate.}
For integers $n>1/w(t,z)$ put $\tau_n:=\inf\{v\ge0:\ w(t+v,Y_v)\le1/n\}$ 
	and $\theta_n:=\tau_n\wedge\rho$.
For $v\le\theta_n$ the point $(t+v,Y_v)$ lies in the set
 	$K_n:=\{(s',y)\in[t_-,t_+]\times[L,R]:\ w(s',y)\ge1/n\}$, 
	which is a compact subset of $\mathcal C\cap\mathcal N$
 	(note $y>\beta(s')>L$ whenever $w(s',y)>0$).
Since $u\in C^{1,2}$ near $K_n$, It\^o's formula and \eqref{eq:LL-pde} give
	 \begin{align*}
  		 u(t,z)=\E\bigl[u(t+\theta_n,Y_{\theta_n})\bigr]+\E\int_0^{\theta_n}M_z(t+v,Y_v)\dd v .
 	\end{align*}
On the event $\{\rho<r_+,\ \rho<\tau_n\}$ we have $\theta_n=\rho$ and $Y_\rho=R$, so
 	$u(t+\theta_n,Y_{\theta_n})\ge\kappa_J$; on the complement $u(t+\theta_n,Y_{\theta_n})\ge0$.
Moreover $M_z\ge\alpha$ along the path by \eqref{eq:LL-M}. Hence
 	\begin{align*}
  		 w_z(t,z)\ge\kappa_J\,\Pp(\rho<r_+,\ \rho<\tau_n)+\alpha\,\E[\theta_n].
 	\end{align*}
The times $\tau_n$ are nondecreasing and bounded by $\tau_0$; their limit $\tau_\infty$ satisfies
 	$w(t+\tau_\infty,Y_{\tau_\infty})=\lim_n1/n=0$ by continuity, 
	so $\tau_\infty\ge\tau_0$, i.e.\ $\tau_n\uparrow\tau_0$.
Monotone convergence gives $\E[\theta_n]\uparrow\E[\theta]$ and
 	$\Pp(\rho<r_+,\rho<\tau_n)\uparrow\Pp(A_R)$. 
Therefore
 	\begin{equation}\label{eq:LL-denominator}
  		 w_z(t,z)\ge c_J\,\Xi,\qquad c_J:=\min\{\alpha,\kappa_J\}>0 .
	 \end{equation}
In particular, $w_z>0$ at every point of $\mathcal C\cap(J\times(L,R))$.

\medskip\noindent\emph{Step 2: the terminal trace.}
Let $a_0\ge0$ and $h>0$, and put
 $\varkappa:=(a_0+h)\wedge\inf\{u\ge a_0:\ Y_u\notin(L,R)\}$.
Let $A\in\mathcal F_{a_0}$ with $A\subset\{Y_{a_0}\in(L,R)\}$. Then
 	\begin{enumerate}[label={\rm(\roman*)}]
  	\item $\E\bigl[\one_A\bigl(H(Y_\varkappa)-H(Y_{a_0})\bigr)\bigr]\ge-q_\infty\,h\,\Pp(A)$;
  	\item if, in addition, $A\subset A'$ for some $A'\in\mathcal F_{a'}$ 
		with $0\le a'\le a_0-v_0$ and $v_0>0$, then
  		 \begin{align*}
    			 \E\bigl[\one_A\bigl(H(Y_\varkappa)-H(Y_{a_0})\bigr)\bigr]
    			 \le\Bigl(q_\infty+\frac{\nu_{\rm tot}}{\sigma\sqrt{2\pi v_0}}\Bigr)h\,\Pp(A').
   			\end{align*}
 	\end{enumerate}
\emph{Proof.}
Let $\varpi_n(y)=n\varpi(ny)$ be a standard mollifier supported in $[-1/n,1/n]$ and
 	$H_n:=H*\varpi_n$, defined on a neighborhood of $[L,R]$ for $n$ large.
Because $\widetilde\cL$ has constant coefficients, it commutes with the convolution, so on $[L,R]$
 	\begin{align*}
  		 \widetilde\cL H_n=\nu_n+q_n,\qquad \nu_n:=\nu*\varpi_n\ge0,\quad q_n:=q*\varpi_n,\quad
   		\|q_n\|_\infty\le q_\infty,\quad \int_L^R\nu_n\le\nu_{\rm tot}.
 	\end{align*}
Also $H_n\to H$ uniformly on $[L,R]$, because $H$ is continuous.
On $A$ the path $(Y_u)_{a_0\le u\le\varkappa}$ stays in $[L,R]$.
Applying It\^o's formula to $H_n$ (replaced by $H_n$ times a smooth cutoff equal to one near $[L,R]$)
	 on $[a_0,\varkappa]$ and conditioning on $\mathcal F_{a_0}$ gives
 	\begin{align}\label{eq:LL-dynkin-n}
   		\E\bigl[\one_A(H_n(Y_\varkappa)-H_n(Y_{a_0}))\bigr]
   			=\E\Bigl[\one_A\int_{a_0}^\varkappa(\nu_n+q_n)(Y_u)\dd u\Bigr].
 	\end{align}
Since $\nu_n\ge0$ and $\varkappa-a_0\le h$, the right-hand side is at least $-q_\infty h\Pp(A)$.
Letting $n\to\infty$ proves~(i).
For (ii), the $q_n$-part of \eqref{eq:LL-dynkin-n} is at most $q_\infty h\Pp(A)\le q_\infty h\Pp(A')$.
For the $\nu_n$-part we use $\nu_n\ge0$, $A\subset A'$ and $Y_u\in[L,R]$ 
	on $[a_0,\varkappa]$, then the Markov property at time $a'$:
 	\begin{align*}
  		\E\Bigl[\one_A\int_{a_0}^\varkappa\nu_n(Y_u)\dd u\Bigr]
  		&\le\E\Bigl[\one_{A'}\int_{a_0}^{a_0+h}(\nu_n\one_{[L,R]})(Y_u)\dd u\Bigr]\\
  		 &=\E\Bigl[\one_{A'}\int_{a_0-a'}^{a_0-a'+h}\widetilde P_v(\nu_n\one_{[L,R]})(Y_{a'})\dd v\Bigr]\\
   		&\le\frac{\nu_{\rm tot}}{\sigma\sqrt{2\pi v_0}}\,h\,\Pp(A'),
 	\end{align*}
	 because $v\ge a_0-a'\ge v_0$ in the integral and $p_v\le(\sigma\sqrt{2\pi v_0})^{-1}$ for such $v$.
Letting $n\to\infty$ proves (ii). \hfill$\square$

\medskip\noindent\emph{Step 3: an upper bound for $w_t$.}
Let $h\in(0,h_0]$; then $t-h\ge t_-$.
Put $F:=\{\tau^*\ge a\}=\{\tau^*=a\}$. 
It belongs to $\mathcal F_a$, it is contained in $\{Y_a\in(L,R)\}$,
 	and $F\subset\{\tau^*\ge r_+\}$ since $a>r_+$.
On $F$ we have $\eta=a$, and $F$ is the event that $Y$ is neither absorbed in the zero set of $w$
 	nor leaves $(L,R)$ before time $a$.
Define the stopping time $\hat\tau:=\tau^*\one_{F^c}+\infty\,\one_F$, and let
 	$\eta':=\eta_{t-h}^z=\inf\{u\ge0:\ u\ge a+h\ \text{or}\ Y_u\notin(L,R)\}$.
By \eqref{eq:LL-dpp} at $(t,z)$ and at $(t-h,z)$,
 	\begin{equation}\label{eq:LL-back}
   		w(t,z)-w(t-h,z)\le\mathcal J_t^z(\tau^*)-\mathcal J_{t-h}^z(\hat\tau)=:D .
 	\end{equation}
We evaluate both functionals. Recall that the payoff at the exit time $\eta$ is
 	$\phi(t+\eta)$ on a lateral exit through $R$, $0$ on a lateral exit 
	through $L$, and $H(Y_a)$ at the time face.
Hence
 	\begin{align*}
   		\mathcal J_t^z(\tau^*)=\E\Bigl[\int_0^{\tau^*}M(t+u,Y_u)\dd u
      		+\one_{G_R}\phi(t+\tau^*)+\one_FH(Y_a)\Bigr].
 	\end{align*}
On $F^c$ we have $\hat\tau=\tau^*<\eta'$ unless $\tau^*=\eta<a$ is a lateral exit, 
	in which case $\eta'=\eta=\tau^*$.
On $F$, $\hat\tau=\infty$, so the process runs up to $\eta'\in[a,a+h]$.
Thus
 \begin{align*}
   \mathcal J_{t-h}^z(\hat\tau)=\E\Bigl[&\int_0^{\tau^*}M(t-h+u,Y_u)\dd u
     +\one_{G_R}\phi(t-h+\tau^*)\\
     &+\one_F\Bigl\{\int_a^{\eta'}M(t-h+u,Y_u)\dd u+w(t-h+\eta',Y_{\eta'})\Bigr\}\Bigr].
 \end{align*}
Subtracting,
 	\begin{align*}
   	D&=\E\int_0^{\tau^*}\bigl[M(t+u,Y_u)-M(t-h+u,Y_u)\bigr]\dd u
      		+\E\bigl[\one_{G_R}(\phi(t+\tau^*)-\phi(t-h+\tau^*))\bigr]\\
    		&\quad+\E\Bigl[\one_F\Bigl\{H(Y_a)-w(t-h+\eta',Y_{\eta'})-\int_a^{\eta'}M(t-h+u,Y_u)\dd u\Bigr\}\Bigr].
 	\end{align*}
By \eqref{eq:LL-M} the first term is at most $m_1h\,\E[\tau^*]$; 
	the second is at most $\lambda h\Pp(G_R)$; and $\int_a^{\eta'}|M|\le m_0h$.
It remains to treat $w(t-h+\eta',Y_{\eta'})$ on $F$.
If $\eta'=a+h$ then $t-h+\eta'=t_2$ and $w(t-h+\eta',Y_{\eta'})=H(Y_{\eta'})$.
If $Y_{\eta'}=L$ then both $w(t-h+\eta',L)$ and $H(L)$ vanish.
If $Y_{\eta'}=R$ with $\eta'<a+h$, then $t_2-h\le t-h+\eta'<t_2$, so
 	$w(t-h+\eta',R)=\phi(t-h+\eta')\ge\phi(t_2)-\lambda h=H(R)-\lambda h$ by \eqref{eq:LL-HLR}.
In all cases $w(t-h+\eta',Y_{\eta'})\ge H(Y_{\eta'})-\lambda h$.
Now $\eta'=\varkappa$ in the notation of Step~2 with $a_0=a$ and $A=F$, so Step~2(i) gives
	 $\E[\one_F(H(Y_a)-H(Y_{\eta'}))]\le q_\infty h\Pp(F)$.
Altogether,
	 \begin{align*}
   		D\le h\bigl\{m_1\E[\tau^*]+\lambda\Pp(G_R)+(m_0+\lambda+q_\infty)\Pp(F)\bigr\}.
 	\end{align*}
Combining \eqref{eq:LL-back} with this bound on $D$, 
	the inclusion $F\subset\{\tau^*\ge r_+\}$ and \eqref{eq:LL-E-sum},
 	\begin{align*}
   		\frac{w(t,z)-w(t-h,z)}{h}\le C_3\,C_0\,\Xi,\qquad
   		C_3:=m_0+m_1+2\lambda+q_\infty+\frac{\nu_{\rm tot}}{\sigma\sqrt{2\pi s_*}} .
 	\end{align*}
Since $(t,z)\in\mathcal C\cap\mathcal N$ and $w\in C^{1,2}$ there, letting $h\downarrow0$ gives
 	\begin{equation}\label{eq:LL-upper}
   		w_t(t,z)\le C_0C_3\,\Xi .
 	\end{equation}
(Here $\tau^*$ is the optimal time of the problem started at $(t,z)$; it does not depend on $h$.
This is why the one-sided quotient is taken in the direction in which the optimal time of the base point can be used.)

\medskip\noindent\emph{Step 4: a lower bound for $w_t$.}
Let again $h\in(0,h_0]$ and put $s_h:=a-h=t_2-t-h$.
Then $s_h-r_+=t_2-t_+-h\ge s_*>0$.
Let $F_h:=\{\tau^*\ge s_h\}\in\mathcal F_{s_h}$; then $F_h\subset\{Y_{s_h}\in(L,R)\}$ and
 	$F_h\subset\{\tau^*\ge r_+\}\in\mathcal F_{r_+}$.
Define $\check\tau:=\tau^*\one_{\{\tau^*<s_h\}}+\infty\,\one_{F_h}$ and
 	$\eta'':=\eta_{t+h}^z=\inf\{u\ge0:\ u\ge s_h\ \text{or}\ Y_u\notin(L,R)\}$.
By \eqref{eq:LL-dpp} at $(t+h,z)$ and $(t,z)$,
	 \begin{equation}\label{eq:LL-fwd}
  		 w(t+h,z)-w(t,z)\ge\mathcal J_{t+h}^z(\check\tau)-\mathcal J_t^z(\tau^*)=:E .
 	\end{equation}
On $\{\tau^*<s_h\}$ we have $\check\tau=\tau^*$, and $\eta''=\tau^*$ only for a lateral exit;
 	on $F_h$ no lateral exit occurs before $s_h$, so $\eta''=s_h$ and the payoff is $H(Y_{s_h})$
 	because $t+h+s_h=t_2$. 
Hence
 \begin{align*}
  	 \mathcal J_{t+h}^z(\check\tau)=\E\Bigl[&\one_{\{\tau^*<s_h\}}
	 	\Bigl\{\int_0^{\tau^*}M(t+h+u,Y_u)\dd u
       	 +\one_{G_R}\phi(t+h+\tau^*)\Bigr\}\\
     		&+\one_{F_h}\Bigl\{\int_0^{s_h}M(t+h+u,Y_u)\dd u+H(Y_{s_h})\Bigr\}\Bigr].
 \end{align*}
Put $\Pi:=\one_{G_R}\phi(t+\tau^*)+\one_{\{\tau^*=\eta=a\}}H(Y_a)$, 
	the terminal payoff in $\mathcal J_t^z(\tau^*)$.
Then
 	\begin{align*}
   		E&=\E\Bigl[\int_0^{\tau^*\wedge s_h}\bigl[M(t+h+u,Y_u)-M(t+u,Y_u)\bigr]\dd u\Bigr]\\
    		&\quad+\E\bigl[\one_{\{\tau^*<s_h\}}\one_{G_R}(\phi(t+h+\tau^*)-\phi(t+\tau^*))\bigr]\\
    		&\quad+\E\Bigl[\one_{F_h}\Bigl\{-\int_{s_h}^{\tau^*}M(t+u,Y_u)\dd u+H(Y_{s_h})-\Pi\Bigr\}\Bigr].
	 \end{align*}
The first term is at least $-m_1h\E[\tau^*]$, the second at least $-\lambda h\Pp(G_R)$, and
 	$\int_{s_h}^{\tau^*}|M|\le m_0h$ on $F_h$ because $\tau^*\le a=s_h+h$.
Let $\varkappa:=a\wedge\inf\{u\ge s_h:\ Y_u\notin(L,R)\}$; this is the stopping time of Step~2 with $a_0=s_h$.
We claim that on $F_h$
 	\begin{equation}\label{eq:LL-T-compare}
   		\Pi\le H(Y_\varkappa)+\lambda h .
 	\end{equation}
There are three cases.
If $\tau^*=\eta=a$ (exit through the time face), then no lateral exit occurs before $a$, 
	so $\varkappa=a$ and $\Pi=H(Y_\varkappa)$.
If $G_R$ occurs, then $\varkappa=\tau^*\in[s_h,a)$, $H(Y_\varkappa)=H(R)=\phi(t_2)$ and
 	$\Pi=\phi(t+\tau^*)\le\phi(t_2)+\lambda(t_2-t-\tau^*)\le H(Y_\varkappa)+\lambda h$.
In all other cases $\tau^*$ is an absorption in the zero set or an exit 
	through $L$ and $\Pi=0\le H(Y_\varkappa)$ by \eqref{eq:LL-HLR}.
Using \eqref{eq:LL-T-compare} and Step~2(ii) with $a_0=s_h$, $A=F_h$, 
	$A'=\{\tau^*\ge r_+\}$, $a'=r_+$, $v_0=s_*$,
 	\begin{align*}
   		\E\bigl[\one_{F_h}(H(Y_{s_h})-\Pi)\bigr]
   		&\ge-\E\bigl[\one_{F_h}(H(Y_\varkappa)-H(Y_{s_h}))\bigr]-\lambda h\Pp(F_h)\\
   		&\ge-\Bigl(q_\infty+\frac{\nu_{\rm tot}}{\sigma\sqrt{2\pi s_*}}+\lambda\Bigr)h\,\Pp(\tau^*\ge r_+).
 	\end{align*}
Collecting the terms and using $\Pp(F_h)\le\Pp(\tau^*\ge r_+)$ and \eqref{eq:LL-E-sum},
	\begin{align*}
  		 \frac{w(t+h,z)-w(t,z)}{h}\ge\frac{E}{h}\ge-C_3\,C_0\,\Xi .
 	\end{align*}
Letting $h\downarrow0$ yields $w_t(t,z)\ge-C_0C_3\,\Xi$. 
Together with \eqref{eq:LL-upper} and \eqref{eq:LL-denominator},
 	\begin{equation}\label{eq:LL-ratio}
  		 |w_t(t,z)|\le C_0C_3\,\Xi\le\Lambda\,w_z(t,z),\qquad
  		 \Lambda:=\frac{C_0C_3}{c_J},\qquad (t,z)\in\mathcal C\cap(J\times(L,R)).
	 \end{equation}

\medskip\noindent\emph{Step 5: level curves and conclusion.}
Let $J_0\subset I$ be a compact interval and choose $\iota>0$ with $J:=[\inf J_0-\iota,\sup J_0+\iota]\subset I$.
Apply Steps~0--4 to $J$, and let $\Lambda$ be the constant in \eqref{eq:LL-ratio}.
Put $\varepsilon_0:=\phi_0>0$ and fix $\varepsilon\in(0,\varepsilon_0)$.
For $t\in\mathrm{int}\,J$ we have $w(t,\beta(t))=0<\varepsilon<w(t,R)$, 
	so some $z\in(\beta(t),R)$ satisfies $w(t,z)=\varepsilon$.
It is unique, because $w_z>0$ on $(\beta(t),R)$ by \eqref{eq:LL-denominator}; 
	denote it by $\beta_\varepsilon(t)$.
Since $w\in C^1$ near $(t,\beta_\varepsilon(t))\in\mathcal C\cap\mathcal N$ 
	and $w_z>0$ there, the implicit function theorem shows
 	$\beta_\varepsilon\in C^1(\mathrm{int}\,J)$ with
 	\begin{align*}
  		\beta_\varepsilon'(t)=-\frac{w_t(t,\beta_\varepsilon(t))}{w_z(t,\beta_\varepsilon(t))},\qquad
   		|\beta_\varepsilon'(t)|\le\Lambda
	 \end{align*}
	 by \eqref{eq:LL-ratio}. 
Hence $\beta_\varepsilon$ is $\Lambda$-Lipschitz on $\mathrm{int}\,J$, uniformly in $\varepsilon$.
As $w(t,\cdot)$ is strictly increasing on $(\beta(t),R]$, 
	$\beta_\varepsilon(t)$ decreases as $\varepsilon\downarrow0$;
	 its limit $\bar\beta(t)\ge\beta(t)$ satisfies $w(t,\bar\beta(t))=\lim_{\varepsilon\downarrow0}\varepsilon=0$,
 	so $\bar\beta(t)\le\beta(t)$ by the description of the zero set. 
Thus $\beta_\varepsilon(t)\downarrow\beta(t)$.
A pointwise limit of $\Lambda$-Lipschitz functions is $\Lambda$-Lipschitz, so
 	$|\beta(t)-\beta(s)|\le\Lambda|t-s|$ for $s,t\in J_0$.
Since $J_0$ was arbitrary, $\beta\in C^{0,1}_{\rm loc}(I)$.
\end{proof}

\subsection{Proof of Theorem~\ref{th:finite-level-regularity}}
\label{app:finite-level-proof}
\begin{proof}
We proceed by induction on $k$.
Throughout the proof, $\Phi$ denotes the standard normal distribution function, and we use the log-price notation
	\begin{align*}
 		\beta_k(t):=\log b^{[k]}(t),\qquad
 		\widetilde\Gamma_k(t,z):=\Gamma_\delta^{[k]}(t,e^z),\qquad \mu:=r-\tfrac12\sigma^2.
	\end{align*}
Write
	\begin{align*}
		 \varphi(u):=\frac{1}{\sqrt{2\pi}}e^{-u^2/2},
			 \qquad
		 \rho_\delta(u):=\frac{1}{\sigma\sqrt{2\pi\delta}}
		 \exp\!\left(-\frac{u^2}{2\sigma^2\delta}\right).
	\end{align*}
The one-right boundary is classically locally smooth away from maturity;
	see, for example, \citet{ChenChadam2007}.  
Proposition~\ref{prop:R-one} gives the corresponding delayed regularity and
	\begin{align*}
 		\widetilde\Gamma_1(t,z) =rKe^{-r\delta}\Phi\!\left(
		 \frac{\beta_1(t+\delta)-z-\mu\delta}{\sigma\sqrt\delta}\right),
	\end{align*}
	so
	\begin{align}
 		-\partial_z\widetilde\Gamma_1(t,z)=\frac{rKe^{-r\delta}}{\sigma\sqrt\delta}\,
		 \varphi\!\left(\frac{\beta_1(t+\delta)-z-\mu\delta}{\sigma\sqrt\delta}\right)>0.
		 \label{eq:Gamma1-strict-new}
	\end{align}
The local Lipschitz regularity of $\beta_1$ also gives local time-Lipschitz regularity of
	$\widetilde\Gamma_1$.

Assume now that the assertions have been proved through level $k-1$, with $k\ge2$, 
	and suppose first that $d_{k-1}>0$. Fix a compact interval $J\subset(0,d_{k-1})$.  
Choose
 	\begin{align*}
  		 0<t_-<\inf J\le\sup J<t_+<t_2<d_{k-1}.
 	\end{align*}
Because $0<b^{[1]}\le b^{[k]}<K$, the one-right boundary is continuous and positive,
 	and $b^{[k]}$ is upper semicontinuous, there are constants $0<\underline b<\overline b<K$ such that
 	\begin{align*}
  		\underline b<b^{[k]}(t)<\overline b, \qquad t\in[t_-,t_2].
 	\end{align*}
Put $L:=\log\underline b$ and $R:=\log\overline b$, fix $R_*\in(R,\log K)$, 
	and let $\mathcal N$ be an open neighborhood of
 	$[t_-,t_2]\times[L,R]$ contained in $(0,d_{k-1})\times(-\infty,R_*)$.
We verify the hypotheses of Theorem~\ref{th:localized-levelset} for $I:=(0,d_{k-1})$ and
 	\begin{align*}
   		w_k(t,z):=e^{-rt}\bigl(V^{[k]}-g_\delta^{[k]}\bigr)(t,e^z),\qquad \beta_k:=\log b^{[k]}.
 	\end{align*}

\emph{Continuity, sign and zero set.}
For $t<d_{k-1}$ we have $t+\delta<d_{k-2}$, so $t+\delta$ lies in the first strip of level $k-1$.
Hence $\Psi_\delta^{[k-1]}$, and with it $g_\delta^{[k]}$, is continuous on $I\times(0,\infty)$;
 	$V^{[k]}$ is continuous there by Theorem~\ref{th:Vprops}(vii).
Thus $w_k$ is continuous and, by \eqref{eq:Vm-rec}, nonnegative.
By Theorem~\ref{th:boundary-m}, $\{w_k(t,\cdot)=0\}=(-\infty,\beta_k(t)]$ with $\beta_k(t)$ finite.

\emph{Band (1).}
$L<\beta_k<R$ on $[t_-,t_2]$ by the choice of $L,R$.
For $x=e^z<K$ we have $w_k=e^{-rt}G_t^{[k]}(x)$ in the notation of Lemma~\ref{lem:singlecross}
 	(the case $k=1$ being classical), so the single-crossing lemma shows 
	that $w_k(t,\cdot)$ is nondecreasing on
 	$(-\infty,\log K)\supset(-\infty,R_*]$.

First consider an interior point of the continuation region.  
A relatively compact continuation cylinder can be chosen away from all deadlines.  
The local dynamic-programming principle identifies the value there with 
	the Feynman--Kac solution of the homogeneous 
	backward Black--Scholes equation having the continuous boundary
	trace supplied by $V^{[k]}$.  
Passing to log-price coordinates gives a uniformly parabolic constant-coefficient
	problem; standard interior regularity therefore yields
	\begin{align}
 		(\partial_t+\cL-r)V^{[k]}=0,\qquad V^{[k]}\in C^{1,2}
 		\label{eq:continuation-classical-new}
	\end{align}
	in the continuation interior.

\emph{Regularity (2) and running reward (3).}
Put $\widetilde\Psi(t,z):=\Psi_\delta^{[k-1]}(t,e^z)$ and
 	$\widehat G_k(t,z):=e^{-rt}\bigl(K-e^z+\widetilde\Psi(t,z)\bigr)$, 
	so that $w_k=e^{-rt}V^{[k]}(t,e^z)-\widehat G_k$ on $\mathcal N$.
The space derivatives of $\widetilde\Psi$ of every order are obtained from 
	the fixed-lag Gaussian kernel and are jointly continuous
	 (Theorem~\ref{th:R-weak}), and $\partial_t\widetilde\Psi=(r-\widetilde\cL)\widetilde\Psi-\widetilde\Gamma_{k-1}$.
By the induction hypothesis and the explicit convolution formula for $\widetilde\Gamma_{k-1}$ 
	(differentiated in $z$ under the integral sign;
 	$\beta_{k-1}$ is continuous on $[t_-+\delta,t_2+\delta]\subset(\delta,d_{k-2})$), 
	the functions $\widetilde\Gamma_{k-1}$ and $\partial_z\widetilde\Gamma_{k-1}$ are jointly continuous.
Hence $\widetilde\Psi$ and $\partial_z\widetilde\Psi$ are of class $C^{1,2}$ on $\mathcal N$, 
	and the same holds for $\widehat G_k$ and $\partial_z\widehat G_k$.
Since $(\partial_t+\widetilde\cL-r)(K-e^z)=-rK$ 
	and $(r-\partial_t-\widetilde\cL)\widetilde\Psi=\widetilde\Gamma_{k-1}$,
 	\begin{align}
   		M_k(t,z):=(\partial_t+\widetilde\cL)\widehat G_k(t,z)
    			=-e^{-rt}\bigl(rK+\widetilde\Gamma_{k-1}(t,z)\bigr).
  			 \label{eq:Mk-new}
	 \end{align}
The function $M_k$ and $\partial_zM_k=-e^{-rt}\partial_z\widetilde\Gamma_{k-1}$ 
	are continuous on $\mathcal N$; $M_k$ is bounded and
 Lipschitz in time on $[t_-,t_2]\times[L,R]$ by the induction time-Lipschitz estimate 
 	for $\widetilde\Gamma_{k-1}$
	 (note $t+\delta\in[t_-+\delta,t_2+\delta]$, a compact subset of the open strip);
	 and by the induction representation
 	\begin{align}
  		 -\partial_z\widetilde\Gamma_{k-1}(t,z)\ge e^{-r\delta}rK\,
   		\rho_\delta\bigl(\beta_{k-1}(t+\delta)-z-\mu\delta\bigr)>0
  		\label{eq:Gamma-strict-induction-new}
 	\end{align}
 	and the compactness of $[t_-,t_2]\times[L,R]$, 
	$\partial_zM_k\ge\alpha$ for some $\alpha>0$ there.
In the continuation region the function $\widehat V(t,z):=e^{-rt}V^{[k]}(t,e^z)$ solves 
	the homogeneous constant-coefficient equation
 	$(\partial_t+\widetilde\cL)\widehat V=0$ (see \eqref{eq:continuation-classical-new}), 
	hence is $C^\infty$ there.
Therefore $w_k=\widehat V-\widehat G_k$ and $\partial_zw_k$ belong to 
	$C^{1,2}(\mathcal C\cap\mathcal N)$, and
 	$(\partial_t+\widetilde\cL)w_k=-M_k$; differentiating in $z$ 
	(the operator $\widetilde\cL$ has constant coefficients) gives
 	$(\partial_t+\widetilde\cL)\partial_zw_k=-\partial_zM_k$.
This is (2) and (3) with $M=M_k$.

\emph{Local dynamic-programming principle (4).}
Let $s\in[t_-,t_2)$, $z\in(L,R)$, $Y=Y^z$ and $\eta=\eta_s^z$.
On $\mathcal N$ we have $e^{-rs'}g_\delta^{[k]}(s',e^y)=\widehat G_k(s',y)$, because $e^y<K$ there.
For every stopping time $\tau$, the strong supermartingale property of Lemma~\ref{lem:supermart}(i) 
	at $\tau\wedge\eta$ and $V^{[k]}\ge g_\delta^{[k]}$ give
	 \begin{align*}
  		\E\bigl[\widehat G_k(s+\tau,Y_\tau)\one_{\{\tau<\eta\}}
			+\widehat V(s+\eta,Y_\eta)\one_{\{\eta\le\tau\}}\bigr]
   		\le\E\bigl[\widehat V(s+\tau\wedge\eta,Y_{\tau\wedge\eta})\bigr]\le\widehat V(s,z).
 	\end{align*}
Conversely, for $\varepsilon>0$ let $\tau_\varepsilon$ be an $\varepsilon$-optimal first exercise time 
	for the recursive problem \eqref{eq:Vm-rec} started at $(s,z)$. 
By the strong Markov property, the conditional value of the reward collected after $\eta$ is at most
	 $\widehat V(s+\eta,Y_\eta)$ on $\{\tau_\varepsilon\ge\eta\}$, 
	 so the left-hand side above, evaluated at $\tau_\varepsilon$, is at least
 	$\widehat V(s,z)-\varepsilon$. 
Hence $\widehat V(s,z)$ equals the supremum of the left-hand side over all $\tau$.
Dynkin's formula for the $C^{1,2}$ function $\widehat G_k$ on $\mathcal N$ gives
 	$\E[\widehat G_k(s+\tau\wedge\eta,Y_{\tau\wedge\eta})]-\widehat G_k(s,z)
		=\E\int_0^{\tau\wedge\eta}M_k(s+u,Y_u)\dd u$.
Subtracting $\widehat G_k(s,z)$ from the supremum above yields exactly 
	$w_k(s,z)=\sup_\tau\mathcal J_s^z(\tau)$ with $M=M_k$.
The supremum is attained at the first contact time $\tau_s^z$, 
	because the discounted value process is a martingale up to the first contact
	 with the stopping reward (the Snell-envelope argument in Step~1 
	 of the proof of Theorem~\ref{th:optimal-first}),
 	and on contact $\widehat V=\widehat G_k$ by value matching.

\emph{Terminal trace (5).}
Put $H_k:=w_k(t_2,\cdot)$ and $(L_1,R_1):=(L-1,R_*)$.
On $(L_1,R_1)$ we have $H_k=e^{-rt_2}\widetilde V-\widehat G_k(t_2,\cdot)$ 
	with $\widetilde V(z):=V^{[k]}(t_2,e^z)$.
The function $\widehat G_k(t_2,\cdot)$ is of class $C^2$, 
	so $\widetilde\cL\widehat G_k(t_2,\cdot)$ is bounded on $(L_1,R_1)$.
Since $x\mapsto V^{[k]}(t_2,x)$ is convex and $k$-Lipschitz (Theorem~\ref{th:Vprops}),
	its second derivative in the sense of distributions is a nonnegative Radon measure; write
	\begin{align*}
		\lambda_k:=D_{xx}V^{[k]}(t_2,\cdot).
	\end{align*}
Moreover $|V_x^{[k]}(t_2,\cdot)|\le k$ almost everywhere. Since $\sigma^2/2+\mu=r$,
	the change of variables $x=e^z$ gives, in $\mathcal D'(L_1,R_1)$,
	\begin{align*}
		\widetilde\cL\widetilde V
		&=r e^z V_x^{[k]}(t_2,e^z)
		  +\frac{\sigma^2}{2}(\log)_\#\bigl[x\,\lambda_k(\dd x)\bigr].
	\end{align*}
Here $(\log)_\#$ denotes push-forward under $x\mapsto\log x$; when $V^{[k]}$ is twice
	differentiable, the measure term has density $e^{2z}V_{xx}^{[k]}(t_2,e^z)$ with respect to $\dd z$.
Therefore $\widetilde\cL H_k=\nu_k+q_k$ on $(L_1,R_1)$, where
	\begin{align*}
		\nu_k&:=\frac{\sigma^2}{2}e^{-rt_2}
		  (\log)_\#\bigl[x\,\lambda_k(\dd x)\bigr]\big|_{(L_1,R_1)},\\
		q_k(z)&:=e^{-rt_2}r e^z V_x^{[k]}(t_2,e^z)
		  -\widetilde\cL\widehat G_k(t_2,z).
	\end{align*}
The measure $\nu_k$ is finite and nonnegative because $\lambda_k$ is locally finite on
	$(0,\infty)$ and $[e^{L_1},e^{R_1}]$ is compactly contained in $(0,\infty)$;
	the function $q_k$ is bounded on $(L_1,R_1)$.

All hypotheses of Theorem~\ref{th:localized-levelset} hold, and therefore
	 \begin{align*}
  		 \beta_k:=\log b^{[k]}\in C_{\rm loc}^{0,1}((0,d_{k-1})),
 	\end{align*}
 	and hence \eqref{eq:finite-b-lip} follows.

We now identify the weak source.  
On the continuation interior it vanishes, whereas on the stopping interior
	\begin{align*}
 		V^{[k]}=K-x+\Psi_\delta^{[k-1]}.
	\end{align*}
The graph of the locally Lipschitz boundary has two-dimensional Lebesgue measure zero.  
Uniqueness of the weak density in Theorem~\ref{th:R-weak} therefore gives
	\begin{align}
		 a^{[k]}(t,x)=\bigl(rK+\Gamma_\delta^{[k-1]}(t,x)\bigr) \one_{\{x<b^{[k]}(t)\}}
 		\quad\text{a.e.}
 		\label{eq:ak-explicit-new}
	\end{align}
	on the main strip. 
On every later open strip the same identity follows from stagewise reduction,
	using the corresponding lower-right obstacle and boundary identities.  
Applying $\mathcal P_\delta$ and using \eqref{eq:Gamma-weak} yields \eqref{eq:Gamma-general-rec} 
	on each open delayed strip where the lagged time does not cross a deadline. 
The boundary and the lower-level $\Gamma$ are continuous, and the lognormal
 	transition law has no atom at the moving boundary, so the right-hand side is jointly continuous.  
Hence
	\begin{align*}
 		\partial_t\Psi_\delta^{[k]}=(r-\cL)\Psi_\delta^{[k]}-\Gamma_\delta^{[k]}
	\end{align*}
	has a continuous representative.  
Together with the spatial regularity from Theorem~\ref{th:R-weak}, this
	proves $\Psi_\delta^{[k]}\in C^{1,2}$.

Finally, in log-price coordinates
	\begin{align*}
 		\widetilde\Gamma_k(t,z) =e^{-r\delta}\int_{-\infty}^{\beta_k(t+\delta)}
 		\bigl(rK+\widetilde\Gamma_{k-1}(t+\delta,y)\bigr)
		 \rho_\delta(y-z-\mu\delta)\dd y.
	\end{align*}
The integrand multiplied by the boundary indicator is decreasing and has a downward jump of size at least
	$rK$ at $\beta_k(t+\delta)$.  
Gaussian--Stieltjes differentiation therefore gives
	\begin{align}
 		-\partial_z\widetilde\Gamma_k(t,z)\ge e^{-r\delta}rK\,
 		\rho_\delta\bigl(\beta_k(t+\delta)-z-\mu\delta\bigr)>0.
	 \label{eq:Gamma-k-strict-new}
	\end{align}
Comparing the same convolution formula at two times, 
	using the local Lipschitz continuity of $\beta_k$ and
	of $\widetilde\Gamma_{k-1}$, gives local time-Lipschitz continuity of $\widetilde\Gamma_k$, 
	uniformly in $z$.
Differentiating the convolution formula in $z$ places the derivative on the Gaussian kernel, so the same
	comparison shows that $\partial_z\widetilde\Gamma_k$ is locally Lipschitz in time as well, uniformly in $z$.
This propagates all induction hypotheses.  
Later strips reduce to the corresponding lower-right problems by
	Proposition~\ref{prop:stage}, completing the induction.
\end{proof}

\bigskip
\noindent Katsunori Ano\\
Department of Human-centered Data Science\\
Bunkyo Gakuin University\\
Tokyo, Japan\\
E-mail: \texttt{k-ano@bgu.ac.jp}

\end{document}